\documentclass{aastex631}

\shorttitle{Reconstructing Intrinsic FRB Properties from Incomplete Observations} \shortauthors{Hu et al.}
\graphicspath{{./}{figures/}}
\usepackage{CJK}
\usepackage{amsmath}
\let\mathell\ell
\usepackage{threeparttable}
\usepackage{longtable}
\usepackage{booktabs}
\usepackage{hyperref}
\usepackage[all]{hypcap}
\usepackage{CJK}
\makeatletter

\newcommand{\Rmnum}[1]{\expandafter\@slowromancap\romannumeral #1@}
\makeatother

\begin{document}

\begin{CJK*}{UTF8}{gbsn}

\title{Decoding FRB Energetics and Frequency Features Hidden by
Observational Incompleteness}

\author[0000-0002-5238-8997]{Chen-Ran Hu (胡宸然)}
\affiliation{School of Astronomy and Space Science, Nanjing
University, Nanjing 210023, China}

\author[0000-0001-7199-2906]{Yong-Feng Huang (黄永锋)}
\thanks{Email: hyf@nju.edu.cn}
\affiliation{School of Astronomy and Space Science, Nanjing
University, Nanjing 210023, China}
 \affiliation{Key Laboratory of
Modern Astronomy and Astrophysics (Nanjing University), Ministry
of Education, China}

\author[0000-0001-9648-7295]{Jin-Jun Geng (耿金军)}
\affiliation{Purple Mountain Observatory, Chinese Academy of
Sciences,  Nanjing 210023, China}

\author[0000-0002-2191-7286]{Chen Deng (邓晨)}
\affiliation{School of Astronomy and Space Science, Nanjing
University, Nanjing 210023, China}

\author[0000-0002-6189-8307]{Ze-Cheng Zou (邹泽城)}
\affiliation{School of Astronomy and Space Science, Nanjing
University, Nanjing 210023, China}

\author[0009-0000-0467-0050]{Xiao-Fei Dong (董小飞)}
\affiliation{School of Astronomy and Space Science, Nanjing
University, Nanjing 210023,  China}

\author[0000-0002-7372-4160]{Yi-Dan Wang (王一丹)}
\affiliation{State Key Laboratory of Radio Astronomy and Technology, 
NAOC, Chinese Academy of Sciences, Beijing 100101, China}
 \affiliation{University of Chinese Academy of Sciences, Beijing 100049,
China}

\author[0000-0002-3386-7159]{Pei Wang (王培)}
\thanks{Email: wangpei@nao.cas.cn}
\affiliation{State Key Laboratory of Radio Astronomy and Technology, 
NAOC, Chinese Academy of Sciences, Beijing 100101, China}
 \affiliation{Institute for Frontiers in Astronomy and Astrophysics, Beijing
Normal University, Beijing 102206, China}

\author[0000-0001-7943-4685]{Fan Xu (许帆)}
\affiliation{Institute of Space Weather, School of Atmospheric Physics,
Nanjing University of Information Science and Technology, Nanjing 210044,
China}

\author[0000-0003-0721-5509]{Lang Cui (崔朗)}
\affiliation{State Key Laboratory of Radio Astronomy and Technology, 
Xinjiang Astronomical Observatory, CAS, 150 Science 1-Street, Urumqi, 
Xinjiang, 830011, China}
 \affiliation{Xinjiang Key Laboratory of Radio Astrophysics, 150 Science
1-Street, Urumqi 830011, China}

\author[0000-0003-2366-219X]{Song-Bo Zhang (张松波)}
\affiliation{Purple Mountain Observatory, Chinese Academy of Sciences,
Nanjing 210023, China}

\author[0000-0002-6299-1263]{Xue-Feng Wu (吴雪峰)}
\thanks{Email: xfwu@pmo.ac.cn}
\affiliation{Purple Mountain Observatory, Chinese Academy of Sciences,
Nanjing 210023, China}
 \affiliation{School of Astronomy and Space Sciences, University of Science
and Technology of China, Hefei 230026, China}

\begin{abstract}

Fast radio bursts (FRBs) are millisecond-duration extragalactic
radio flashes likely powered by magnetars, yet their radiation mechanism
remains unknown. Limited sensitivity and finite observing bandwidth
inevitably lead to observational truncation, biasing our understanding of
intrinsic burst properties. Assuming Gaussian-like spectra, we present a general inverse-modeling
framework that reconstructs the intrinsic frequency and energy
characteristics of repeating FRBs directly from truncated data, without
spectral profile fitting. In our approach, detected bursts are classified
as in-band (affected only by the sensitivity cutoff) or band-chipped
(affected by both sensitivity and operating-band cutoffs) events. For in-band
events, observed and intrinsic quantities are linked through a set of
equations. For band-chipped bursts, with
spectral peaks possibly outside the telescope's operating band, a
population-based method is used to infer individual burst properties from the
statistical properties of the entire sample. Applied to 2,223 bursts from
FRB 20121102A, it is found that
intrinsically energetic bursts tend to have narrower spectra than weak
ones. We further quantify, for the first time, the number of out-of-band
bursts, and reveal distinct frequency-evolution behaviors across active
periods and frequency bands. Comparisons between reconstructed and
original samples show that the sensitivity cutoff barely affects burst
energy but biases the observed bandwidth, whereas the operating-band
cutoff may cause severe energy leakage and bandwidth underestimation,
suggesting that the energy release of some repeaters may be
underestimated, with potential implications for the energy supply beyond
the magnetar magnetosphere. Our methodology transforms incomplete
archival observations into physically meaningful probes, bridging
instrumental readouts and intrinsic FRB physics.

\end{abstract}



\section{Introduction}
\label{sec1:Introduction}

Fast radio bursts (FRBs) are cosmic explosions that release
immense energy at radio wavelengths on a millisecond timescale.
They are commonly classified as either non-repeating (one-off) or
repeating sources. Among the latter, FRB 20121102A, the first
reported repeating FRB source \citep{2014ApJ...790..101S, 2},
has the longest monitoring baseline and has been observed extensively
\citep{11,4,2018ApJ...863....2G,7,40,9,6,5,1,3,2023MNRAS.519..666J}.
However, during the observing process,
the detected bursts are often incompletely recorded due to
the limited telescope sensitivity and the finite operating
band \citep{3,10}, which seriously restricts our exploration of the
underlying physics. The large number of bursts from the same
repeater provides us with an opportunity to reconstruct a
``complete'' sample. Utilizing classical observational data from the 305-m
William E. Gordon Telescope at the Arecibo Observatory
(Arecibo) \citep{3}, the Robert C. Byrd Green Bank Telescope
(GBT) \citep{9} and the Five-hundred-meter Aperture Spherical radio
Telescope (FAST) \citep{1}, we have tried to correct for the
observational cutoff effects on the bursts from FRB 20121102A
through an inverse modeling process. Our reconstruction involves
using mathematical and statistical techniques to recover the
genuine burst properties from the observational data. Based on the
reconstructed bursts, intrinsic energetics and frequency features
are then analyzed.

The methodology presented in this work, reconstructing
intrinsic burst parameters from imperfect observational
characteristics, can be broadly applied to a wide range of FRB
samples. It is not only applicable to incomplete bursts that are
detected, but also offers a way to estimate the number
of bursts that remain entirely unobserved (i.e., those falling
outside the telescope's operating band during monitoring). This
approach has implications for refining our understanding of FRB
energetics, event rates, waiting times, and pulse morphology,
thereby bringing us closer to the underlying burst mechanisms and
emission processes.

\setcounter{figure}{0}
\makeatletter
\renewcommand*{\fnum@figure}{{\normalfont\bfseries \figurename~\thefigure}}
\makeatother

\begin{figure}[!htbp]
\centering
\includegraphics[width=1\textwidth]{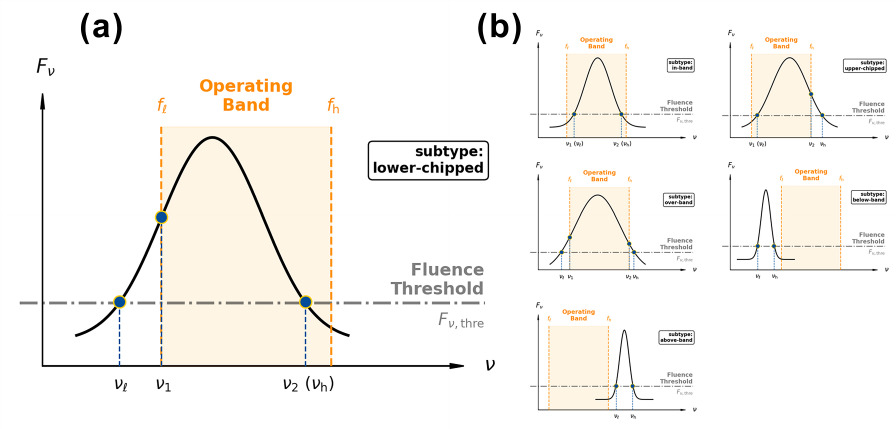}
\caption{(a) A schematic illustration of the observational cutoff
of FRBs on the $F_\nu$-$\nu$ plane, where $F_\nu$ is the monochromatic
fluence at frequency $\nu$. The solid black curve represents the
intrinsic spectrum of the burst. The dash-dotted gray line represents the fluence
threshold $F_{\nu,{\rm{thre}}}$ of the telescope. The orange
shaded region shows the operating band of the telescope, with lower and
upper boundaries denoted as $f_{\rm{\mathell}}$ and $f_{\rm{h}}$,
respectively. The spectrum of the burst intersects with
$F_\nu=F_{\nu, {\rm{thre}}}$ at two frequencies,
$\nu_{\rm{\mathell}}$ and $\nu_{\rm{h}}$, which mark proper lower and
upper spectral boundaries of the burst. Under the influence of both the
sensitivity cutoff and operating-band cutoff, the observed lower and 
upper spectral boundarie are denoted as $\nu_1$ and $\nu_2$. 
(b) Other five possible cases, depending on the
relative ordering of $\nu_{\rm{\mathell}}$, $\nu_{\rm{h}}$,
$f_{\rm{\mathell}}$, and $f_{\rm{h}}$. }
 \label{Figure1}
\end{figure}


\section{Observational Cutoff Effect}
\label{sec2:Observational Cutoff Effect}

Observational cutoff primarily arises from the fluence threshold
and band limitations \citep{2021ApJ...920L..18A, 2026ApJ...998...84L}
of the telescope, leading to incompleteness in data acquisition.
To investigate such observational cutoff effect, we focus on FRB
20121102A, the first known repeating FRB source.
Since its discovery, more than two thousand
bursts have been detected across a frequency range of $\sim$1--8 GHz
\citep{24}. The data sets analyzed in this work include 478 bursts from
the Arecibo telescope at 1.15--1.73 GHz \citep{3}, 93 bursts from GBT
at 4--8 GHz \citep{9}, and 1652 bursts from the FAST telescope at 1--1.5
GHz \citep{1}. The source resides in a low-metallicity irregular dwarf
galaxy at a redshift of $z = 0.193$, corresponding to a luminosity
distance $D_L = 972$ Mpc \citep{2017ApJ...834L...7T}, and is
spatially coincident with a luminous persistent radio source
\citep{2017Natur.541...58C, 2017ApJ...844...95K, 2017ApJ...834L...8M, 4}.
The spectra of the bursts from FRB 20121102A
are found to be well described by the Gaussian function
\citep{3,7,11,12,13,25}. The assumption of Gaussian spectra has
been adopted in previous studies to quantify the energy loss of FRBs
caused by the limited observing bandwidth, either through simulations
\citep{2021ApJ...920L..18A} or direct spectral fitting
\citep{2026ApJ...998...84L}.
Our quantitative analysis based on the
Bayesian Information Criterion (BIC) \citep{43} also shows that the
Gaussian model provides a better fit to the FRB spectra over the
power-law and cutoff power-law models (Appendix \ref{appendix A:Spectrum Profile of FRBs}).
Figure \ref{Figure1} illustrates the impact of observational cutoff on
the burst, which can be caused by either sensitivity cutoff or
operating-band cutoff. Due to the existence of the fluence
threshold $F_{\nu,\rm{thre}}$, the lower wings of the burst
spectrum will be completely omitted in observations since they are
beneath the threshold (i.e., sensitivity cutoff). Also, the
operating-band cutoff leads to inaccurate estimations of the burst
energy and bandwidth (e.g., the proper spectral boundaries
$\nu_{\rm{\mathell}}$ and $\nu_{\rm{h}}$ are cutoff to the
observed spectral boundaries $\nu_1$ and $\nu_2$).

Based on Figure \ref{Figure1}, we can perform theoretical modeling of the
effect of observational cutoff on the FRB spectrum. The frequencies at which
the spectrum intersects with the horizontal line of $F_{\nu,\rm{thre}}$ are
denoted as $\nu_{\rm{\mathell}}$ and $\nu_{\rm{h}}$. The existence of the
operating-band cutoff implies that the observed spectrum will be
in a frequency range of $\nu_1 < \nu < \nu_2$, with the boundary
frequencies determined by
\begin{eqnarray}
\label{eq1}
\nu_1=\max\left(f_{\rm{\mathell}}, \nu_{\rm{\mathell}}\right),
\end{eqnarray}
\begin{eqnarray}
\label{eq2}
\nu_2=\min\left(f_{\rm{h}}, \nu_{\rm{h}}\right).
\end{eqnarray}
Here $f_{\rm{\mathell}}$ and $f_{\rm{h}}$ represent the lower and
upper limits of the operating band, respectively.

The redshifted intrinsic burst spectrum can be described
by a Gaussian function as \citep{13}
\begin{eqnarray}
\label{eq3}
F_\nu=\frac{F}{\sqrt{2\pi}\sigma_\nu}{\rm{e}}^{-\frac{\left(\nu-\nu_{\rm{p}}\right)^2}{2\sigma_\nu^2}},
\end{eqnarray}
where $F_\nu$ is the monochromatic fluence at frequency $\nu$, $F
= \int_{0}^{\infty} F_{\nu} \, \rm{d}\nu$ is the integrated
fluence, $\nu_{\rm{p}}$ is the peak frequency, and
$\sigma_\nu$ characterizes the full width at half
maximum (FWHM) of the spectrum.

According to Figure \ref{Figure1}, we can easily derive the following
equations that connect the observed spectral parameters with the
intrinsic ones,
\begin{eqnarray}
\label{eq4}
\nu_{\rm{p}}=\frac{\nu_{\rm{\mathell}}+\nu_{\rm{h}}}{2},
\end{eqnarray}
\begin{eqnarray}
\label{eq5} F_\nu\left(\nu_{\rm{\mathell}}\right) =
F_\nu\left(\nu_{\rm{h}}\right) = F_{\nu,{\rm{thre}}},
\end{eqnarray}
\begin{eqnarray}
\label{eq6}
F_{\nu,{\rm{obs}}}^{\rm{equiv}}\delta\nu_{\rm{obs}} =
\int_{\nu_1}^{\nu_2}\frac{F}{\sqrt{2\pi}
\sigma_\nu}{\rm{e}}^{-\frac{\left(\nu-\nu_{\rm{p}}\right)^2}{2\sigma_\nu^2}}{\rm{d}}
\nu.
\end{eqnarray}
Here $F_{\nu,{\rm{obs}}}^{\rm{equiv}}$ represents the
frequency-equivalent fluence derived from the observation, and
$\delta\nu_{\rm{obs}}= \nu_2-\nu_1$ represents the observed
bandwidth of the burst. For convenience, the observed burst energy
is often averaged over the observed burst bandwidth (e.g.,
\citet{3}) or the telescope's operating band (e.g., \citet{1}),
yielding the commonly used frequency-equivalent fluence of
$F_{\nu, {\rm{obs}}}^{\rm{equiv}}$. Equation (\ref{eq6})
delineates that only a fraction of $F$ is detectable under the
impact of sensitivity and operating-band cutoffs. Note that
Equation (\ref{eq6}) allows for the substitution of
$\delta\nu_{\rm{obs}}$ with the width of the operating band
$\delta f$, depending on whether $F_{\nu,{\rm{obs}}}^{\rm{equiv}}$
is derived from $\delta\nu_{\rm{obs}}$ or $\delta f$ \citep{14}.

Combining Equations (\ref{eq3}), (\ref{eq4}),
(\ref{eq5}) and (\ref{eq6}), we obtain
\begin{eqnarray}
\label{eq7}
\sqrt{2\pi}F_{\nu,{\rm{thre}}}{\sigma_\nu}{\rm{e}}^{\left(\frac{\delta\nu}{2\sqrt2\sigma_\nu}\right)^2}\left[\Phi\left(\frac{\nu_2-\nu_{\rm{p}}}{\sigma_\nu}\right)-\Phi\left(\frac{\nu_1-\nu_{\rm{p}}}{\sigma_\nu}\right)\right]=F_{\nu,{\rm{obs}}}^{\rm{equiv}}\delta\nu_{\rm{obs}},
\end{eqnarray}
where $\Phi$ is the cumulative function of a standard Gaussian distribution,
and $\delta\nu=\nu_{\rm{h}}-\nu_{\rm{\mathell}}$ stands for the proper
bandwidth. It is worth noting that both $\delta\nu$ and
$\delta\nu_{\rm{obs}}$ depend on the telescopes used in the
observation. Once $\nu_{\rm{p}}$ is known, the remaining spectral parameters
of the intrinsic spectrum can be derived from observations by solving the
above equation. We have utilized the algorithm of adaptive gradient
descent (AGD) \citep{15,16,17} to numerically solve the above
equation (Appendix \ref{appendix B:Numerical Solution of the Gaussian Spectrum}). From Equations
(\ref{eq3}), (\ref{eq4}), and (\ref{eq5}), we can also calculate the
integrated fluence as
\begin{eqnarray}
\label{eq8}
 F = \int_{0}^{\infty} F_{\nu} \, \rm{d}\nu = \sqrt{2\pi}
F_{\nu,{\rm{thre}}}\sigma_\nu{\rm{e}}^{\left(\frac{\delta\nu}{2\sqrt2\sigma_\nu}\right)^2}.
\end{eqnarray}
By substituting the solution of Equation (\ref{eq7}) into Equation
(\ref{eq8}), we can further calculate the proper energy of the burst as
\begin{eqnarray}
\label{eq9} E = ({10}^{-20}{\rm{erg}})\frac{4\pi}{1+z}
\left(\frac{F}{{\rm{Jy}}\cdot{\rm{ms}}\cdot{\rm{MHz}}} \right)
\left(\frac{D_L}{\rm{cm}}\right)^2,
\end{eqnarray}
where $z$ and $D_L$ correspond to the redshift and luminosity
distance of the FRB source, respectively.
For FRB 20121102A, we adopt $z = 0.193$ and $D_L = 972$ Mpc
\citep{2017ApJ...834L...7T}.

In brief, we choose to establish a set of equations
involving the directly observed quantities to reconstruct the
intrinsic burst parameters, rather than performing spectral
fitting on the raw observational data. Spectral fitting loses
robustness when dealing with weak bursts. Additionally, for those
bursts with spectral peaks falling outside the operating band, it
is impossible to reconstruct the intrinsic energy through normal
spectral fitting. However, we can estimate their energetics
through population-based methods, which will be discussed in
Section \ref{sec4:Operating-band Cutoff}.


\newpage
\section{Sensitivity Cutoff}
\subsection{Reconstruction and Analysis of In-band Bursts}
\label{sec3.1:Reconstruction and Analysis of In-band Bursts}

\begin{figure}[!htbp]
\centering
\includegraphics[width=1\textwidth]{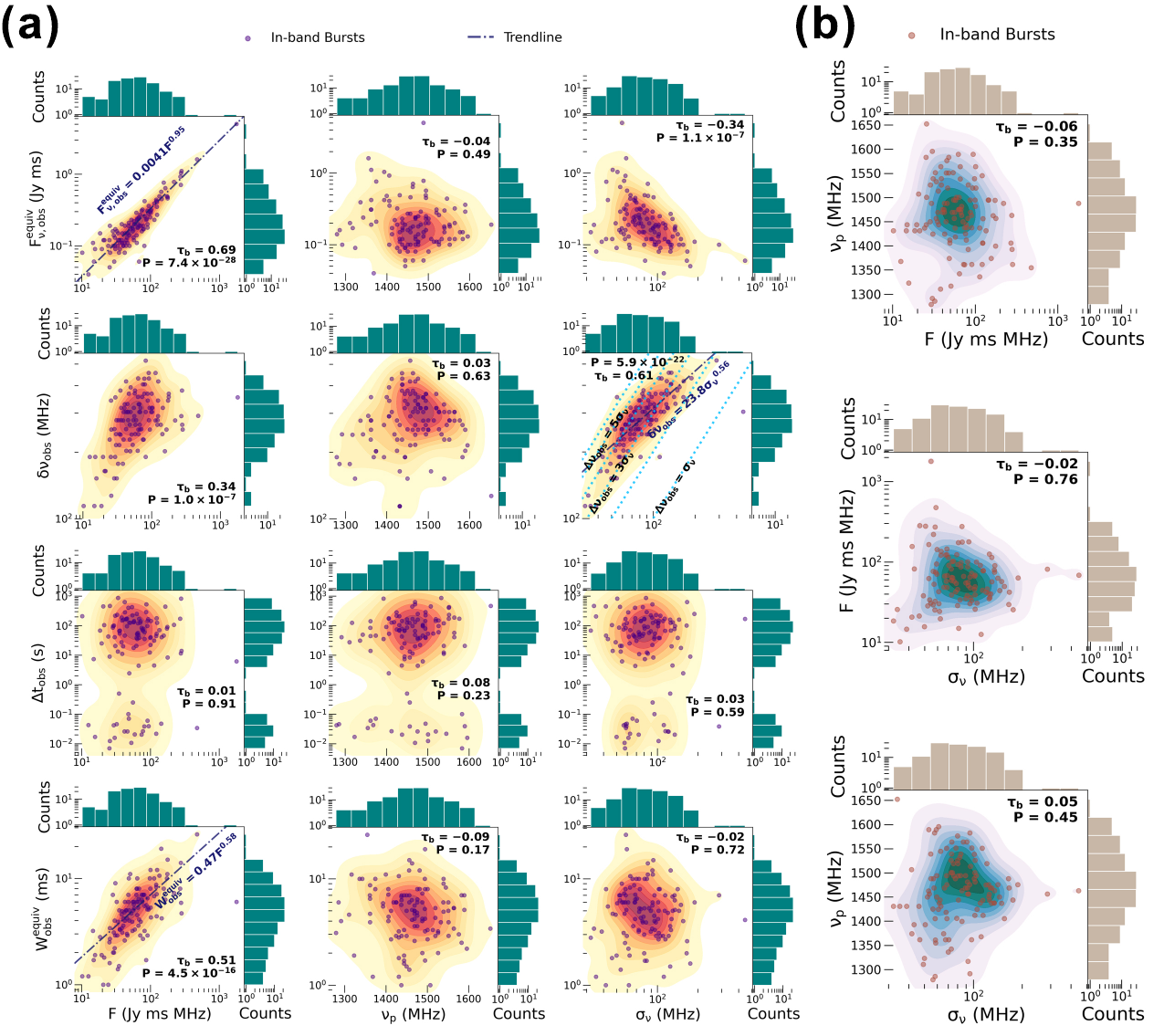}
\caption{Comparison between the observed quantities and the
derived spectral quantities  for the in-band bursts of FRB 20121102A
observed by Arecibo between December 2015 and
October 2016 (data from \citealt{3}). Here, the observed
quantities, i.e. the
frequency-equivalent fluence $(F_{\nu, {\rm{obs}}}^{\rm{equiv}} )$,
observed bandwidth $( \delta\nu_{\rm{obs}})$,
waiting time $(\Delta{t}_{\rm{obs}})$, and
boxcar-equivalent pulse width of the burst $ ( W_{\rm{obs}}^{\rm{equiv}})$ are adopted from \citet{3},
while the derived quantities i.e., the intrinsic integrated fluence $(F)$,
true peak frequency $(\nu_{\rm{p}})$, and Gaussian
spectral width (full width at half maximum, FWHM, $\sigma_\nu $) of the burst are
obtained through our spectral reconstruction
method (Section \ref{sec3.1:Reconstruction and Analysis of In-band
Bursts}). (a) The pairwise comparison between
$\left(F_{\nu, {\rm{obs}}}^{\rm{equiv}}, \delta\nu_{\rm{obs}},
\Delta{t}_{\rm{obs}}, W_{\rm{obs}}^{\rm{equiv}}\right)$ and
$\left(F, \nu_{\rm{p}}, \sigma_\nu\right)$. Each main plot
illustrates the 2D distribution of the in-band bursts. The
contour lines correspond to the kernel density estimation (KDE) of
the data points. 1D histograms are also plotted along the axes.
The correlation coefficient ($\tau_{\rm{b}}$) is calculated for
each pair of quantities by using the Kendall tau method. The
corresponding coincidence probability ($P$) is also shown.
A trendline (dash-dotted line) is plotted when the
correlation is statistically significant (i.e., $|\tau_{\rm{b}}| >
0.5$ and $P < 0.05$). (b) A pairwise plot
for the three quantities of $\left(F, \nu_{\rm{p}},
\sigma_\nu\right)$.}
 \label{Figure2}
\end{figure}

During nearly 59 hours of observations between December 2015 and
October 2016, Arecibo detected 478 bursts from FRB 20121102A. The
observations were performed with a fluence threshold of $\sim$
0.057 Jy ms in an operating band of 1.15--1.73 GHz, providing the key
measurements of the observed lower and upper spectral boundaries $\nu_1$
and $\nu_2$, the frequency-equivalent fluence $F_{\nu,
{\rm{obs}}}^{\rm{equiv}}$, and the boxcar-equivalent pulse width
${W_{\rm{obs}}^{\rm{equiv}}}$ for the bursts \citep{3}.
For a specific burst, because the fluence sensitivity depends on
the pulse width, we have $F_{\nu,{\rm{thre}}}=
0.057\sqrt{{W_{\rm{obs}}^{\rm{equiv}}}/4\, \rm{ms}}$ Jy ms
\citep{29,30,3,13}.

We have reconstructed the intrinsic burst properties of this
Arecibo sample. All bursts are inevitably affected by the sensitivity
cutoff. Among them, 117 events (25\% of the sample)
are in-band, i.e., their observed spectral
boundaries satisfy $f_{\rm{\mathell}} < \nu_1 < \nu_2 < f_{\rm{h}}$ due to $f_{\rm{\mathell}} <
\nu_{\rm{\mathell}} < \nu_{\rm{h}} < f_{\rm{h}}$, meaning that they are free from the
operating-band cutoff (as illustrated in Figure \ref{Figure1}). 
These in-band bursts will be corrected for the
sensitivity cutoff effect in this section. The other 361 bursts (75\% of
the sample) are band-chipped,
meaning that their observed spectral boundaries reach one
or both edges of the telescope's operating band. They can be further
classified into three subtypes: lower-chipped ($\nu_1 = f_{\rm{\mathell}}$, since
$\nu_{\rm{\mathell}} < f_{\rm{\mathell}} < \nu_{\rm{h}} < f_{\rm{h}}$), upper-chipped ($\nu_2 = f_{\rm{h}}$, since
$f_{\rm{\mathell}} < \nu_{\rm{\mathell}} < f_{\rm{h}} < \nu_{\rm{h}}$), and over-band ($\nu_1 = f_{\rm{\mathell}}$
and $\nu_2 = f_{\rm{h}}$, with $\nu_{\rm{\mathell}} < f_{\rm{\mathell}} < f_{\rm{h}} < \nu_{\rm{h}}$). Again, for
these band-chipped bursts, a direct estimation of their intrinsic energy
is challenging, especially when the spectral peak falls outside the
operating band. Nevertheless, their energies can be estimated through
population-based methods. Moreover, for bursts that lie completely outside
the operating band (i.e., out-of-band bursts: below-band, $\nu_{\rm{\mathell}} <\nu_{\rm{h}}
< f_{\rm{\mathell}} < f_{\rm{h}}$; above-band, $f_{\rm{\mathell}} < f_{\rm{h}} < \nu_{\rm{\mathell}} < \nu_{\rm{h}}$), the same
population-based approach allows us to estimate their number and thereby
correct the burst rate of the source. These population-based analyses are
presented in Section \ref{sec4:Operating-band Cutoff}.

For in-band bursts, we have $\nu_{\rm{p}} =
\nu_{\rm{p,obs}} = \frac{\nu_1+\nu_2}{2}$. Utilizing Equations
(\ref{eq7}) and (\ref{eq8}), we can obtain the spectral
solution of $\left(F,\nu_{\rm{p}},\sigma_\nu\right)$ for each
in-band burst in the Arecibo sample.
The derived $\left(F,\nu_{\rm{p}},\sigma_\nu\right)$ and the
observed physical quantities
$\left(F_{\nu,{\rm{obs}}}^{\rm{equiv}}, \delta\nu_{\rm{obs}},
\Delta{t}_{\rm{obs}},W_{\rm{obs}}^{\rm{equiv}}\right)$ are
compared pairwise in Figure \ref{Figure2}(a). Here,
$\Delta{t}_{\rm{obs}}$ is the waiting time and
$W_{\rm{obs}}^{\rm{equiv}}$ corresponds to the boxcar-equivalent
pulse width. The Kendall tau method \citep{19,20,21,22} is employed
to calculate the correlation coefficient ($\tau_{\rm{b}}$) between
the quantities and the corresponding hypothesis test results
($P$). A correlation is deemed to exist when the absolute value of
$\tau_{\rm{b}}$ exceeds 0.5, with positive and negative values
indicating positive and negative correlations, respectively. The
correlation is significant when $P$ is below 0.05. For the
statistically significant correlations, corresponding trendlines
(dash-dotted lines) are plotted in the panels. In Figure
\ref{Figure2}(a), the correlation between
$F_{\nu,{\rm{obs}}}^{\rm{equiv}}$ and $F$ can be approximated as
$F_{\nu,{\rm{obs}}}^{\rm{equiv}} = 0.0041F^{0.95}$, which
essentially indicates $F_{\nu,{\rm{obs}}}^{\rm{equiv}} \propto
\frac{F} {\delta f}$. It means that for in-band bursts, the
method of estimating the burst energy by adopting the width of the
operating band as the burst bandwidth  is viable \citep{14}.
Interestingly, on the $W_{\rm{obs}}^{\rm{equiv}}$ vs. $F$ plane,
the correlation is approximately $F \propto
{W_{\rm{obs}}^{\rm{equiv}}}^2$. Considering that
$F_{\nu,{\rm{obs}}}^{\rm{equiv}} \propto \frac{F}{\delta f}$ and
$F_{\nu,{\rm{obs}}}^{\rm{equiv}} = S_\nu
W_{\rm{obs}}^{\rm{equiv}}$ for in-band bursts, where
$S_\nu$ represents the peak flux density, we have $S_\nu \propto
W_{\rm{obs}}^{\rm{equiv}}$. It indicates that bursts with a higher
energy tend to have a longer duration.

\begin{figure}[!htbp]
\centering
\includegraphics[width=1\textwidth]{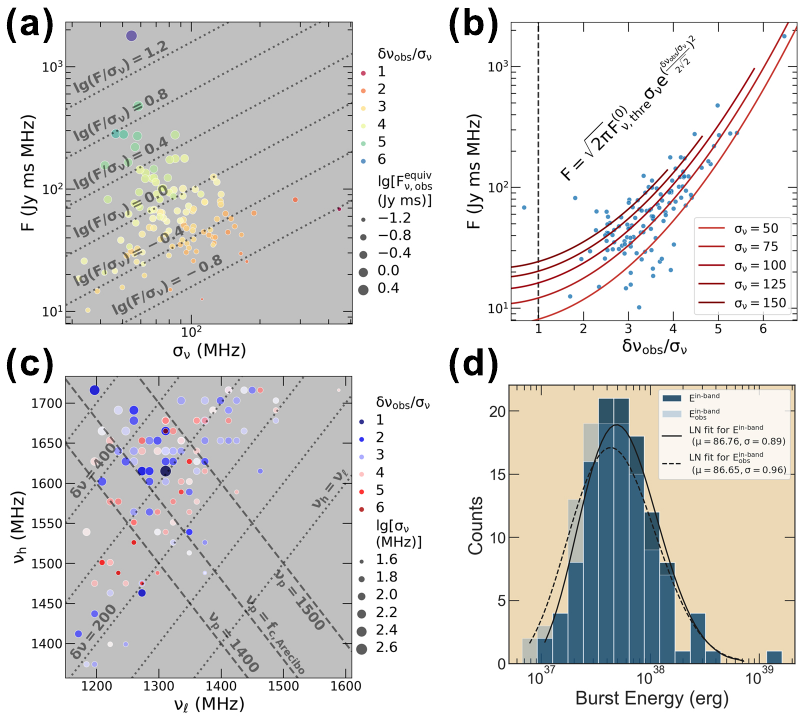}
\caption{Effects of sensitivity cutoff and the characteristics of
reconstructed in-band bursts of FRB 20121102A observed by
Arecibo  between December 2015 and October 2016 (data from
\citealt{3}). The observed quantities such as $F_{\nu,
{\rm{obs}}}^{\rm{equiv}}$ and $\delta\nu_{\rm{obs}}$, which denote
the frequency-equivalent fluence and observed bandwidth of the
burst, respectively, are adopted from \citet{3}. The derived
quantities such as $F$ and $\sigma_\nu$, which denote the
intrinsic integrated fluence and Gaussian spectral width (FWHM) of
the burst, respectively, are obtained through our spectral
reconstruction method (Section \ref{sec3.1:Reconstruction and
Analysis of In-band Bursts}). (a) The distribution of in-band
bursts on the $F$-$\sigma_\nu$ plane. The data points are
color-coded based on $\frac{\delta\nu_{\rm{obs}}}{\sigma_\nu}$ and
their sizes are proportional to
$\lg\left[F_{\nu,{\rm{obs}}}^{\rm{equiv}} \left(\rm{Jy\
ms}\right)\right]$. Two dot sequences serving as legends are 
shown on the right of the main plot, one for the color encoding and the 
other for the size encoding. (b) The distribution of in-band bursts on 
the $F$-$\frac{\delta\nu_{\rm{obs}}}{\sigma_\nu}$ plane.
$F_{\nu,\rm{thre}}^{(0)}$ denotes a fiducial value of the
fluence threshold $F_{\nu,\rm{thre}}$ (e.g., based on the typical
pulse width of the Arecibo sample). (c) The distribution of
in-band bursts on the $\nu_{\rm{h}}$-$\nu_{\rm{\mathell}}$ plane,
with data points color-coded based on
$\frac{\delta\nu_{\rm{obs}}}{\sigma_\nu}$ and scaled in size
proportional to $\lg\left[\sigma_\nu
\left(\rm{MHz}\right)\right]$. The color and size encodings are 
indicated by two dot sequences on the right of the main plot, 
respectively. For in-band bursts, the
proper spectral boundaries equal the observed ones
($\nu_{\rm{\mathell}} = \nu_1$ and $\nu_{\rm{h}} = \nu_2$), and
hence the true peak frequency and proper bandwidth are simply
$\nu_{\rm{p}} = \nu_{\rm{p,obs}} = \left(\nu_1+\nu_2\right)/2$ and
$\delta\nu = \delta\nu_{\rm{obs}} = \nu_2-\nu_1$.
$f_{\rm{c,Arecibo}}=1440$ MHz stands for the central frequency of
the Arecibo operating band. (d) Observed energy
distribution ($E^{\rm{in\text{-}band}}_{\rm{obs}}$) as compared
with the intrinsic energy distribution ($E^{\rm{in\text{-}band}}$)
for in-band bursts. Each distribution is fitted with a shifted
log-normal function (with fitted parameters $\mu$ and $\sigma$).
$E^{\rm{in\text{-}band}}$ is computed from the reconstructed
Gaussian spectra, while $E^{\rm{in\text{-}band}}_{\rm{obs}}$ is
derived directly from $F_{\nu,{\rm{obs}}}^{\rm{equiv}}$ and
$\delta\nu_{\rm{obs}}$.}
 \label{Figure3}
\end{figure}

We have further analyzed the reconstructed in-band bursts.
The results are shown in Figure \ref{Figure3}. We define
$\frac{F}{\sigma_\nu}$ as the intrinsic frequency-averaged fluence and
$\frac{\delta\nu_{\rm{obs}}}{\sigma_\nu}$ as the
observed-to-intrinsic bandwidth ratio. In Figure \ref{Figure3}(a), both
$F_{\nu,{\rm{obs}}}^{\rm{equiv}}$ and
$\frac{\delta\nu_{\rm{obs}}}{\sigma_\nu}$ increase as $\frac{F}
{\sigma_\nu}$ increases. This is because a higher intrinsic
frequency-averaged fluence of the burst results in an elevated
fraction of its spectrum emerging above the telescope's fluence threshold,
leading to a higher observed-to-intrinsic bandwidth ratio. In
fact, from Equation (\ref{eq8}), adopting a fiducial value
$F_{\nu,\rm{thre}}^{(0)}$ for $F_{\nu,\rm{thre}}$ (e.g., based on
the typical pulse width of the Arecibo sample; \citealt{3}), we can easily derive
\begin{eqnarray}
\label{eq10}
F/\sigma_\nu=\sqrt{2\pi}F_{\nu,{\rm{thre}}}^{(0)}
{\rm{e}}^{\left(\frac{\delta\nu/\sigma_\nu}{2\sqrt2}\right)^2}.
\end{eqnarray}
For in-band bursts, $\delta\nu_{\rm{obs}}$ is identical to
$\delta\nu$. The correlation between $F$ and
$\frac{\delta\nu_{\rm{obs}}} {\sigma_\nu}$ (Figure
\ref{Figure3}(b)) can then be well interpreted by Equation
(\ref{eq10}). In Figure \ref{Figure3}(b),
intrinsically high-energy bursts appear predominantly on the
curves with smaller $\sigma_\nu$, suggesting that intrinsically
energetic bursts tend to have narrower spectra. Intrinsically
low-energy bursts, by contrast, are scattered across a wide range
of $\sigma_\nu$, showing no clear dependence on the spectral
width. When $\frac{\delta\nu_{\rm{obs}}} {\sigma_\nu} < 1$ (the region 
to the left of the dashed line), bursts with lower energies are difficult
to observe, which explains why this region is devoid of detected
intrinsically low-energy bursts.

In fact, intrinsically energetic bursts, which tend to
possess narrow intrinsic spectral widths $\sigma_\nu$ (i.e.,
FWHM), can nevertheless exhibit a broad observed bandwidth
$\delta\nu_{\rm{obs}}$. This is directly illustrated in Figure
\ref{Figure3}(c), where the small and red dots (which mark the
high-energy bursts in the upper part of Figure \ref{Figure3}(a))
can appear near the dotted lines that correspond to larger
$\delta\nu_{\rm{obs}}$. This occurs because these bursts have a
higher spectral energy (i.e., more energy per unit frequency), so
the lower wings of their spectra remain above the telescope's
fluence threshold out to frequencies far from the spectral peak,
making the observed bandwidth significantly larger than the
intrinsic spectral width. It can explain the finding that the
observed bandwidths of repeating FRBs are generally narrower than
those of non-repeating FRBs \citep{25,26}, especially considering
the possibility that non-repeating FRBs could be high-energy
events from repeaters \citep{31}. Particularly, when a
non-repeating FRB represents the rarest and most extreme
high-energy burst from a repeater, even if the intrinsic spectral
width is narrow, the observed bandwidth can still be significantly
broadened. Based on this, we argue that the intrinsic spectral
width $\sigma_\nu$ is a more meaningful measure of the
concentration level of energy per unit frequency, compared to the
observed bandwidth $\delta\nu_{\rm{obs}}$, as the latter
phenomenologically characterizes the observed spectral range well.
This is somewhat similar to our definition of the temporal width
of a burst (i.e., the spectral width is the FWHM of the burst in
the frequency domain, while the pulse width is the FWHM in the
time domain). The observed narrow bandwidth can also be attributed
to the operating-band cutoff, a point that will be discussed in
detail in subsequent sections.

In Figure \ref{Figure3}(c), we see that bursts with
a larger $\sigma_\nu$ are predominantly distributed around the
dashed line of $\nu_{\rm{p}}=f_{\rm{c,Arecibo}}$ (the central
frequency of Arecibo operating band). This is because the bursts
with much larger $\sigma_\nu$ that are far from the dashed line of
$\nu_{\rm{p}}=f_{\rm{c,Arecibo}}$ are subject to the
operating-band cutoff and therefore are not included in the sample
of in-band bursts.

In Figure \ref{Figure3}(c), bursts with a larger $\sigma_\nu$ 
are predominantly distributed around $\nu_{\rm{p}}=f_{\rm{c,Arecibo}}$, 
because those far from this line are affected by the operating-band 
cutoff and thus excluded from the in-band sample.

Adopting the shifted log-normal distribution function
(Appendix \ref{appendix C:Standard/Shifted Log-normal Distribution
Function}), the distributions of intrinsic energy
and observed energy of the in-band bursts are fitted through Markov Chain
Monte Carlo (MCMC) simulations (Figure \ref{Figure3}(d)). We see that the
energy loss caused by the sensitivity cutoff leads to distortion
in the energy distribution. Specifically, compared to the
intrinsic energy, the distribution of the observed energy
presents a peak shift toward a lower value, along with a wider
range in the lower region.
With a total energy loss of
$1-\left(\frac{\Sigma E_{\rm{obs}}}{\Sigma
E}\right)_{\rm{in\text{-}band}}=0.06$, the energy loss due to the
sensitivity cutoff in in-band bursts of the Arecibo sample
is small (particularly for those intrinsically energetic bursts).
Naturally, for telescopes with a better fluence
sensitivity, the energy loss due to the sensitivity cutoff will be
even smaller. In general, sensitivity cutoff predominantly
influences the observed bandwidth of the burst.

\subsection{Bridging Observed and Intrinsic Quantities}
\label{sec3.2:Bridging Observed and Intrinsic Quantities}

To systematically interpret the reconstructed properties presented above, we
establish a quantitative bridge from incomplete observations to the
understanding of their intrinsic physical properties, enabling a direct
comparison between observed and intrinsic parameters. We
define $\frac{F}{\sigma_\nu}$ as the intrinsic frequency-averaged
fluence, and use $\frac{\delta\nu_{\rm{obs}}}{\sigma_\nu}$ to
denote the observed-to-intrinsic bandwidth ratio. According to Equation
(\ref{eq3}), the monochromatic fluence at the peak frequency can be
calculated as $F_{\nu_{\rm{p}}} = \frac{1}{\sqrt{2\pi}}\frac{F}
{\sigma_\nu}$. The observed-to-intrinsic fluence ratio is
$\frac{F_{\nu,{\rm{obs}}}^{\rm{equiv}}}{F/\sigma_\nu}$. The
observed-to-intrinsic energy ratio, denoted as
$\frac{E_{\rm{obs}}}{E}$, is expressed as $\frac{F_{\nu,
{\rm{obs}}}^{\rm{equiv}}\times\delta\nu_{\rm{obs}}}{F}$, where
$E_{\rm{obs}}$ represents the observed burst energy. Combining
Equations (\ref{eq7}) and (\ref{eq8}), we get
\begin{eqnarray}
\label{eq11} \frac{E_{\rm{obs}}}{E} = \Phi\left(\frac{\nu_2 -
\nu_{\rm{p}}}{\sigma_\nu}\right) - \Phi\left(\frac{\nu_1 -
\nu_{\rm{p}}}{\sigma_\nu}\right).
\end{eqnarray}
For in-band bursts, we have $\nu_1=\nu_{\rm{\mathell}}$ and
$\nu_2=\nu_{\rm{h}}$. Equation (\ref{eq11}) can be further
expressed as
\begin{eqnarray}
\label{eq12}
\left(\frac{E_{\rm{obs}}}{E}\right)_{\rm{in\text{-}band}} =
{\rm{erf}}\left(\frac{\delta\nu/\sigma_\nu}{2\sqrt2}\right),
\end{eqnarray}
where ${\rm{erf}}\left(x\right) = \frac{2}{\sqrt\pi}
\int_{0}^{x}{{\rm{e}}^{- \eta^2}{\rm{d}}\eta}$ stands for the
Gauss error function. According to the definition of
$\frac{E_{\rm{obs}}}{E}$, we can further have
\begin{eqnarray}
\label{eq13} \frac{E_{\rm{obs}}}{E} =
\frac{F_{\nu,{\rm{obs}}}^{\rm{equiv}}}{F/\sigma_\nu} \times
\frac{\delta\nu_{\rm{obs}}}{\sigma_\nu}.
\end{eqnarray}
Consequently, from Equations (\ref{eq12}) and
(\ref{eq13}), we derive
\begin{eqnarray}
\label{eq14} \left( \frac{F_{\nu,{\rm{obs}}}^{\rm{equiv}}}{F
/ \sigma_\nu} \right)_{\rm{in\text{-}band}} = \frac{{\rm{erf}}
\left( \frac{\delta\nu/\sigma_\nu}{2\sqrt2}\right)}
{\delta\nu/\sigma_\nu}.
\end{eqnarray}

\begin{figure}[!htbp]
\centering
\includegraphics[width=1\textwidth]{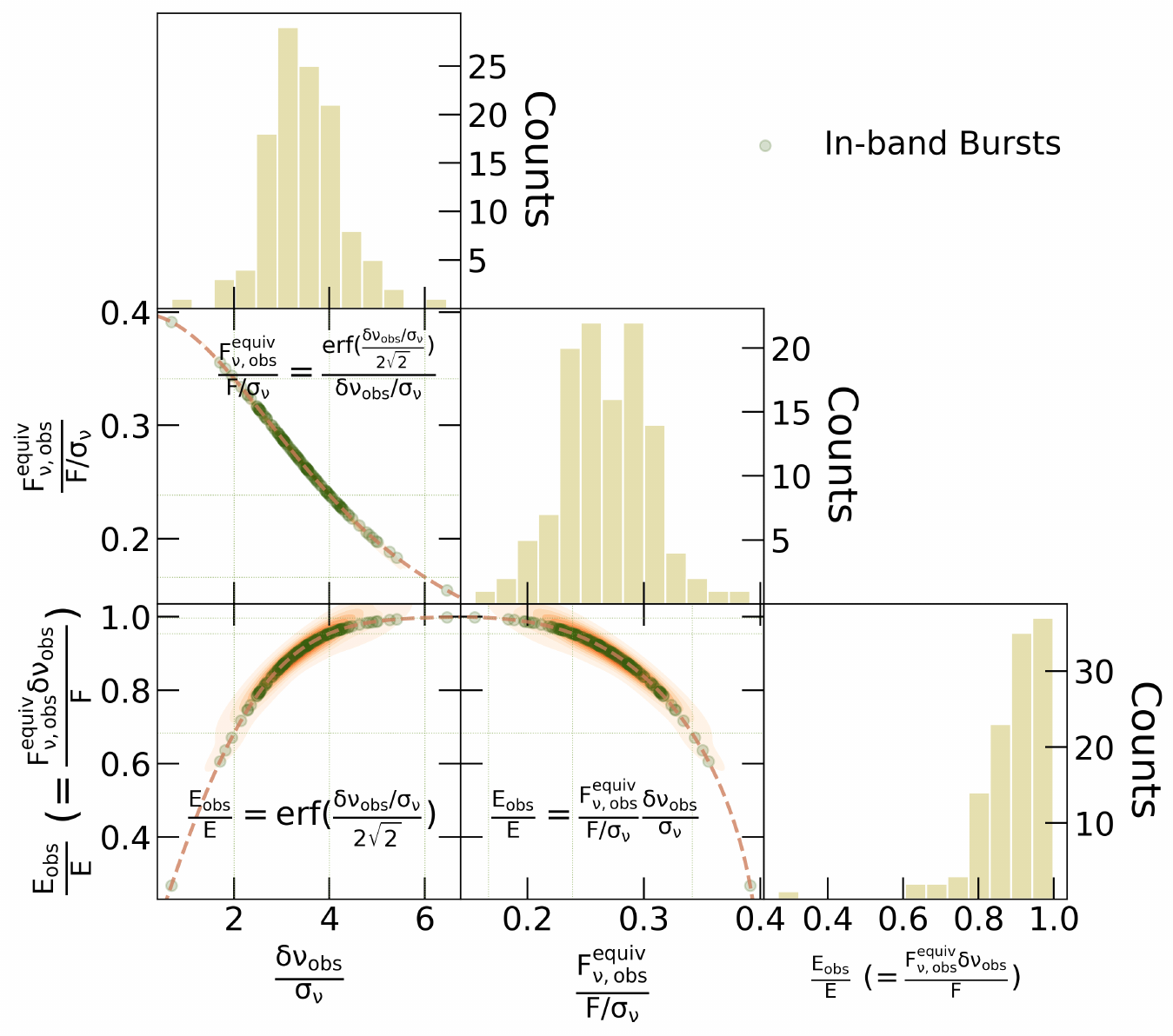}
\caption{The distribution of three observed-to-intrinsic ratios and their
correlations for the in-band bursts of FRB 20121102A observed by
Arecibo between December 2015 and October 2016 (data from \citealt{3}).
The three ratios are the observed-to-intrinsic bandwidth ratio
${\delta\nu_{\rm{obs}}}/{\sigma_\nu}$, the observed-to-intrinsic fluence
ratio ${F_{\nu,{\rm{obs}}}^{\rm{equiv}}}/\left({F/\sigma_\nu}\right)$ and
the observed-to-intrinsic energy ratio ${E_{\rm{obs}}}/{E}$. Here,
$\delta\nu_{\rm{obs}}$ and $\sigma_\nu$ are the observed
bandwidth and Gaussian spectral width (FWHM) of the burst,
$F_{\nu,{\rm{obs}}}^{\rm{equiv}}$ and $F$ are the observed
frequency-equivalent fluence and intrinsic integrated fluence,
and $E_{\rm{obs}}$ and $E$ are the observed and intrinsic burst
energies, respectively. The observed quantities are adopted from
\citet{3}, while the intrinsic ones are derived through our
spectral reconstruction method (Section
\ref{sec3.1:Reconstruction and Analysis of In-band Bursts}).
Each off-diagonal panel shows a pairwise comparison between two ratios,
with contour lines representing the KDE of the data points. The diagonal
panels display the 1D distribution of each ratio.}
 \label{Figure4}
\end{figure}

From Figure \ref{Figure4}, we see that the
vast majority of in-band Arecibo bursts have a
$\frac{E_{\rm{obs}}}{E}$ value above 0.68, corresponding to
$\frac{\delta\nu_{\rm{obs}}}{\sigma_\nu}>2$ (also see the
$\delta\nu_{\rm{obs}}$-$\sigma_\nu$ plot in Figure \ref{Figure2}(a)). The
distribution of $\frac{\delta\nu_{\rm{obs}}}{\sigma_\nu}$ exhibits
a prominent peak around 3, which aligns with
$\frac{E_{\rm{obs}}}{E}=0.87$. It should be emphasized that as
$\frac{E_{\rm{obs}}}{E}$ and
$\frac{\delta\nu_{\rm{obs}}}{\sigma_\nu}$ increase simultaneously,
$\frac{F_{\nu,{\rm{obs}}}^{\rm{equiv}}} {F/\sigma_\nu}$ decreases
in the range of $\left(0,0.4\right]$, indicating a finite energy
being spread over a broader observed bandwidth.

\clearpage


\section{Operating-band Cutoff}
\label{sec4:Operating-band Cutoff}

In the Arecibo sample, nearly 75\% of the bursts are band-chipped,
indicating that they are affected by both the sensitivity cutoff
and the operating-band cutoff (see Figure \ref{Figure1}).
Given the Gaussian-like spectra of FRBs, the occurrence of
the operating band can lead to significant variation of the
spectral index across bursts, which may range from negative to
positive values \citep{2,23}. This can also explain why some
bursts can be well fitted by a power-law spectrum, while others
cannot \citep{26,11}. Additionally, operating-band cutoff can
cause distortion of bursts on the
$\nu_{\rm{h}}$-$\nu_{\rm{\mathell}}$ plane (i.e., the upper
boundary of intrinsic spectrum plotted versus the lower boundary
of intrinsic spectrum). This 2D distribution of
$(\nu_{\rm{\mathell}}, \nu_{\rm{h}})$ is crucial because it
directly characterizes the frequency range of each burst
($\delta\nu=\nu_{\rm{h}}-\nu_{\rm{\mathell}}$). Based on the
observed, distorted $(\nu_{\rm{\mathell}}, \nu_{\rm{h}})$
distribution, we can reconstruct the intrinsic
$(\nu_{\rm{\mathell}}, \nu_{\rm{h}})$ distribution via inverse
modeling. This reconstruction, in turn, enables us to estimate the
intrinsic energies (more precisely, the expected values of the
intrinsic energies) of band-chipped bursts, especially for those
whose spectral peaks fall outside the operating band. Such a
population-based approach, which relies on the statistical
properties of the entire burst sample to infer the properties of
individual band-chipped events, will be elaborated in detail in
the remainder of this section.

\begin{figure}[!htbp]
\centering
\includegraphics[width=0.92\textwidth]{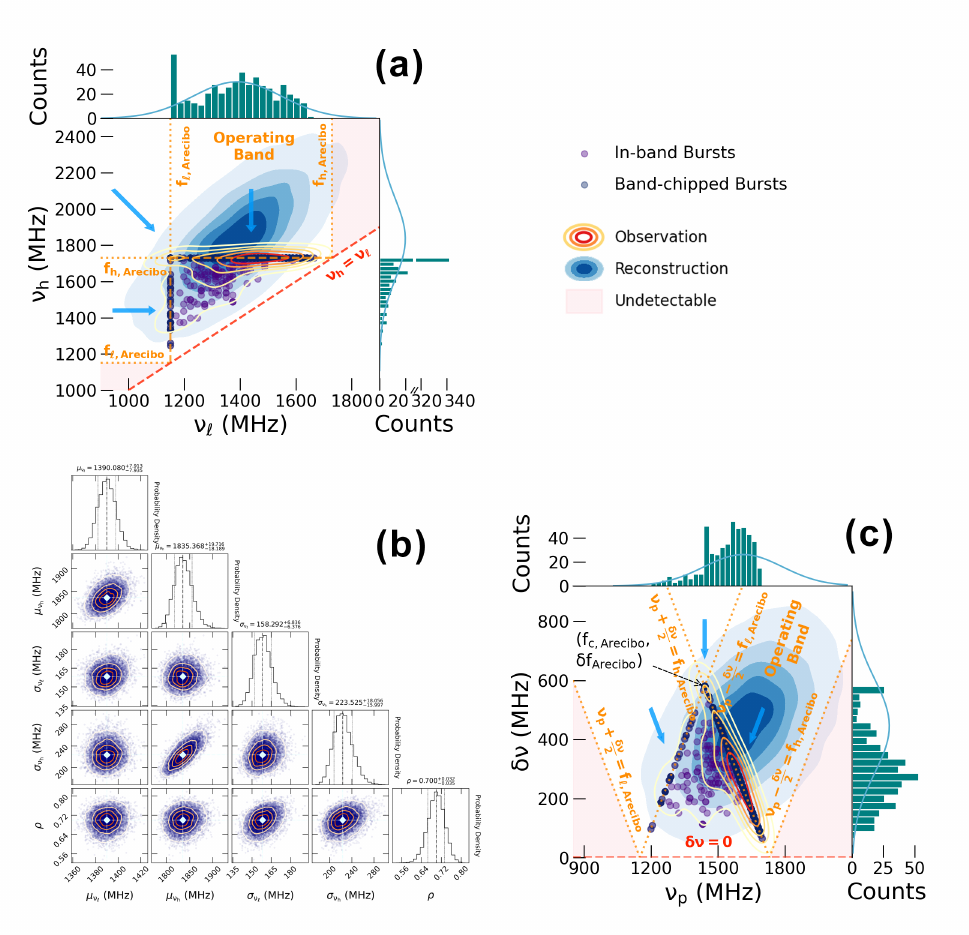}
\caption{Operating-band cutoff effects on the observed 2D
distribution of bursts on the $\nu_{\rm{h}}$-$\nu_{\rm{\mathell}}$ plane
and the reconstructed intrinsic 2D distribution for FRB 20121102A observed
by Arecibo between December 2015 and October 2016 (data from \citealt{3}).
$\nu_{\rm{\mathell}}$ and $\nu_{\rm{h}}$ represent
the proper lower and upper spectral boundaries of the burst,
respectively, while $\nu_{\rm{p}}$ and $\delta\nu$ denote the
true peak frequency and proper bandwidth of the burst.
(a) The distribution of the observed bursts on the
$\nu_{\rm{h}}$-$\nu_{\rm{\mathell}}$ plane. The yellow-orange-red contour
lines represent the KDE of the observed $\left(\nu_1, \nu_2\right)$
distribution, where $\nu_1$ and $\nu_2$ are the observed lower and upper
spectral boundaries of the burst. The blue contour lines show the
corresponding intrinsic $\left(\nu_{\rm{\mathell}}, \nu_{\rm{h}}\right)$
distribution reconstructed via our population-based inverse modeling
method based under the Bayesian inference framework (Section
\ref{sec4:Operating-band Cutoff}).
1D histograms of the observed bursts are displayed along the axes
(green histograms), along with the marginal distributions of the
reconstructed intrinsic $\left(\nu_{\rm{\mathell}}, \nu_{\rm{h}}\right)$
distribution (blue solid lines). The dashed orange lines mark the
operating band limits of Arecibo, $f_{\rm{\mathell,Arecibo}}=1150$ MHz and
$f_{\rm{h,Arecibo}}=1730$ MHz. The region below the red dashed line
($\nu_{\rm{h}} < \nu_{\rm{\mathell}}$) is non-physical, while the pink
region is undetectable ($\nu_{\rm{h}} < f_{\rm{\mathell,Arecibo}}$ or
$\nu_{\rm{\mathell}} > f_{\rm{h,Arecibo}}$).
(b) Constraints on the model parameters of the reconstructed intrinsic
distribution function $N\left(\nu_{\rm{\mathell}}, \nu_{\rm{h}};
\mu_{\nu_{\rm{\mathell}}}, \mu_{\nu_{\rm{h}}},
\sigma_{\nu_{\rm{\mathell}}}, \sigma_{\nu_{\rm{h}}}, \rho\right)$.
(c) The distribution of the observed bursts on the
$\delta\nu$-$\nu_{\rm{p}}$ plane, with graphical elements following
those in panel (a). Compared with panel (a), in the observed 1D
histograms of $\nu_{\rm{p}}$ and $\delta\nu$, we see that the anomalous
accumulation caused by the operating-band cutoff is not
restricted to the edge bin, but can occur in any bin.
This behavior is a natural consequence of the intrinsic 2D
distribution being squeezed into the observed one under the
operating-band cutoff, following the directions indicated by the
blue arrows, which is the same physical effect as in panel (a) but
manifested here on the $\delta\nu$-$\nu_{\rm{p}}$ plane.
The point $\left(f_{\rm{c,Arecibo}},\delta f_{\rm{Arecibo}}\right)$
represents the central frequency and width of Arecibo operating band.}
\label{Figure5}
\end{figure}

Based on Bayesian inference, using all bursts in the Arecibo
sample (i.e., both in-band and band-chipped events),
we have reconstructed the intrinsic 2D Gaussian distribution of
$\left(\nu_{\rm{\mathell}},\nu_{\rm{h}}\right)$
(Appendix \ref{appendix D:Reconstructing the 2D Gaussian Distribution}).
The blue contour lines in Figure \ref{Figure5}(a) depict the
reconstructed intrinsic $\left(\nu_{\rm{\mathell}},\nu_{\rm{h}}\right)$
distribution, while the yellow-orange-red contours represent the
corresponding observed distribution, i.e., the intrinsic distribution
squeezed by the operating-band cutoff effect.
Specifically, the band-chipped bursts are classified into three
categories: lower-chipped ($\nu_{\rm{\mathell}}<f_{\rm{\mathell}}
<\nu_{\rm{h}}<f_{\rm{h}}$), upper-chipped ($f_{\rm{\mathell}}
<\nu_{\rm{\mathell}}<f_{\rm{h}}<\nu_{\rm{h}}$) and over-band
($\nu_{\rm{\mathell}}<f_{\rm{\mathell}}<f_{\rm{h}}<\nu_{\rm{h}}$),
comprising 6\%, 64\%, and 5\% of the Arecibo sample, respectively. These
three types of bursts inherently occupy their respective regions on the
$\nu_{\rm{h}}$-$\nu_{\rm{\mathell}}$ plane, but undergo squeezing along
the directions indicated by the arrows toward the orange dashed lines
(i.e., the operating band limits) and cluster there.
Such a phenomenon arises because the operating-band cutoff
truncates the proper spectral boundaries $\nu_{\rm{\mathell}}$ and/or
$\nu_{\rm{h}}$ into the observed boundaries $\nu_1 = f_{\rm{\mathell}}$
and/or $\nu_2=f_{\rm{h}}$. This effect can
also be observed in the subplots of Figure \ref{Figure5}(a), where the
intrinsic 1D distributions of $\nu_{\rm{\mathell}}$ and
$\nu_{\rm{h}}$ (the blue solid lines) are squeezed and distorted
into the observed 1D distributions (green histograms). 
The reconstructed intrinsic distribution of $\left(\nu_{\rm{\mathell}},
\nu_{\rm{h}}\right)$, denoted as $f_N\left(\nu_{\rm{\mathell}},
\nu_{\rm{h}}\right) \sim{N} \left(\nu_{\rm{\mathell}},
\nu_{\rm{h}}; \mu_{\nu_{\rm{\mathell}}}, \mu_{\nu_{\rm{h}}},
\sigma_{\nu_{\rm{\mathell}}}, \sigma_{\nu_{\rm{h}}}, \rho\right)$,
can be described by
\begin{eqnarray}
\label{eq15} f_N\left(\nu_{\rm{\mathell}},\nu_{\rm{h}}\right)
= \frac{1}{2\pi\sigma_{\nu_{\rm{\mathell}}} \sigma_{\nu_{\rm{h}}}
\sqrt{1-\rho^2}} \exp\left\{-\frac{1}{2\left(1-\rho^2\right)}
\left[\frac{\left(\nu_{\rm{\mathell}} -
\mu_{\nu_{\rm{\mathell}}}\right)^2}
{\sigma_{\nu_{\rm{\mathell}}}^2} -
\frac{2\rho\left(\nu_{\rm{\mathell}} -
\mu_{\nu_{\rm{\mathell}}}\right)\left(\nu_{\rm{h}} -
\mu_{\nu_{\rm{h}}}\right)} {\sigma_{\nu_{\rm{\mathell}}}
\sigma_{\nu_{\rm{h}}}}+\frac{\left(\nu_{\rm{h}} -
\mu_{\nu_{\rm{h}}}\right)^2}{\sigma_{\nu_{\rm{h}}}^2}\right]\right\}.
\end{eqnarray}

The model parameters $\left(\mu_{\nu_{\rm{\mathell}}}, \mu_{\nu_{\rm{h}}},
\sigma_{\nu_{\rm{\mathell}}}, \sigma_{\nu_{\rm{h}}}, \rho\right)$ of
$N\left(\nu_{\rm{\mathell}},\nu_{\rm{h}}; \mu_{\nu_{\rm{\mathell}}},
\mu_{\nu_{\rm{h}}},\sigma_{\nu_{\rm{\mathell}}}, \sigma_{\nu_{\rm{h}}},
\rho\right)$ are well constrained, as shown in Figure \ref{Figure5}(b).
According to Equation (\ref{eq15}), there should also exist
completely out-of-band bursts, i.e., below-band ($\nu_{\rm{\mathell}}
<\nu_{\rm{h}}<f_{\rm{\mathell}}<f_{\rm{h}}$) and above-band
($f_{\rm{\mathell}}<f_{\rm{h}}<\nu_{\rm{\mathell}}<\nu_{\rm{h}}$)
events, corresponding to the pink region on the
$\nu_{\rm{h}}$-$\nu_{\rm{\mathell}}$ plane
in the main plot of Figure \ref{Figure5}(a).
By numerically integrating Equation (\ref{eq15}) over the undetectable
region employing the optimal model parameters provided in Figure
\ref{Figure5}(b), we derived a $1.67\%$ probability of undetected
bursts for the Arecibo sample, corresponding to approximately eight
completely out-of-band events, which are inferred to be mainly above-band
cases. For the band-chipped bursts, given the condition that the burst
can be detected, the theoretical probabilities derived from
Equation (\ref{eq15}) (i.e., those predicted by the reconstructed
$\left(\nu_{\rm{\mathell}},\nu_{\rm{h}}\right)$ distribution)
corresponding to the lower-chipped, upper-chipped, and
over-band cases are 5.71\%, 67.07\%, and 0.78\%, respectively.
These probabilities indicate that the band-chipped bursts are dominated by
upper-chipped cases, which is fully consistent with the
observations, where upper-chipped events account for $\sim$64\%
of the Arecibo sample. This agreement supports the validity of our
reconstructed intrinsic $\left(\nu_{\rm{\mathell}},\nu_{\rm{h}}\right)$
distribution. Moreover, the predominance of upper-chipped bursts
implies that the intrinsic frequency ranges of the Arecibo bursts
tend to extend above the telescope's operating band. This naturally
explains why the completely out-of-band bursts are inferred to be
predominantly above-band events.

Note that the effect of operating-band cutoff in a sample depends not only
on the telescope's operating band range but also on the intrinsic
frequency ranges of the bursts, which may vary across different periods.
Therefore, even for the same repeating FRB source, the impact
of the operating-band cutoff on the observed burst population could
vary dramatically depending on the telescope and the observing
epoch. This highlights the necessity of a sample-specific analysis
when correcting for such observational biases.

Since $\left(\nu_{\rm{p}}, \delta \nu \right)$ (i.e., the peak
frequency and the proper bandwidth of the burst) is a linear transformation
of $\left(\nu_{\rm{\mathell}}, \nu_{\rm{h}}\right)$, we can derive
the $\left(\nu_{\rm{p}}, \delta \nu \right)$ distribution
$N\left(\nu_{\rm{p}}, \delta\nu; \mu_{\nu_{\rm{p}}},
\mu_{\delta\nu}, \sigma_{\nu_{\rm{p}}}, \sigma_{\delta\nu}, k
\right)$ from the $\left(\nu_{\rm{\mathell}}, \nu_{\rm{h}}\right)$
distribution as (Appendix \ref{appendix E:Transforming})
\begin{eqnarray}
\label{eq16} f_N \left(\nu_{\rm{p}}, \delta\nu\right) =
\frac{1}{2\pi\sigma_{\nu_{\rm{p}}} \sigma_{\delta\nu}
\sqrt{1-k^2}} \exp\left\{-\frac{1}{2\left(1-k^2\right)}
\left[\frac{\left(\nu_{\rm{p}} - \mu_{\nu_{\rm{p}}}
\right)^2}{\sigma_{\nu_{\rm{p}}}^2} - \frac{2k\left(\nu_{\rm{p}} -
\mu_{\nu_{\rm{p}}} \right) \left(\delta\nu - \mu_{\delta\nu}
\right)}{\sigma_{\nu_{\rm{p}}} \sigma_{\delta\nu}} +
\frac{\left(\delta\nu - \mu_{\delta\nu}
\right)^2}{\sigma_{\delta\nu}^2} \right]\right\}.
\end{eqnarray}
Here $\mu_{\nu_{\rm{p}}} = \frac{\mu_{\nu_{\rm{\mathell}}} +
\mu_{\nu_{\rm{h}}}}{2}$, $\mu_{\delta\nu} = \mu_{\nu_{\rm{h}}} -
\mu_{\nu_{\rm{\mathell}}}$, $\sigma_{\nu_{\rm{p}}} =
\sqrt{\left(\sigma_{\nu_{\rm{\mathell}}}^2 +
\sigma_{\nu_{\rm{h}}}^2 + 2 \rho \sigma_{\nu_{\rm{\mathell}}}
\sigma_{\nu_{\rm{h}}} \right) / 4}$, $\sigma_{\delta\nu} =
\sqrt{\sigma_{\nu_{\rm{\mathell}}}^2 + \sigma_{\nu_{\rm{h}}}^2 - 2
\rho \sigma_{\nu_{\rm{\mathell}}} \sigma_{\nu_{\rm{h}}}}$ and $k =
\left( \sigma_{\nu_{\rm{h}}}^2 - \sigma_{\nu_{\rm{\mathell}}}^2
\right) / \sqrt{\left(\sigma_{\nu_{\rm{\mathell}}}^2 +
\sigma_{\nu_{\rm{h}}}^2\right)^2 - 4 \rho^2
\sigma_{\nu_{\rm{\mathell}}}^2 \sigma_{\nu_{\rm{h}}}^2}$. By
substituting the optimal model parameters
$\left(\mu_{\nu_{\rm{\mathell}}}, \mu_{\nu_{\rm{h}}},
\sigma_{\nu_{\rm{\mathell}}}, \sigma_{\nu_{\rm{h}}}, \rho\right)$ of the
$\left(\nu_{\rm{\mathell}}, \nu_{\rm{h}}\right)$
distribution from Figure \ref{Figure5}(b) into Equation (\ref{eq16}),
we obtained the intrinsic distribution of
$\left(\nu_{\rm{p}},\delta\nu\right)$ as demonstrated
in Figure \ref{Figure5}(c).
In essence, for the intrinsic $\left(\nu_{\rm{p}},\delta\nu\right)$
distribution (blue contour lines), it is interesting to note that
$\delta\nu$ exhibits a gradual increase with the rise of
$\nu_{\rm{p}}$, with a correlation coefficient of $k=0.44$, which is not
evident in the observed distribution (yellow-orange-red contours) due to
the presence of the operating-band cutoff.
As shown in the right subplot of Figure \ref{Figure5}(c),
distortion due to operating-band cutoff leads to an
underestimation of $\delta\nu$, as the peak of its intrinsic
distribution (blue solid line) should be nearly twice that of
the observed distribution (green histogram).
As discussed in Section
\ref{sec3.1:Reconstruction and Analysis of In-band Bursts},
the observed bandwidth can be significantly biased by the
sensitivity cutoff, while the operating-band cutoff analyzed here
introduces another type of distortion to the observed bandwidth.

\begin{figure}[!htbp]
\centering
\includegraphics[width=1\textwidth]{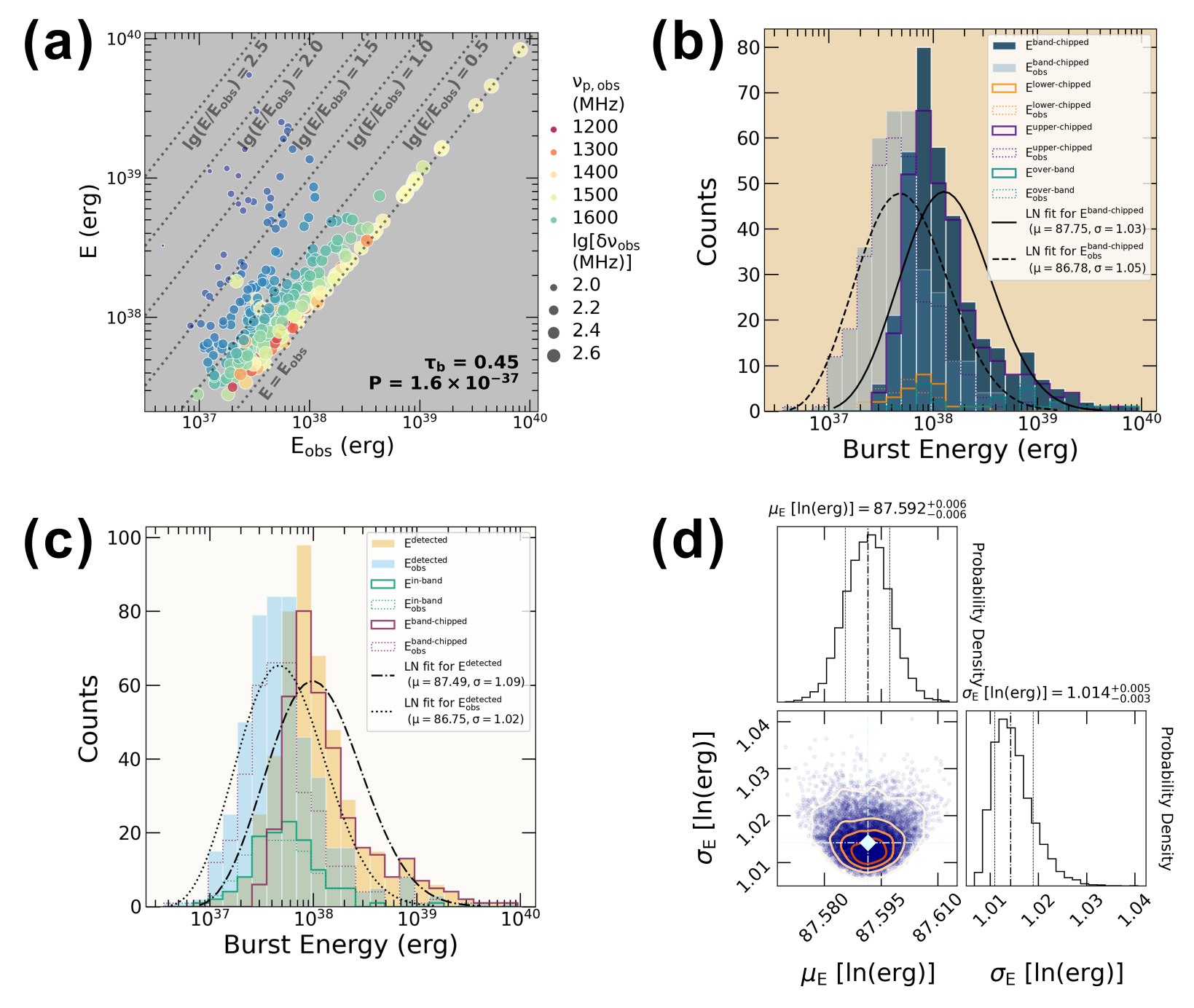}
\caption{Effects of operating-band cutoff on FRB energetics for
FRB 20121102A observed by Arecibo between December 2015 and October 2016
(data from \citealt{3}). The observed energy $E_{\rm{obs}}$ is derived
from the observed fluence and bandwidth, while the intrinsic energy $E$ is
estimated through our population-based method (Section
\ref{sec4:Operating-band Cutoff}). (a) The distribution of band-chipped
bursts on the $E$-$E_{\rm{obs}}$ plane. The data points are color-coded
by the observed peak frequency $\nu_{\rm{p,obs}}$ and scaled in size
proportional to $\lg\left(\delta\nu_{\rm{obs}}\right)$, where
$\delta\nu_{\rm{obs}}$ is the observed bandwidth of the burst.
The Kendall tau correlation coefficient $\tau_{\rm{b}}$ and its
corresponding $P$-value are shown for the data points.
(b) Comparison between the observed energy distribution
($E_{\rm{obs}}^{\rm{band\text{-}chipped}}$) and the intrinsic energy
distribution ($E^{\rm{band\text{-}chipped}}$). Each distribution is
fitted with a shifted log-normal function (with fitted parameters
$\mu$ and $\sigma$). The observed and intrinsic energy
distributions for the lower-chipped
($E_{\rm{obs}}^{\rm{lower\text{-}chipped}}$,
$E^{\rm{lower\text{-}chipped}}$), upper-chipped
($E_{\rm{obs}}^{\rm{upper\text{-}chipped}}$,
$E^{\rm{upper\text{-}chipped}}$), and over-band
($E_{\rm{obs}}^{\rm{over\text{-}chipped}}$,
$E^{\rm{over\text{-}chipped}}$) bursts are also shown separately.
(c) Same as panel (b), but for all detected bursts (in-band and 
band-chipped combined).
(d) Constraints on the parameters of the
standard log-normal distribution for the intrinsic energy,
obtained by fitting the recovered intrinsic energy distribution of
the detected bursts in panel (c). The standard log-normal 
function accommodates undetectable low-energy bursts by extrapolating 
the distribution toward lower energies.}
\label{Figure6}
\end{figure}

Based on the reconstructed intrinsic distribution of
$\left(\nu_{\rm{\mathell}},\nu_{\rm{h}}\right)$,
we have estimated the intrinsic energy of these band-chipped
bursts. For band-chipped bursts, although the proper spectral boundaries
$\nu_{\rm{\mathell}}$ and/or $\nu_{\rm{h}}$ of a specific burst are
unknown, the intrinsic distribution of $\left(\nu_{\rm{\mathell}},
\nu_{\rm{h}}\right)$ for the
entire sample provides the probability densities of different
possible values of the unknown spectral boundaries.
Specifically, for a band-chipped burst, for all possible combinations of
$\left(\nu_{\rm{\mathell}}, \nu_{\rm{h}}\right)$, we can compute the
integrated fluence $F$ for each combination and then convolve
the resulting integrated fluences with the corresponding probability
densities to obtain the expected value of the integrated fluence
$E\left(F\right)$ for this burst as (see Appendix \ref{appendix
F:Estimation of Integrated Fluence for Band-chipped Bursts})
\begin{eqnarray}
\label{eq17}
 E \left( F \right) = \left\{
    \begin{array}{lc}
         \frac{\iint_{\nu_{\rm{\mathell}} < f_{\rm{\mathell}}
            < f_{\rm{h}}<\nu_{\rm{h}}}{F \left(\nu_{\rm{\mathell}},
            \nu_{\rm{h}}, F_{\nu,{\rm{obs}}}^{\rm{equiv}}, F_{\nu,{\rm{thre}}}\right)
            \times f_N\left(\nu_{\rm{\mathell}}, \nu_{\rm{h}} \right){\rm{d}}
            \nu_{\rm{\mathell}}{\rm{d}} \nu_{\rm{h}}}}{\iint_{\nu_{\rm{\mathell}}
            < f_{\rm{\mathell}} < f_{\rm{h}} < \nu_{\rm{h}}}{f_N\left(\nu_{\rm{\mathell}},
            \nu_{\rm{h}}\right){\rm{d}}\nu_{\rm{\mathell}}{\rm{d}}\nu_{\rm{h}}}},
            \quad\nu_{\rm{\mathell}} < f_{\rm{\mathell}}<f_{\rm{h}}<\nu_{\rm{h}}, \\
         \frac{\int_{\nu_{\rm{\mathell}}<f_{\rm{\mathell}}}{F\left(\nu_{\rm{\mathell}},
            \nu_{\rm{h}},F_{\nu,{\rm{obs}}}^{\rm{equiv}}, F_{\nu,{\rm{thre}}}\right)
            \times\left[f_N\left(\nu_{\rm{\mathell}},\nu_{\rm{h}}\right)/
            \varphi\left(\frac{\nu_{\rm{h}}-\mu_{\nu_{\rm{h}}}}{\sigma_{\nu_{\rm{h}}}}
            \right)\right]{\rm{d}}\nu_{\rm{\mathell}}}}{\int_{\nu_{\rm{\mathell}}
            < f_{\rm{\mathell}}}{\left[f_N\left(\nu_{\rm{\mathell}},
            \nu_{\rm{h}}\right)/\varphi\left(\frac{\nu_{\rm{h}}
            -\mu_{\nu_{\rm{h}}}}{\sigma_{\nu_{\rm{h}}}}\right)\right]{\rm{d}}
            \nu_{\rm{\mathell}}}},
            \quad\nu_{\rm{\mathell}}<f_{\rm{\mathell}}
            <\nu_{\rm{h}}<f_{\rm{h}}, \\
         \frac{\int_{\nu_{\rm{h}}>f_{\rm{h}}}{F\left(\nu_{\rm{\mathell}},
            \nu_{\rm{h}},F_{\nu,{\rm{obs}}}^{\rm{equiv}},F_{\nu,{\rm{thre}}}\right)
            \times\left[f_N\left(\nu_{\rm{\mathell}},\nu_{\rm{h}}\right)/
            \varphi\left(\frac{\nu_{\rm{\mathell}}-
            \mu_{\nu_{\rm{\mathell}}}}{\sigma_{\nu_{\rm{\mathell}}}}\right)\right]
            {\rm{d}}\nu_{\rm{h}}}}{\int_{\nu_{\rm{h}}
            >f_{\rm{h}}}{\left[f_N\left(\nu_{\rm{\mathell}},
            \nu_{\rm{h}}\right)/\varphi\left(\frac{\nu_{\rm{\mathell}}
            -\mu_{\nu_{\rm{\mathell}}}}{\sigma_{\nu_{\rm{\mathell}}}}\right)\right]{\rm{d}}
            \nu_{\rm{h}}}},
            \quad{f}_{\rm{\mathell}}<\nu_{\rm{\mathell}}<f_{\rm{h}}<\nu_{\rm{h}}.
    \end{array}
\right.
\end{eqnarray}
Here the function $\varphi$ is the probability density function
(PDF) of the standard Gaussian distribution, and the integrated
fluence $F\left(\nu_{\rm{\mathell}},\nu_{\rm{h}}, F_{\nu,
{\rm{obs}}}^{\rm{equiv}}, F_{\nu,{\rm{thre}}}\right)$ is
obtained by substituting the solution of Equation (\ref{eq7})
into Equation (\ref{eq8}). Combining the above
equation with Equation (\ref{eq9}), a reasonable estimation
of the intrinsic energy of the band-chipped burst can be obtained.

In Figure \ref{Figure6}(a), where the band-chipped bursts are
plotted on the $E$-$E_{\rm{obs}}$ plane (i.e., intrinsic energy
versus observed energy), we see that the decrease of the observed
bandwidth $\delta\nu_{\rm{obs}}$ (represented by the size of the
data points) leads to a decrease in $E_{\rm{obs}}/E$, indicating a
more severe energy loss stemming from a stronger operating-band
cutoff. In the most extreme case, $E_{\rm{obs}}/E$ can be as
severe as $\sim$0.005. Additionally, band-chipped bursts with a 
smaller $\delta\nu_{\rm{obs}}$ tend to have a higher $\nu_{\rm{p,obs}}$, 
again pointing to the fact that the intrinsic 1D distribution of 
$\nu_{\rm{h}}$ extends above Arecibo's operating band. On the other 
hand, bursts with 
$\nu_{\rm{p,obs}}$ smaller than $f_{\rm{c,Arecibo}}$ (1440 MHz), which 
correspond to the lower-chipped case, tend to align closely with the 
$E_{\rm{obs}}=E$ line, indicating that the intrinsic 1D distribution of 
$\nu_{\rm{\mathell}}$ almost coincides with Arecibo's operating band. 
For band-chipped bursts with the same $E$, the energy loss 
(${E_{\rm{obs}}}/{E}$) is mainly influenced by $\nu_{\rm{p,obs}}$, which 
essentially reflects the positional relation between 
$(\nu_{\rm{\mathell}},\nu_{\rm{h}})$ and the telescope's operating band.
From Figure \ref{Figure6}(a), we can also see that intrinsically 
low-energy bursts cannot possess excessively small values of
${E_{\rm{obs}}}/{E}$, as such bursts would fall below the
detection threshold and thus be absent from the sample.
Intrinsically high-energy bursts exhibit a bifurcation in
${E_{\rm{obs}}}/{E}$, with values tending to be either very large
or very small. As mentioned in Section \ref{sec3.1:Reconstruction
and Analysis of In-band Bursts}, intrinsically energetic bursts
tend to concentrate more energy within a narrower spectrum,
leading to a higher energy per unit frequency that is particularly
pronounced in the central portion of the Gaussian-like spectrum.
Hence, they are more sensitive to the operating-band cutoff. When
subjected to the operating-band cutoff, such bursts either lose
only a negligible portion of energy (if the spectrum wing is
truncated) or suffer a significant energy loss (if the central
portion of the spectrum is cut off).

Figure \ref{Figure6}(b) compares the intrinsic and observed energy
distributions of the band-chipped bursts, each fitted with a shifted
log-normal function. Under the influence of the operating-band
cutoff, the intrinsic energy distribution is shifted toward lower
energies and manifests as the observed distribution. This shift is
primarily driven by the energy loss of upper-chipped bursts
(represented by the purple solid and dashed histograms for their intrinsic
and observed energy distributions, respectively),
the dominant subgroup among band-chipped events.
We note that, as discussed in Section
\ref{sec3.1:Reconstruction and Analysis of In-band Bursts},
the sensitivity cutoff has a minor effect on the energy loss and
mainly biases the observed bandwidth. Therefore, although
band-chipped bursts are subject to both sensitivity and
operating-band cutoffs, the energy loss seen here is predominantly
caused by the operating-band cutoff.

In Figure \ref{Figure6}(c), the intrinsic energy and observed
energy distributions of all the detected Arecibo bursts, which include
both the band-chipped and the in-band events, are compared.
The distributions are each fitted with a shifted log-normal function.
Compared with the intrinsic energy distribution, the observed
energy distribution lies systematically at lower values, owing to the
observational cutoff effect mainly induced by the operating-band cutoff.
The high-energy end of the observed distribution
profile displays some anomalous substructures, which arise because
intrinsically high-energy bursts experience very different degrees of
energy loss, causing their observed energies to be scattered broadly
(as reflected by the bifurcation in $E_{\rm{obs}}/E$ seen in the
upper part of Figure \ref{Figure6}(a)).
In the Arecibo sample, for the detected bursts, the observational cutoff
effect leads to a total energy loss of $1-\left(\frac{\Sigma
E_{\rm{obs}}}{\Sigma E}\right)_{\rm{detected}}=0.49$,
i.e., nearly 50\%,
which means the observational incompleteness is really significant.
This highlights that the energy release of some repeating FRB
sources may have been previously underestimated, and underscores the
necessity of a sample-by-sample assessment, as the degree of observational
incompleteness can vary considerably among different samples.
To meet the requirement of energy budget of radio and
X-ray emissions from the central engine, additional energy
mechanisms beyond the magnetar magnetosphere energy supply may be
required, particularly for highly active sources \citep{44,45,46,47}.

In the Arecibo sample, as mentioned before, the operating-band
cutoff renders some bursts completely out-of-band and thus undetectable
(about eight events).
Moreover, there should be additional undetectable
bursts whose intensity falls below the fluence threshold.
For the completely out-of-band bursts, only
their number can be estimated, which affects the burst rate but
provides little information about their individual intrinsic
energies. In contrast, for the undetectable low-energy bursts,
their total intrinsic energy can still be estimated by
extrapolating the profile of the intrinsic energy distribution
of the detected bursts toward the low-energy end.
To consider these undetectable low-energy bursts, we employed the
standard log-normal distribution function to fit the intrinsic energy
distribution of the detected bursts through MCMC methods. While the
standard log-normal distribution function may not fit
the distribution profile as perfectly as the shifted log-normal
distribution function, it allows for the potential presence of
undetectable low-energy bursts (Appendix \ref{appendix C:Standard/Shifted
Log-normal Distribution Function}). Accordingly, the fitted
parameters are shown in Figure \ref{Figure6}(d).

Distinguishing from the out-of-band bursts and the undetectable
low-energy bursts, there may also exist
unidentified bursts in the current observational data (e.g.,
extremely short-duration bursts; see \citealt{27} and \citealt{28}),
which need to be identified using advanced data processing methods
\citep{3,9} and would update our understanding of FRB energy,
waiting time and event rate.


\section{GBT and FAST Samples}
\label{sec5:GBT and FAST Samples}

Focusing on FRB 20121102A, the prototypical repeating FRB,
we have also applied the reconstruction methodology to other
samples, such as the GBT and observations.

\begin{figure}[!htbp]
\centering
\includegraphics[width=0.74\textwidth]{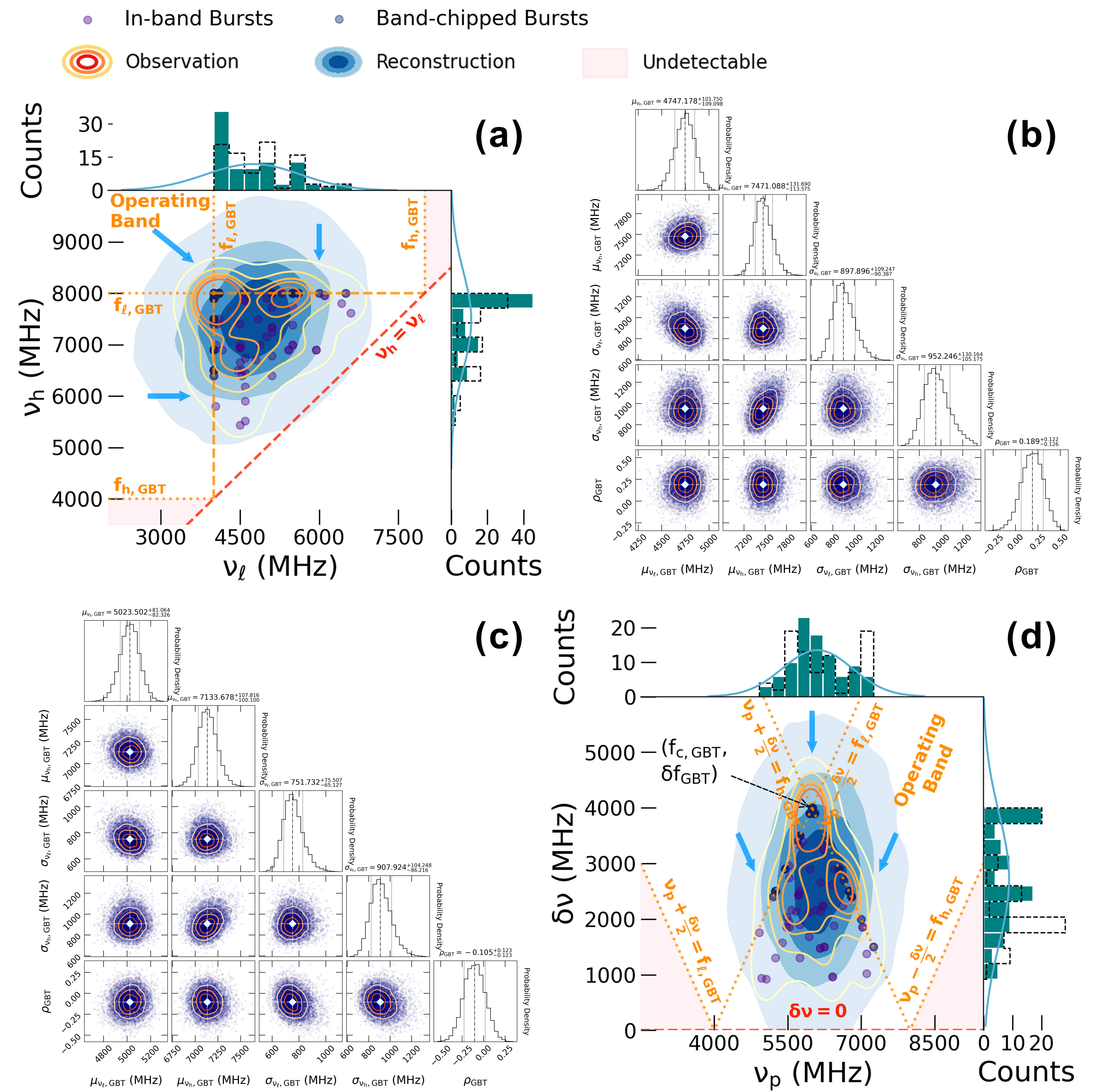}
\caption{Operating-band cutoff effects on the observed 2D
distribution of bursts on the $\nu_{\rm{h}}$-$\nu_{\rm{\mathell}}$ plane
and the reconstructed intrinsic 2D distribution for FRB 20121102A observed
by GBT on August 26, 2017 (data from \citealt{9}). $\nu_{\rm{\mathell}}$
and $\nu_{\rm{h}}$ represent the proper lower and upper spectral
boundaries of the burst, respectively, while $\nu_{\rm{p}}$ and
$\delta\nu$ denote the true peak frequency and proper bandwidth of the
burst. (a) The distribution of the observed bursts on the
$\nu_{\rm{h}}$-$\nu_{\rm{\mathell}}$ plane.
The yellow-orange-red contour lines represent the KDE of the
observed $\left(\nu_1, \nu_2\right)$ distribution, where $\nu_1$ and
$\nu_2$ are the observed lower and upper spectral boundaries of
the burst. The blue contour lines show the corresponding intrinsic
$\left(\nu_{\rm{\mathell}}, \nu_{\rm{h}}\right)$ distribution
reconstructed via our population-based inverse modeling method
under the Bayesian inference framework (Section
\ref{sec4:Operating-band Cutoff}). 1D histograms of the observed bursts
are displayed along the axes (green histograms), along with the marginal
distributions of the reconstructed intrinsic $\left(\nu_{\rm{\mathell}},
\nu_{\rm{h}}\right)$ distribution (blue solid lines). The dashed orange
lines mark the operating band limits of GBT, $f_{\rm{\mathell,GBT}}=4000$
MHz and $f_{\rm{h,GBT}}=8000$ MHz. The region below the red dashed line
($\nu_{\rm{h}} < \nu_{\rm{\mathell}}$) is non-physical, while the
pink region is undetectable ($\nu_{\rm{h}} < f_{\rm{\mathell,GBT}}$
or $\nu_{\rm{\mathell}} > f_{\rm{h,GBT}}$). Because many GBT 
bursts were identified via machine-learning methods,
$\nu_1$ and $\nu_2$ are often asymmetric about the true peak frequency
$\nu_{\rm{p}}$. Assuming intrinsic Gaussian symmetry, we adjust them as
$\nu_1^{\prime} = \max[f_{\rm{\mathell,GBT}}, \nu_{\rm{p}}
- \max(\nu_{\rm{p}} - \nu_1, \nu_2 - \nu_{\rm{p}})]$, $\nu_2^{\prime} =
\min[f_{\rm{h,GBT}}, \nu_{\rm{p}} + \max(\nu_{\rm{p}} - \nu_1,
\nu_2 - \nu_{\rm{p}})]$. 
In the subplots, histograms delineated
by the dashed outline correspond to bursts before adjustment. (b)
Constraints on the model parameters of the reconstructed
intrinsic distribution function $N\left(\nu_{\rm{\mathell}},
\nu_{\rm{h}}; \mu_{\nu_{\rm{\mathell}}}, \mu_{\nu_{\rm{h}}},
\sigma_{\nu_{\rm{\mathell}}}, \sigma_{\nu_{\rm{h}}}, \rho\right)$.
(c) Constraints on the model parameters of the reconstructed
intrinsic distribution function $N\left(\nu_{\rm{\mathell}},
\nu_{\rm{h}}; \mu_{\nu_{\rm{\mathell}}}, \mu_{\nu_{\rm{h}}},
\sigma_{\nu_{\rm{\mathell}}}, \sigma_{\nu_{\rm{h}}}, \rho\right)$, but
derived from the sample before adjustment. The results are broadly
consistent with those derived from the adjusted sample. (d) The
distribution of the observed bursts on the $\delta\nu$-$\nu_{\rm{p}}$
plane, with graphical elements following those in panel (a).
Although the true peak frequency $\nu_{\rm{p}}$ of
the burst is available, for reference the data points are plotted at
values $\delta\nu_{\rm{obs}} = \nu_2^{\prime} - \nu_1^{\prime}$ and
$\nu_{\rm{p,obs}} = \frac{\nu_1^{\prime} + \nu_2^{\prime}}{2}$.
The dashed-outlined histogram on the $\nu_{\rm{p}}$ axis
corresponds to the literature-provided
$\nu_{\rm{p}}$ values, and the dashed-outlined histogram on the
$\delta\nu$ axis represents the $\delta\nu$ values of the sample before
adjustment. The point $\left(f_{\rm{c,GBT}},\delta f_{\rm{GBT}}\right)$
represents the central frequency and width of the GBT operating band.}
 \label{Figure7}
\end{figure}

GBT detected 93 bursts from FRB 20121102A in the 4--8 GHz
operating band, during a 6-hour observation on August 26, 2017,
providing the key measurements of of the observed spectral
boundaries $\nu_1$ and $\nu_2$, the frequency-equivalent fluence
$F_{\nu, {\rm{obs}}}^{\rm{equiv}}$, and the pulse width
${W_{\rm{obs}}^{\rm{equiv}}}$ for each burst \citep{9}. From the
observed $\nu_1$ and $\nu_2$ of the bursts, we have reconstructed
the intrinsic 2D distribution of the proper spectral boundaries
$\left(\nu_{\rm{\mathell}},\nu_{\rm{h}}\right)$ for the GBT
sample, as shown in Figure \ref{Figure7}(a). The model parameters
are well constrained and shown in Figure \ref{Figure7}(b).
Accordingly, the intrinsic 2D distribution of $\left(\nu_{\rm{p}},
\delta\nu\right)$, i.e., the peak frequency and the proper
bandwidth of the burst, was derived from the reconstructed
$\left(\nu_{\rm{\mathell}},\nu_{\rm{h}}\right)$ distribution using
Equation (\ref{eq16}) and is shown in Figure \ref{Figure7}(d). In
contrast to the Arecibo sample (L-band, 1--2 GHz), where
$\delta\nu$ slowly increases with $\nu_{\rm{p}}$ (see Figure
\ref{Figure5}(c)), for the GBT bursts in the C-band (4--8 GHz),
$\delta\nu$ does not increase with $\nu_{\rm{p}}$ (blue contour
lines in Figure \ref{Figure7}(d)).

Note that in the GBT sample, many weak bursts
were identified by utilizing machine learning method, which complicates
the definition of their fluence thresholds (i.e., the fluence sensitivity)
\citep{29,30}. Consequently, although $F_{\nu,{\rm{obs}}}^{\rm{equiv}}$
and $W_{\rm{obs}}^{\rm{equiv}}$ are available for these bursts,
reconstructing their intrinsic energies is challenging. However, the
reconstruction of the intrinsic
$\left(\nu_{\rm{\mathell}},\nu_{\rm{h}}\right)$ and $\left(\nu_{\rm{p}},
\delta\nu\right)$ distributions is not affected, because our 2D
distribution reconstruction method has an advantage of adapting to the
fluence threshold variation in a limited range across bursts (Appendix
\ref{appendix D:Reconstructing the 2D Gaussian Distribution}).

\begin{figure}[!htbp]
\centering
\includegraphics[width=0.84\textwidth]{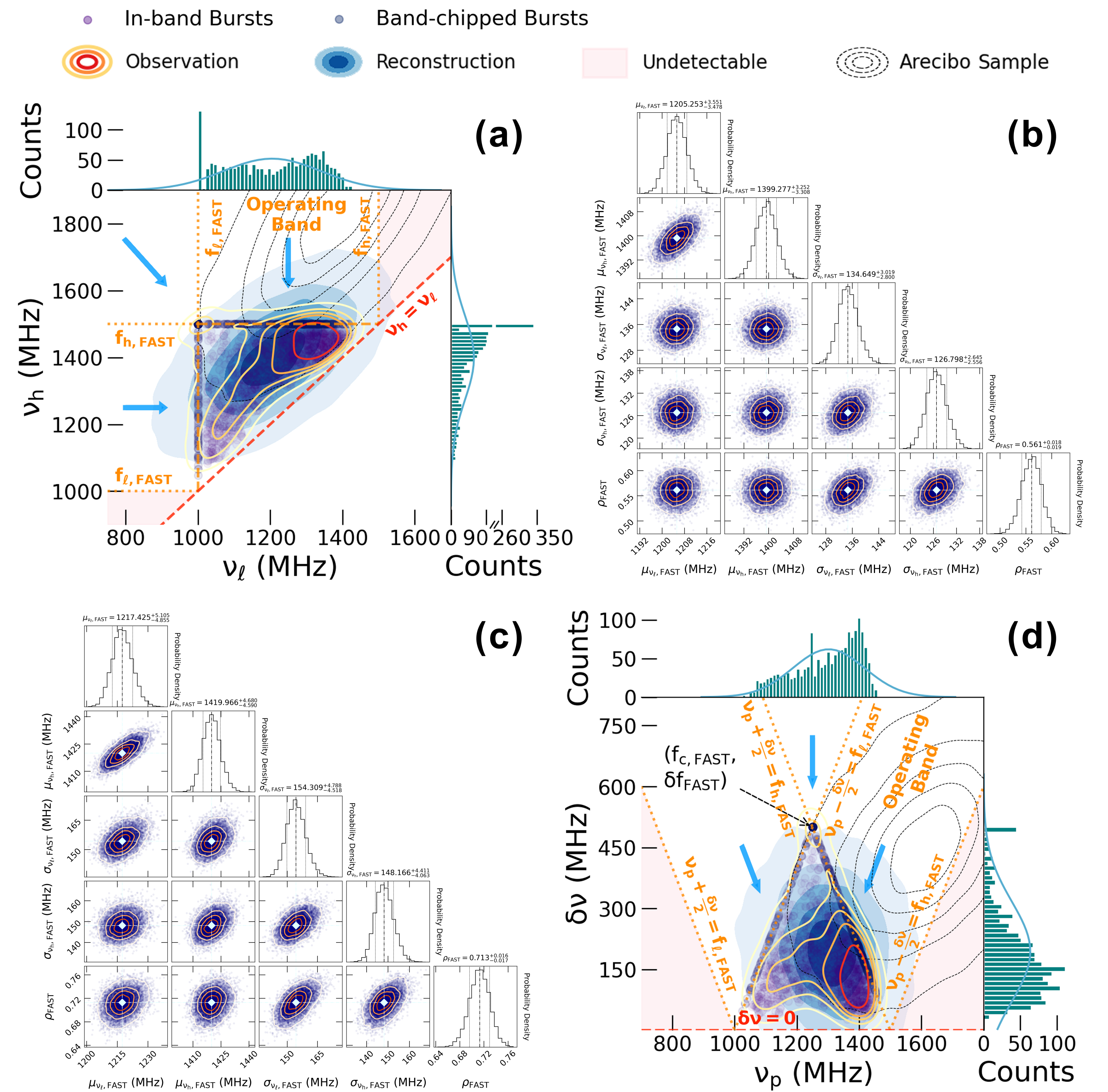}
\caption{Operating-band cutoff effects on the observed 2D
distribution of bursts on the $\nu_{\rm{h}}$-$\nu_{\rm{\mathell}}$ plane
and the reconstructed intrinsic 2D distribution for FRB 20121102A observed
by FAST between August 29 and October 29, 2019. The burst sample is taken
from \citet{1}, but we reprocessed the data to more precisely extract the
observed lower and upper spectral boundaries $\nu_1$ and $\nu_2$ of the
burst. $\nu_{\rm{\mathell}}$ and $\nu_{\rm{h}}$ represent
the proper lower and upper spectral boundaries of the burst,
respectively, while $\nu_{\rm{p}}$ and $\delta\nu$ denote the
true peak frequency and proper bandwidth of the burst.
(a) The distribution of the observed bursts on the
$\nu_{\rm{h}}$-$\nu_{\rm{\mathell}}$ plane. The yellow-orange-red
contour lines represent the KDE of the observed
$\left(\nu_1, \nu_2\right)$ distribution.
The blue contour lines show the corresponding intrinsic
$\left(\nu_{\rm{\mathell}}, \nu_{\rm{h}}\right)$ distribution
reconstructed via our population-based inverse modeling method
under the Bayesian inference framework (Section
\ref{sec4:Operating-band Cutoff}). The black dashed contours
indicate the intrinsic $\left(\nu_{\rm{\mathell}}, \nu_{\rm{h}}\right)$
distribution of the Arecibo sample, taken from Figure \ref{Figure5}(a),
for comparison. 1D histograms of the observed bursts are displayed along
the axes (green histograms), along with the marginal distributions of the
reconstructed intrinsic $\left(\nu_{\rm{\mathell}}, \nu_{\rm{h}}\right)$
distribution (blue solid lines). The dashed orange lines mark the
operating band limits of FAST, $f_{\rm{\mathell,FAST}}=1000$ MHz
and $f_{\rm{h,FAST}}=1500$ MHz. The region below the red dashed
line ($\nu_{\rm{h}} < \nu_{\rm{\mathell}}$) is non-physical, while
the pink region is undetectable ($\nu_{\rm{h}} < f_{\rm{\mathell,FAST}}$
or $\nu_{\rm{\mathell}} > f_{\rm{h,FAST}}$).
(b) Constraints on the  model parameters of the reconstructed
intrinsic distribution function $N\left(\nu_{\rm{\mathell}},
\nu_{\rm{h}}; \mu_{\nu_{\rm{\mathell}}}, \mu_{\nu_{\rm{h}}},
\sigma_{\nu_{\rm{\mathell}}}, \sigma_{\nu_{\rm{h}}}, \rho\right)$.
(c) Constraints on the parameters of the reconstructed
intrinsic distribution function $N\left(\nu_{\rm{\mathell}},
\nu_{\rm{h}}; \mu_{\nu_{\rm{\mathell}}}, \mu_{\nu_{\rm{h}}},
\sigma_{\nu_{\rm{\mathell}}}, \sigma_{\nu_{\rm{h}}}, \rho\right)$,
but derived from the sample artificially truncated to the central FAST
band (1.05--1.45 GHz). The results
are largely broadly with that derived from the original
sample. (d) The distribution of the observed bursts on the
$\delta\nu$-$\nu_{\rm{p}}$ plane, with graphical elements following those
in panel (a). The point $\left(f_{\rm{c,FAST}},\delta
f_{\rm{FAST}}\right)$ represents the central frequency and width of the
FAST operating band.}
 \label{Figure8}
\end{figure}

FAST detected 1652 bursts from FRB 20121102A in the 1--1.5
GHz operating band, during the 59.5 hours of observation spanning
from 29 August to 29 October 2019, providing the key measurements
of $F_{\nu,{\rm{obs}}}^{\rm{equiv}}$ and
$W_{\rm{obs}}^{\rm{equiv}}$ for all bursts \citep{1}. We
reprocessed the data and precisely extracted $\nu_1$ and $\nu_2$
of the bursts (Table \ref{Table3:FAST Sample}). For a specific
burst, since the fluence sensitivity depends on the pulse width,
we adopt $F_{\nu,
{\rm{thre}}}=0.015\sqrt{{W_{\rm{obs}}^{\rm{equiv}}}/1\, \rm{ms}}$
Jy ms for the FAST sample \citep{29,1,30,13}. After excluding
indeterminate bursts (approximately 1\% of the total sample) whose
spectral frequency-equivalent fluence (i.e., the observed
integrated fluence divided by the observed bandwidth of the burst)
falls below the burst-specific fluence threshold, or whose
observed bandwidth is narrower than 20 MHz (likely spurious), the
intrinsic 2D distribution of
$\left(\nu_{\rm{\mathell}},\nu_{\rm{h}}\right)$ was reconstructed
(Figure \ref{Figure8}(a)), and the corresponding model parameters
were well constrained (Figure \ref{Figure8}(b)). In Figure
\ref{Figure8}(a), the intrinsic frequency ranges (represented by
the intrinsic $\left(\nu_{\rm{\mathell}},\nu_{\rm{h}}\right)$
distribution) of FAST (blue contour lines) and Arecibo bursts
(black dashed contours) differ markedly. As discussed earlier, the
intrinsic frequency ranges of Arecibo bursts tend to extend above
the telescope's operating band, resulting in a large proportion of
upper-chipped events. If these Arecibo bursts were observed by
FAST, most of them would lie completely above the FAST band (i.e.,
become above-band events), and any detectable bursts would be
upper-chipped. This can be seen from the relative positions of the
black dashed contours and the the triangular region enclosed by
the orange and red dashed lines (which marks the FAST operating
band on the $\nu_{\rm{h}}$-$\nu_{\rm{\mathell}}$ plane). As for
the FAST bursts, the intrinsic frequency ranges (blue contours)
closely match the FAST band, with only a slight extension beyond
it. Consequently, only 429 bursts (26\% of the effective sample)
are band-chipped, including 87 lower-chipped (5\%), 297
upper-chipped (18\%), and 45 over-band events (3\%). In addition,
integrating the PDF of the intrinsic
$\left(\nu_{\rm{\mathell}},\nu_{\rm{h}}\right)$ distribution over
the undetectable (pink) region in Figure \ref{Figure8}(a) yields
an estimate of approximately 16 completely out-of-band bursts for
the FAST sample.

From the reconstructed $\left(\nu_{\rm{\mathell}},
\nu_{\rm{h}}\right)$ distribution, we derived the intrinsic 2D
distribution of $\left(\nu_{\rm{p}},\delta\nu\right)$, as shown in Figure
\ref{Figure8}(d). Also in the L-band, unlike the Arecibo sample (black
dashed contours), $\delta\nu$ of the FAST burst does not increase with the
increasing $\nu_{\rm{p}}$ (blue contours), similar to what is seen in the
GBT sample. This indicates that FRBs exhibit differing behaviors on the
$\delta\nu$-$\nu_{\rm{p}}$ plane
during different active periods and across different frequency ranges.
It suggests that repeaters may reside in complex environments and involve
various magnetospheric activities \citep{32,33,34,35}
or plasma instabilities \citep{36,37,38}. Coordinated observations across
low- and high-frequency radio bands \citep{11,40,41} may provide critical
evidence to diagnose the mechanisms. Otherwise, an intrinsic
evolutionary process of the repeating source \citep{42} itself might
explain such observed diversity, necessitating extensive follow-up
monitoring of well-localized repeating FRBs with high burst rates to
reveal the enigma.

Alternatively, another factor that may
contribute to the observed differences in frequency evolution (i.e., the
intrinsic $\left(\nu_{\rm{p}},\delta\nu\right)$ distribution) among the
GBT, FAST, and Arecibo samples is the vastly different monitoring
baselines: the Arecibo sample has the longest detection span
(approximately 10 months), whereas the GBT sample (6 hours) and the FAST
sample (around 2 months) have considerably shorter coverages. From this
perspective, the repeating FRB source may exhibit long-term frequency
evolution.

\begin{figure}[!htbp]
\centering
\includegraphics[width=0.9\textwidth]{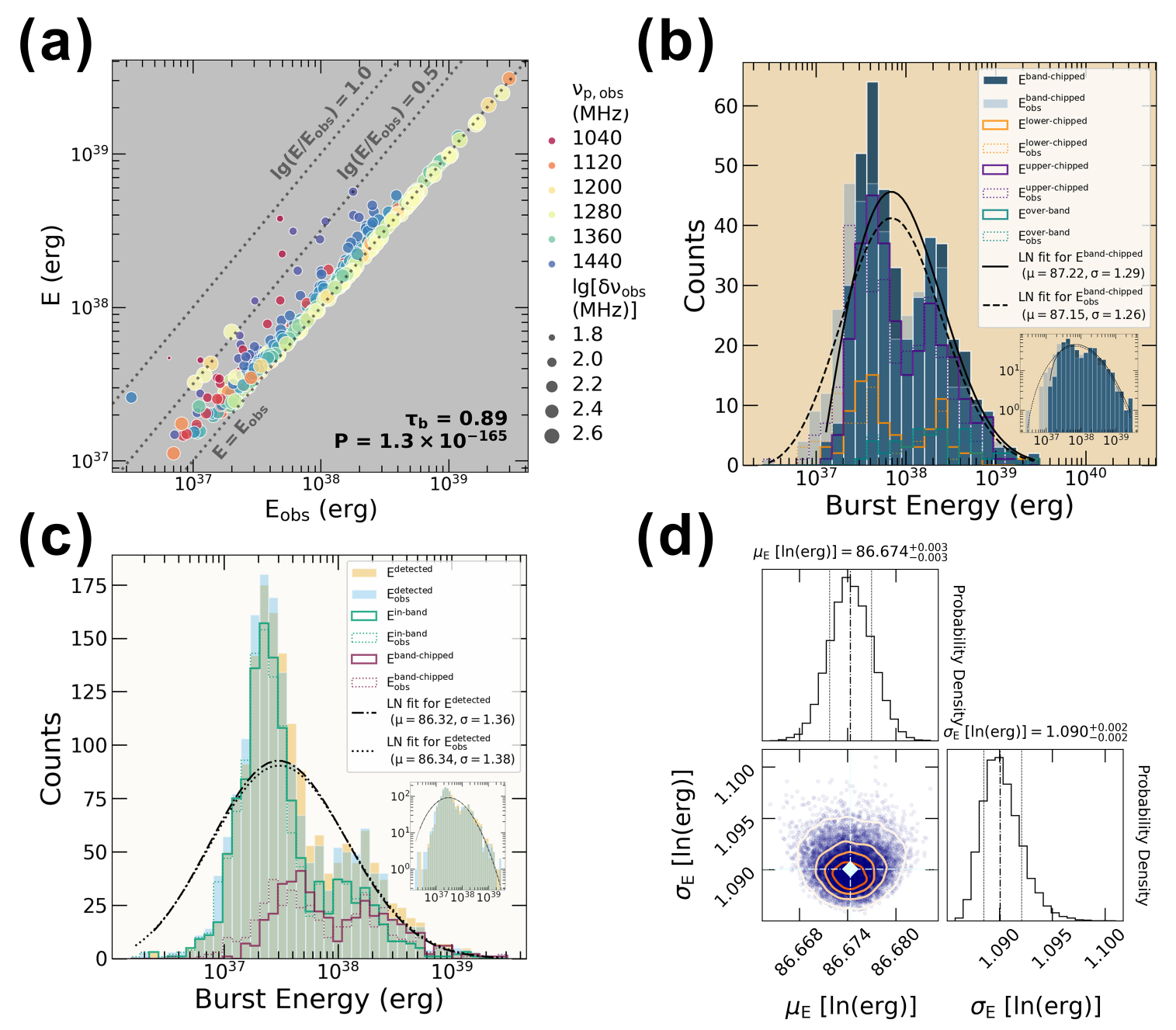}
 \caption{Effects of observational cutoff on FRB energetics for
FRB 20121102A observed by FAST between August 29 and October 29,
2019 (data from \citealt{1}). The observed energy $E_{\rm{obs}}$ is
derived from the observed fluence and the observed bandwidth, the
latter computed from our reprocessed spectral boundaries.
The intrinsic energy $E$ is estimated through our population-based
method (Section \ref{sec4:Operating-band Cutoff}).
(a) The distribution of band-chipped bursts on the
$E$-$E_{\rm{obs}}$ plane. The data points are color-coded by the
observed peak frequency $\nu_{\rm{p,obs}}$ and scaled in size proportional
to $\lg\left(\delta\nu_{\rm{obs}}\right)$, where
$\delta\nu_{\rm{obs}}$ is the observed bandwidth of the burst.
The Kendall tau correlation coefficient $\tau_{\rm{b}}$ and its
corresponding $P$-value are shown for the data points. (b) Comparison
between the observed energy distribution
($E_{\rm{obs}}^{\rm{band\text{-}chipped}}$) and the intrinsic energy
distribution ($E^{\rm{band\text{-}chipped}}$) for the band-chipped
bursts. Each distribution is fitted with a shifted log-normal
function (with fitted parameters $\mu$ and $\sigma$). The inset displays
the same histograms of $E_{\rm{obs}}^{\rm{band\text{-}chipped}}$
and $E^{\rm{band\text{-}chipped}}$, but with the vertical axis in
logarithmic scale, where reasonable log-space fits are clearly seen.
The observed and intrinsic energy distributions for the lower-chipped
($E_{\rm{obs}}^{\rm{lower\text{-}chipped}}$,
$E^{\rm{lower\text{-}chipped}}$), upper-chipped
($E_{\rm{obs}}^{\rm{upper\text{-}chipped}}$,
$E^{\rm{upper\text{-}chipped}}$), and over-band
($E_{\rm{obs}}^{\rm{over\text{-}band}}$,
$E^{\rm{over\text{-}band}}$) bursts are also shown separately.
(c) Same as panel (b), but for all detected bursts (in-band and 
band-chipped combined).
(d) Constraints on the parameters of the standard log-normal distribution
of the intrinsic energy, obtained by fitting the
recovered intrinsic energy distribution of the detected bursts
in panel (c). The standard log-normal function accommodates 
undetectable low-energy bursts by extrapolating the distribution toward 
lower energies.}
 \label{Figure9}
\end{figure}

We also reconstructed the intrinsic burst energies for the
FAST sample. As shown in Figure \ref{Figure9}(a), the majority of
band-chipped bursts lie close to the $E_{\rm{obs}}=E$ line on the
$E$-$E_{\rm{obs}}$ plane, indicating that their energy losses are
generally small, regardless of the observed peak frequency
$\nu_{\rm{p,obs}}$ (represented by the color of the data points)
or the observed bandwidth $\delta\nu_{\rm{obs}}$ (represented by
the size of the data points). This is due to the fact the
intrinsic frequency ranges of the bursts closely match the FAST
band (see Figure \ref{Figure8}(a)), so that even though individual
bursts have different true peak frequencies $\nu_{\rm{p}}$ and
proper bandwidths $\delta\nu$, the energy loss incurred by the
operating-band cutoff remains minimal. This can also be seen in
Figure \ref{Figure9}(b), where the observed and intrinsic energy
distributions show little difference across all subclasses of
band-chipped bursts (i.e., lower-chipped, upper-chipped, and
over-band events). As shown in Figure \ref{Figure9}(c), the total
energy loss caused by observational cutoffs is small for the
detected FAST bursts (including both in-band and band-chipped
events): the contribution of the sensitivity cutoff is negligible
owing to the relatively good sensitivity of FAST, and the
contribution of the operating-band cutoff is also small because
the intrinsic frequency ranges of the bursts coincidentally match
the FAST band well, leading to a total fractional energy loss of
$1-\left(\Sigma E_{\rm{obs}}/\Sigma E\right)_{\rm{detected}} =
0.06$.

To account for the potential contribution of undetectable weak
bursts to the intrinsic energy distribution, we fit the intrinsic energy
distribution of the detected bursts with a standard log-normal function
via MCMC, extrapolating the distribution profile toward the low-energy
end. The resulting fit parameters are presented in
Figure \ref{Figure9}(d).

As a final remark, we briefly discuss the definition of 
a burst adopted for the three samples analyzed in this work. The exact 
operational definition of a multi-peak burst may vary slightly among the Arecibo, 
GBT, and FAST samples, because their data were processed by 
different groups. For the FAST bursts, which were reprocessed by us from 
the raw data, a multi-peak burst is identified by significant bridge emission 
between adjacent pulse peaks exceeding the 5$\sigma$ detection threshold 
(see Appendix \ref{appendix A:Spectrum Profile of FRBs}). This criterion 
is physically motivated: significant bridge emission indicates a 
continuous physical process, which ensures that the sub-peaks of a 
multi-peak burst cover similar frequency ranges, so that modeling the 
time-integrated spectrum of such a burst with a single Gaussian function 
is well justified (see, e.g., the second row, third column of Figure 
\ref{FigureA2}). For those multi-peak bursts defined by the valley of 
the waiting-time distribution, the waiting time between sub-peaks is 
typically shorter than a few milliseconds, which also corresponds to a 
continuous emission process; the time-integrated spectrum of such bursts 
is therefore similarly well described by a single Gaussian function.

The key observed and derived quantities for the Arecibo, GBT, and
FAST burst samples analyzed in this work are summarized in Tables
\ref{Table1:Arecibo Sample}, \ref{Table2:GBT Sample}, and
\ref{Table3:FAST Sample}. These tables are intended to facilitate the
reproduction of our results and to serve as a resource for future
studies of repeating FRBs.

\clearpage

\clearpage

\clearpage


\section{Conclusion}
\label{sec6:Conclusion}

Observational cutoffs can severely bias estimates of the energy
release, bandwidth, burst rate, and waiting time of repeating FRBs. To
overcome these limitations, we propose a new inverse-modeling framework
that reconstructs intrinsic burst properties from incomplete observational
data without resorting to conventional spectral profile fitting.

In our approach, detected bursts are divided into two categories:
in-band events, which are affected only by the sensitivity cutoff, and
band-chipped events, which suffer from both sensitivity and operating-band
cutoffs. For in-band bursts, assuming Gaussian-like spectra, we derive a
set of equations that directly link the observed quantities to the
intrinsic parameters, bypassing any fitting to the raw dynamic spectra.
For band-chipped bursts, some of which may have spectral peaks outside the
telescope's operating band, we adopt a population-based method. In this
method, the intrinsic 2D distribution of
$\left(\nu_{\rm{\mathell}},\nu_{\rm{h}}\right)$ (i.e., the proper lower
and upper spectral boundaries of the burst, characterizing its intrinsic
frequency range) of the entire sample is first recovered from the
observed, distorted 2D distribution of $\left(\nu_1,\nu_2\right)$ (i.e.,
the observed spectral boundaries of the burst) through inverse modeling.
For band-chipped bursts, although the individual spectral boundaries
$\nu_{\rm{\mathell}}$ and/or $\nu_{\rm{h}}$ are unavailable, the
reconstructed intrinsic $(\nu_{\rm{\mathell}}, \nu_{\rm{h}})$ distribution
of the entire sample offers the probability densities of all possible
combinations of $\left(\nu_{\rm{\mathell}}, \nu_{\rm{h}}\right)$. Thus,
for a band-chipped burst, the expected value of the integrated fluence
$E(F)$ is obtained by convolving the integrated fluences $F$ computed for
each possible $\left(\nu_{\rm{\mathell}}, \nu_{\rm{h}}\right)$ pair with
the corresponding probability densities. We take $E(F)$ as the natural
estimator of the intrinsic energy of a band-chipped burst, analogous to
ensemble averages serve as macroscopic observables in equilibrium
statistical mechanics and expectation values give the predicted results of
measurements in quantum mechanics.

Applied to three incomplete samples, our method successfully
reconstructs their intrinsic properties. We find that intrinsically
energetic bursts tend to have narrower spectra than intrinsically weak
bursts. It is also found that the repeating FRB source exhibits different
frequency evolution on the $\delta\nu$-$\nu_{\rm{p}}$ plane (where
$\delta\nu$ and $\nu_{\rm{p}}$ are the proper bandwidth and peak frequency
of the burst, respectively) in different active periods and across
different frequency bands. For instance, in the reconstructed Arecibo
sample (1.15--1.73 GHz), $\delta\nu$ increases with $\nu_{\rm{p}}$,
whereas no such trend is seen in the reconstructed FAST (1--1.5 GHz) and
GBT (4--8 GHz) samples. Nevertheless, the possibility that such difference
is due to different monitoring baselines across samples cannot be ruled out.

By comparing the reconstructed and original burst samples, we find
that the sensitivity cutoff contributes little to the total energy loss
but can distort the observed bandwidth, whereas the operating-band cutoff
may cause substantial energy leakage and bandwidth underestimation,
depending on how well the intrinsic frequency ranges of the bursts match
the telescope's operating band. For the Arecibo bursts, where the
intrinsic frequency ranges tend to extend above the operating band, the
fractional energy loss reaches $\sim$50\% and the proper bandwidths are
about twice the observed values. For the FAST bursts, where the intrinsic
frequency ranges coincidentally match the operating band, the fractional
energy loss is $\sim$6\% and the bandwidth underestimation is less
pronounced, highlighting that such observational biases require
sample-by-sample corrections rather than a one-size-fits-all approach.
These findings point to the possibility that the energy release of some
repeating FRB sources has been considerably underestimated, which
may have implications for the energy supply mechanisms beyond the
magnetar magnetosphere. Moreover, given that the
observed bandwidth is a telescope-dependent quantity, we strongly
recommend adopting the FWHM of the spectrum as an instrument-independent
descriptor of spectral width.

Our methodology converts incomplete archival observations into
physically meaningful diagnostics, facilitates the joint analysis of data
from various telescopes, and would have wide application as more repeating
FRB observations become available in the future.


\newpage

\section*{Data Availability}

The Arecibo data \citep{3} and GBT data \citep{9} can be obtained
from the corresponding references. We reprocessed the FAST data \citep{1}
and precisely extracted the observed spectral boundaries of the bursts
by applying a more precise temporal windowing to minimize noise
contamination. The key observed and derived quantities for the three
burst samples analyzed in this work are organized in Tables
\ref{Table1:Arecibo Sample}, \ref{Table2:GBT Sample}, and
\ref{Table3:FAST Sample}.


\section*{Code Availability}

Results can be fully reproduced using the methodology described in
this paper, based on Python 3.12.4 and the publicly available Python
packages NumPy 1.26.4, SciPy 1.16.3, emcee 3.1.6, pandas 2.2.2,
Seaborn 0.13.2 and Matplotlib 3.8.4. All source codes have been
deposited in Zenodo (\doi{10.5281/zenodo.20679643}) for reproducibility.


\section*{Acknowledgments}

The analysis in this article utilizes data from Arecibo and GBT,
and this work made use of the data from FAST FRB Key Science
Project.

We thank the anonymous referee for useful suggestions that led 
to an overall improvement of this study.
This work is supported by the National Natural Science Foundation
of China (Grant Nos. 12233002, 12273113), by the National Key R\&D
Program of China (2021YFA0718500), and the CAS Project for Young
Scientists in Basic Research (Grant No. YSBR-063). PW acknowledges
support from the National Natural Science Foundation of China
(NSFC) Programs No.11988101, 12041303, the CAS Youth
Interdisciplinary Team, the Youth Innovation Promotion Association
CAS (id. 2021055), and the Cultivation Project for FAST Scientific
Payoff and Research Achievement of CAMS-CAS. YFH acknowledges the
support from the Xinjiang Tianchi Program. JJG acknowledges
support from the Youth Innovation Promotion Association (2023331).

\clearpage


\appendix
\vspace{-6mm}

\section{Spectrum Profile of FRBs}
\label{appendix A:Spectrum Profile of FRBs}

\setcounter{figure}{0}
\numberwithin{figure}{section}
\renewcommand{\thefigure}{\Alph{section}\arabic{figure}}

We use the Bayesian Information Criterion (BIC) to analyze the
spectrum shape of FRBs. For the 1652 FAST bursts from FRB
20121102A observed between August 29 and October 29, 2019 and
reported in \citet{1}, we reprocessed the data by ourselves. In
the time domain, following \citet{1}, adjacent pulse peaks
connected by continuous flux above the 5$\sigma$ noise level
(i.e., bridge emission) are classified as substructure of a single
burst, whereas those separated by flux below this threshold are
considered independent bursts. Different from \citet{1}, however,
we applied a more precise temporal windowing to extract the
spectrum of each burst with minimized noise contamination. This
was achieved by performing Gaussian fit to individual pulse
profile (i.e., the temporal profile), determining the time range
centered on the pulse peak and measure the FWHM. Approximately
13$\%$ of bursts were excluded from the analysis due to their
irregular pulse morphologies that prevent a reliable determination
of their temporal boundaries.

\begin{figure}[!htbp]
\centering
\includegraphics[width=1\textwidth]{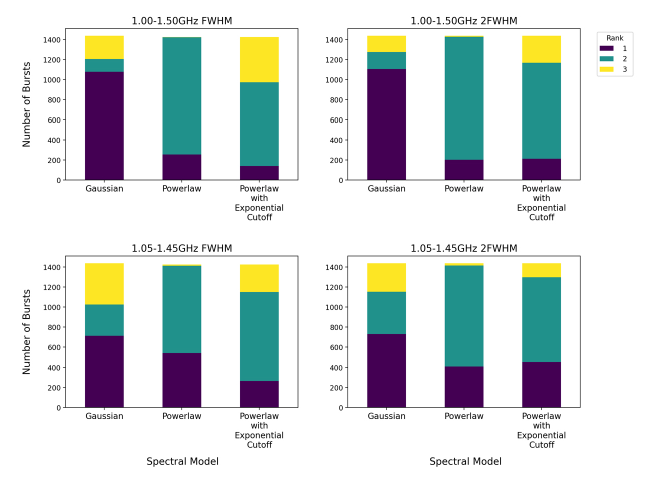}
\caption{Spectral model rank distribution for FRB 20121102A
observed by FAST between August 29 and October 29, 2019. The burst sample
is taken from \citet{1}, but we reprocessed the data and applied a
more precise temporal windowing to extract the spectrum of each
burst with minimized noise contamination. The analysis is
performed on the bursts with well-defined temporal boundaries
(approximately 87\% of the full sample). For each burst, the spectrum is
fitted with three models (Gaussian, power-law, and cutoff power-law), and
the models are ranked by the Bayesian Information Criterion (BIC; rank 1
indicates the best fit). The four panels correspond to different spectral
extraction methods: the upper and lower panels represent frequency ranges
of 1.0--1.5 GHz and 1.05--1.45 GHz, respectively, while the left and
right panels correspond to temporal windows of the pulse FWHM and
twice the pulse FWHM, both centered at the pulse peak. Each panel
shows the distribution of the resulting BIC rankings for the three
spectral models, with a higher count of top rankings (rank 1)
indicating a better model performance. The Gaussian model
consistently receives the highest number of top rankings across
all four extraction methods.}
 \label{FigureA1}
\end{figure}

For the remaining 87$\%$ of the sample, we fit their spectra with
three spectral models (Gaussian, powerlaw, and cutoff powerlaw) and
assessed the goodness of fit by using BIC \citep{43}. Each burst received model rankings
(1st to 3rd) based on ascending BIC values, where a lower BIC value indicates
a better fit. Figure \ref{FigureA1}
presents the statistical distribution of these model rankings for four
different spectral extraction methods. The upper panels demonstrate
the spectra extracted in 1--1.5 GHz, contrasting
temporal windows defined by the pulse FWHM versus twice the FWHM (i.e.,
2FWHM). Both upper panels show that the Gaussian function is better than
the powerlaw and cutoff powerlaw models. To investigate the impact of the operating
bandwidth on the spectrum profile, we have also artificially restricted the spectrum range
in 1.05--1.45 GHz (the lower panels), mimicking observations
taken with a narrower operating band. We see that a narrower band increases
the goodness of both the powerlaw and the cutoff powerlaw models,
particularly for bursts exhibiting incomplete Gaussian profiles where the
spectral peak falls outside the restricted band, resulting in quasi-powerlaw
morphologies. However, the Gaussian function is still the best
spectrum model.

\begin{figure}[!htbp]
\centering
\includegraphics[width=1\textwidth]{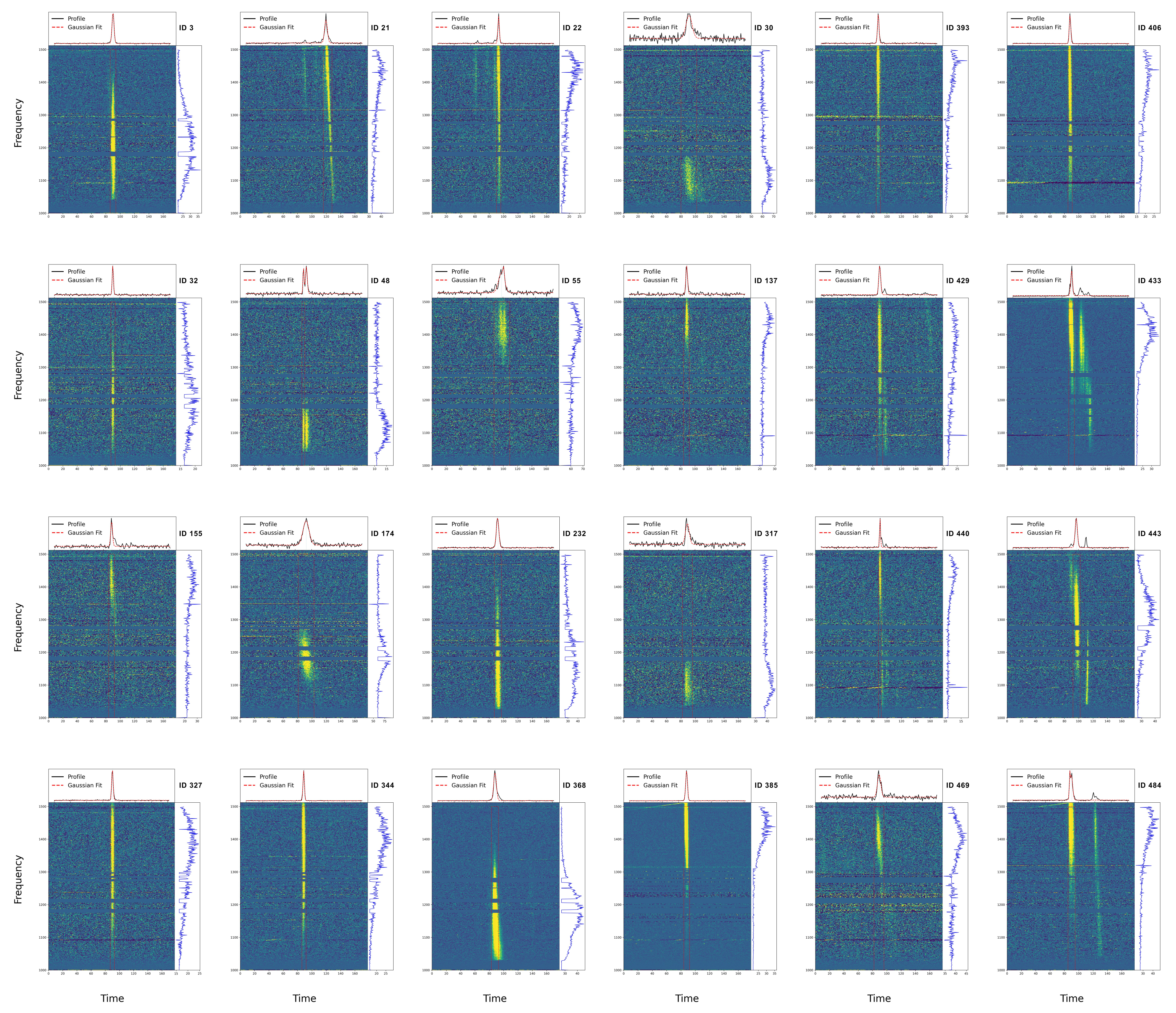}
\caption{Typical bright bursts exhibiting a Gaussian spectral profile
for FRB 20121102A observed by FAST between August 29 and
October 29, 2019. The burst sample is taken from \citet{1} but the data
have been reprocessed with a more precise temporal windowing to
extract the spectrum of each burst with minimized noise
contamination, with the burst ID consistent with the original 
reference. The bursts shown here are selected as representative
examples of bright events. For each burst, the spectral
extraction was performed by applying a temporal window of twice
the pulse FWHM centered at the pulse peak, together with a
frequency range of 1--1.5 GHz. The bursts exhibit either a complete
Gaussian spectral profile or an incomplete (truncated) one, with
their dynamic spectra and time-integrated spectra displayed in each panel.
The Gaussian spectral profile is clearly seen in the complete cases, while
the truncated cases show how the operating band cutoff can diminish the
Gaussian character of the observed spectrum.}
 \label{FigureA2}
\end{figure}

Figure \ref{FigureA2} shows the dynamical spectra of some typical
bright bursts which exhibit a Gaussian spectrum. Those bursts with the central
frequency near the operating band edges display a truncated Gaussian
profile. When restricting the spectral extraction to a narrower frequency
range, these truncated bursts can exhibit a pseudo-powerlaw
spectrum. It clearly demonstrates how the operating band limitation affects the observed
spectral morphologies. Note that for both powerlaw and cutoff powerlaw
models, the absolute value of the powerlaw index should be
greater than 0.1; otherwise, the spectrum would be nearly flat, yielding a
good but physically meaningless fit for powerlaw or cutoff powerlaw models.

\clearpage

\clearpage

\section{Numerical Solution of the Gaussian Spectrum}
\label{appendix B:Numerical Solution of the Gaussian Spectrum}

\numberwithin{equation}{section}
\renewcommand{\theequation}{\Alph{section}\arabic{equation}}

\setcounter{figure}{0}
\numberwithin{figure}{section}
\renewcommand{\thefigure}{\Alph{section}\arabic{figure}}

When $\nu_{\rm{p}}$ is available, Equation (\ref{eq7}) can be
used to determine $\sigma_\nu$ from the observed spectrum,
yielding the unique solution $\left(F, \nu_{\rm{p}},
\sigma_\nu\right)$ for the Gaussian spectrum of the burst. The
algorithm of AGD \citep{15,16,17} is frequently employed for
numerically solving such equations and is also commonly used in
machine learning. In such a process, it is necessary to define a
loss function to measure the discrepancy between the trial
solution and the true solution. From Equation (\ref{eq7}), we
can define the loss function as
\begin{eqnarray}
 \label{eq18}
L = \left( \sqrt{2\pi} F_{\nu,{\rm{thre}}} \sigma_\nu
\left[\rm{\Phi} \left(\frac{\nu_2 - \nu_{\rm{p}}}{\sigma_\nu}
\right) - \rm{\Phi} \left(\frac{\nu_1 -
\nu_{\rm{p}}}{\sigma_\nu} \right) \right] -
F_{\nu,{\rm{obs}}}^{\rm{equiv}} \delta\nu_{\rm{obs}}{\rm{e}}^{{ -
\left( \frac{\delta\nu}{2\sqrt2\sigma_\nu} \right)}^2} \right)^2.
\end{eqnarray}
To avoid encountering excessively large values beyond the
computational capabilities, we transform the exponential term
from Equation (\ref{eq7}) into its reciprocal when constructing the loss
function in Equation (\ref{eq18}). When the loss function
$L\left(\sigma_\nu\right)$ equals 0, the trial solution $\sigma_\nu$
becomes the true solution of Equation (\ref{eq7}).

In accordance with the naming convention for symbols used in
machine learning theory, we denote the value of the trial solution at the
$t$-th step as $w^t$, which is equivalent to ${\sigma_\nu}^t$.
Furthermore, we can derive the gradient of the loss function with
respect to the trial solution at the $t$-th step as
\begin{eqnarray}
 \label{eq19}
\begin{aligned}
g^t = &\frac{\partial L\left(w^t\right)}{\partial w^t} = 2
  \left(\sqrt{2\pi} F_{\nu,{\rm{thre}}}w^t \left[\Phi
  \left(\frac{\nu_2-\nu_{\rm{p}}}{w^t}\right) - \Phi
  \left( \frac{\nu_1-\nu_{\rm{p}}}{w^t}\right)\right]
  - F_{\nu,{\rm{obs}}}^{\rm{equiv}}
  \delta\nu_{\rm{obs}}{\rm{e}}^{{-
  \left(\frac{\delta\nu}{2\sqrt2w^t}\right)}^2}\right) \\
& \left\{\sqrt{2\pi}F_{\nu,{\rm{thre}}}
  \left[\Phi\left(\frac{\nu_2-\nu_{\rm{p}}}{w^t}\right)
  -\Phi\left(\frac{\nu_1-\nu_{\rm{p}}}{w^t}\right)\right]
  +\sqrt{2\pi}F_{\nu,{\rm{thre}}}w^t\left[{\frac{\nu_1
  -\nu_{\rm{p}}}{\sqrt{2\pi}
  \left(w^t\right)^2}}{\rm{e}}^{{-\left(\frac{\nu_1
  -\nu_{\rm{p}}}{\sqrt2w^t}\right)}^2}
  -{\frac{\nu_2-\nu_{\rm{p}}} {\sqrt{2\pi}
  \left(w^t\right)^2}}{\rm{e}}^{{-\left(\frac{\nu_2
  -\nu_{\rm{p}}}{\sqrt2w^t}\right)}^2}\right]\right.\\
& \left.-F_{\nu,{\rm{obs}}}^{\rm{equiv}}
  \delta\nu_{\rm{obs}}{\rm{e}}^{{
  -\left(\frac{\delta\nu}{2\sqrt2w^t}\right)}^2}
  \frac{{\delta\nu}^2}{4\left(w^t\right)^3}\right\}.
\end{aligned}
\end{eqnarray}
During the training process of AGD, the learning rate
$\eta^t=\frac{\eta^0}{\sqrt{t+1}}$ is dynamically adjusted by
considering the root mean square of the previous gradients
$\sigma^t=\sqrt{\frac{1}{t+1}\sum_{i=0}^{t}\left(g^i\right)^2}$.
Consequently, the trial solution $w^t$ undergoes optimal
adjustments at each step, with substantial updates when it
deviates significantly from the true solution and smaller updates
as it approaches the true solution, i.e.,
\begin{eqnarray}
 \label{eq20}
w^{t+1} = w^t - \frac{\eta^t}{\sigma^t}g^t = w^t -
\frac{\eta^0}{\sqrt{\sum_{i=0}^{t} \left(g^i\right)^2}}g^t.
\end{eqnarray}

\begin{figure}[!htbp]
\centering
\includegraphics[width=1\textwidth]{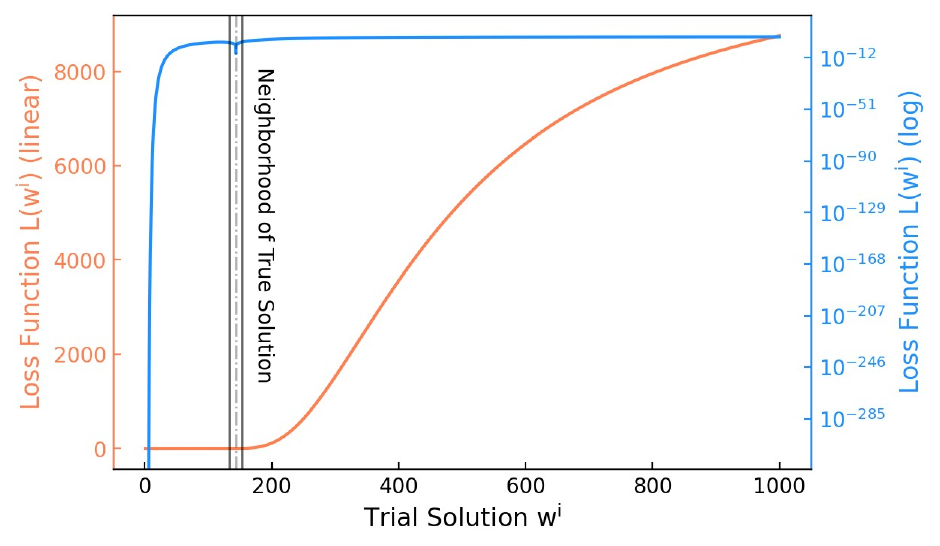}
 \caption{Response of the loss function to different trial solutions.
This figure shows how the loss function $L\left(w^i\right)$
(defined in Equation (\ref{eq18}) of Appendix \ref{appendix B:Numerical
Solution of the Gaussian Spectrum}) behaves with different trial
solutions $w^i$ during the numerical determination of the Gaussian
spectral width $\sigma_\nu$ for a burst. The loss function is shown in
both linear (the orange curve with respect to the orange vertical axis) and
logarithmic (the blue curve with respect to the blue vertical
axis) scales. The true solution corresponds to a neighborhood
where $L(w^i)$ approaches zero, but one that is away from the
non-physical solution at $w^i=0$; this is more readily
identified in the logarithmic representation.}
 \label{FigureB1}
\end{figure}

Throughout the training and equation-solving process, the initial
value of the learning rate, $\eta^0$, plays a crucial role. If
$\eta^0$ is set to a very large value, the training may fail to
converge to the true solution. On the other hand, if it is set too
small, the training time might become unacceptably long although a
convergence is possible. Moreover, since the true solution will be
used in the integral term of Equation (\ref{eq15}) for
convolution (we need to numerically solve the true solution for
every point in the 2D integration region), we should balance the
computational efficiency and accuracy. Thus, we made some
modifications to AGD. Figure \ref{FigureB1}
illustrates the behavior of the loss function $L\left(w^i\right)$
with respect to the trial solution $w^i$, which shows no clear
evidence of the true solution in linear space. It implies that
directly using AGD may fail to find the true solution unless
$\eta^0$ is set extremely small. However, setting $\eta^0$ to such
a small value can result in a significant increase in computation
time. In our modification, we first try to identify the
neighborhood of the true solution quickly based on the slope
changes of $L\left(w^t\right)$ in the logarithmic space. Then,
within this neighborhood, we assign an initial trial solution
$w^0$ and a small $\eta^0$ to efficiently obtain a high-precision
solution. It should be noted that the point $\left(0,0\right)$ in
Figure \ref{FigureB1} corresponds to a
non-physical solution, which particularly affects the training
process when the true solution neighborhood is close to this
point. Therefore, we need to incorporate a boundary condition
during the training.

\clearpage

\section{Standard/Shifted Log-normal Distribution Function}
\label{appendix C:Standard/Shifted Log-normal Distribution Function}

\numberwithin{equation}{section}
\renewcommand{\theequation}{\Alph{section}\arabic{equation}}

The standard log-normal distribution
$LN\left(x;\mu_{LN},\sigma_{LN}\right)$ and the shifted log-normal
distribution $LN\ (x;\delta,$ $\mu_{LN}^{\rm{s}},
\sigma_{LN}^{\rm{s}})$ (where the superscript "s" denotes
"shifted") are used to fit the distribution of many key parameters
in this study. These two distribution function differ only by a
location parameter $\delta$. Their PDFs are expressed as
\begin{eqnarray}
 \label{eq22}
f_{LN}\left(x\right) = \frac{1}{\sqrt{2\pi}
\sigma_{LN}x}{\rm{e}}^{-\frac{ \left(\ln{x} -
\mu_{LN}\right)^2}{2\sigma_{LN}^2}}, \quad{x}>0, \sigma_{LN}>0,
\end{eqnarray}
\begin{eqnarray}
 \label{eq21}
f_{LN}^{\rm{s}} \left(x\right) = \frac{1}{\sqrt{2\pi}
\sigma_{LN}^{\rm{s}} \left(x - \delta\right)}{\rm{e}}^{
-\frac{\left[\ln{\left(x - \delta\right)} -
\mu_{LN}^{\rm{s}}\right]^2}{2 \left(
\sigma_{LN}^{\rm{s}}\right)^2}},
\quad{x}>\delta,\sigma_{LN}^{\rm{s}}>0.
\end{eqnarray}
The introduction of the parameter $\delta$ in the PDF of the
shifted log-normal distribution allows for a better fit to the
observed distribution profile. However, it also restricts the
range of $x$ values and lacks consideration for potential low
values of $x$ that have not been observed. In order to simplify
the fitting process and improve the accuracy, Equation
(\ref{eq21}) can be transformed into
\begin{eqnarray}
 \label{eq23}
f_{LN}^{\rm{s}} \left(z\right) = \frac{1}{\sqrt{2\pi}{M
\sigma}_{LN}^{\rm{s}}z}{\rm{e}}^{- \frac{\ln^2{z}}{2 \left(
\sigma_{LN} \right)^2}},
\end{eqnarray}
where $z=\frac{x-\delta}{M}$ and $M={\rm{e}}^{\mu_{LN}^{\rm{s}}}$.

Considering the influence of statistical fluctuations on the
observed distribution, we fit the log-normal distribution through
MCMC simulations based on Bayesian inference. The logarithmic
formulation of Bayes' theorem is expressed below, which offers
computational simplicity and reduces risk of memory overflow in
machine-based calculations,
\begin{eqnarray}
 \label{eq24}
\ln{ \left[P \left( \theta\,|\,D \right) \right]} = \ln{\left[P
\left(D\,|\,\theta \right) \right]} + \ln{\left[P \left( \theta
\right) \right]} - \ln{\left[P\left(D\right)\right]}.
\end{eqnarray}
Here $\theta$ represents the model parameters of the log-normal
distribution function, and $D$ stands for the observed data.
$P\left(\theta\,|\,D\right)$ is the posterior, which refers to our
updated knowledge about the probability distribution of the
model parameters based on the available data.
$P\left(D\,|\,\theta\right)$ is the likelihood, corresponding to
the distribution that the data will exhibit given the model
parameters. It follows the Poisson distribution of
\begin{eqnarray}
 \label{eq25}
\ln{ \left[P \left(D\,|\, \theta\right)\right]} = \sum_{i} \left[
\ln \left({\lambda_i}^{k_i}\right) - \ln\left(k_i!\right) -
\lambda_i\right],
\end{eqnarray}
\begin{eqnarray}
 \label{eq26}
\lambda_i = \frac{N}{\int_{X_0}^{X_{n+1}}{f_{LN}^{({\rm{s}})}
\left(x; \theta \right){\rm{d}}x}}
\int_{X_i}^{X_{i+1}}{f_{LN}^{({\rm{s}})} \left(x; \theta
\right){\rm{d}}x}.
\end{eqnarray}
To obtain better estimates of the model parameters, it is
necessary to divide the observational data into $n$ bins. The
boundaries of the $i$-th bin are denoted as $X_i$ and $X_{i+1}$.
$\lambda_i$ is the theoretical count of data in the $i$-th bin,
while $k_i = N_i$ represents the corresponding observed count.
Empirically, the number of bins is usually taken as $n =
\left\lceil \sqrt N \right\rceil$ (i.e., rounding up the square
root of $N$), where $N=\sum_{i} N_i$. $f_{LN}^{({\rm{s}})}$
denotes the log-normal distribution function
$f_{LN}^{\rm{s}}\left(x\right)$ or $f_{LN}\left(x\right)$,
depending on which is utilized. The prior $P\left(\theta\right)$
is typically characterized by a uniform distribution,
\begin{eqnarray}
 \label{eq27}
\ln{ \left[P \left( \theta \right) \right]} = \left\{
    \begin{array}{lc}
         -\ln{\left(b-a\right)}, \quad{a} < \theta < b,  \\
         -\infty, \quad{\rm{otherwise}}.
    \end{array}
\right.
\end{eqnarray}
The range $\left[a,b\right]$ of $\theta$ can be set by considering
the distribution of the observed data. For instance,
$\mu_{LN}^{({\rm{s}})}$ should not exceed the maximum and minimum
values of the data, $3\sigma_{LN}^{({\rm{s}})}$ should lie between
0 and the natural logarithm of the maximum value in the data, and
$\delta$ generally should match the minimum value in the data. The
evidence $P\left(D\right)$ refers to the probability of the
observed facts, which should equal a constant $C$. Since the PDF's
global integral is 1, the sum of $P\left(\theta\,|\,D\right)$ over
all conditions should be 1, i.e., $\sum P\left(\theta\,|\,D\right)
= 1$. Furthermore, according to Equation (\ref{eq24}), we
have
$\sum\left[P\left(D\,|\,\theta\right)P\left(\theta\right)\right]=C$,
from which we can obtain the value of $P\left(D\right)$. In
practice, to improve the efficiency, we can treat constant terms
(such as the values of $P\left(\theta\right)$ with $\theta$
between $a$ and $b$) as 1 initially and then scale each
$P\left(\theta\,|\,D\right)$ proportionally so that $\sum
P\left(\theta\,|\,D\right)=1$.

Using this method, we can derive the probability distribution of
the model parameters for the log-normal distribution and determine
the parameters with the highest probability, known as the maximum
likelihood estimation.

\section{Reconstructing the 2D Gaussian Distribution of $\left(\nu_{\rm{\mathell}},\nu_{\MakeLowercase{\rm{h}}}\right)$}
\label{appendix D:Reconstructing the 2D Gaussian Distribution}

\numberwithin{equation}{section}
\renewcommand{\theequation}{\Alph{section}\arabic{equation}}

\setcounter{figure}{0}
\numberwithin{figure}{section}
\renewcommand{\thefigure}{\Alph{section}\arabic{figure}}

\begin{figure}[!htbp]
\centering
\includegraphics[width=0.6\textwidth]{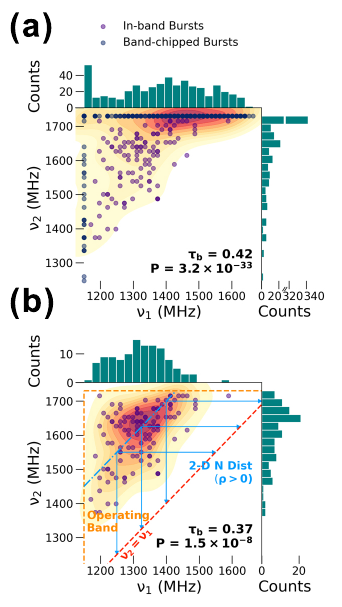}
\caption{Impact of operating-band cutoff on the observed burst
distribution for FRB 20121102A observed by Arecibo between December 2015
and October 2016 (data from \citealt{3}). $\nu_1$ and $\nu_2$ are the
observed lower and upper spectral boundaries of the burst, respectively.
Each panel displays the Kendall tau correlation coefficient
$\tau_{\rm{b}}$ and its associated $P$-value for the data points shown.
(a) The distribution of all Arecibo bursts on the
$\nu_2$-$\nu_1$ plane. The contour lines represent their KDE, and
the corresponding 1D histograms are plotted along the axes.
(b) The distribution of in-band Arecibo bursts on the $\nu_2$-$\nu_1$
plane. The orange dashed lines mark the operating band limits of Arecibo,
$f_{\rm{\mathell}}=1150$ MHz and $f_{\rm{h}}=1730$ MHz. The region
below the red dashed line is non-physical ($\nu_2 < \nu_1$). The KDE of
the data points away from the orange dashed lines (i.e., below the blue
dashed line) hints at an intrinsic 2D Gaussian distribution of
$\left(\nu_{\rm{\mathell}}, \nu_{\rm{h}}\right)$ for the bursts, where
$\nu_{\rm{\mathell}}$ and $\nu_{\rm{h}}$ are the proper lower and
upper spectral boundaries of the burst, respectively.}
 \label{FigureD1}
\end{figure}

As shown in Figure \ref{FigureD1}(a), on the $\nu_2$-$\nu_1$ plane (i.e.,
the upper boundary of observed spectrum plotted versus the lower boundary
of observed spectrum), the
band-chipped bursts exhibit an unnatural accumulation along the
lines $\nu_1=f_{\rm{\mathell}}$ and $\nu_2=f_{\rm{h}}$, which
correspond to the Arecibo operating band limits ($f_{\rm{\mathell}}$ and
$f_{\rm{h}}$ denote the lower and upper boundaries of the
telescope's operating band). According to
Equations (\ref{eq1}) and (\ref{eq2}), the portion of
the spectrum below $f_{\rm{\mathell}}$ or above $f_{\rm{h}}$
cannot be detected, meaning that the proper spectral boundaries
$\nu_{\rm{\mathell}}$
below $f_{\rm{\mathell}}$ and $\nu_{\rm{h}}$ above $f_{\rm{h}}$
are observationally identified as $f_{\rm{\mathell}}$ and
$f_{\rm{h}}$, respectively. It leads to such an accumulation. The
histograms in the subplots of Figure \ref{FigureD1}(a) reveal that the cutoff for
$\nu_{\rm{h}}$ is
significantly greater than that for $\nu_{\rm{\mathell}}$, both of
which could ideally follow normal distributions according to their
observed distribution profiles \citep{24}. In Figure
\ref{FigureD1}(b) which focuses on the in-band
bursts, the region of their kernel density estimation (KDE) that separate
from the operating
band limits (below the blue dash-dotted line) implies a potential 2D
Gaussian distribution of $\left(\nu_{\rm{\mathell}}, \nu_{\rm{h}}\right)$.

Our objective is to reconstruct the intrinsic 2D Gaussian
distribution $N \left(\nu_{\rm{\mathell}}, \nu_{\rm{h}};
\mu_{\nu_{\rm{\mathell}}}, \mu_{\nu_{\rm{h}}},
\sigma_{\nu_{\rm{\mathell}}}, \sigma_{\nu_{\rm{h}}}, \rho\right)$
from the observational data of $\left(\nu_1,\nu_2\right)$ via
inverse modeling. Here,
we denote the five model parameters $\left(\mu_{\nu_{\rm{\mathell}}},
\mu_{\nu_{\rm{h}}},\sigma_{\nu_{\rm{\mathell}}}, \sigma_{\nu_{\rm{h}}},
\rho\right)$ as $\theta$. The
reconstruction is done based on Bayes' theorem (see Equation
(\ref{eq24})). Compared with Equation (\ref{eq25}), the expression of the
likelihood $P\left(D\,|\,\theta\right)$ is slightly revised to adapt the 2D
data,
\begin{eqnarray}
 \label{eq28}
\ln{ \left[P \left( D\,|\, \theta\right) \right]} =
\sum_{i}\sum_{j} \left[\ln \left({\lambda_{ij}}^{k_{ij}}\right) -
\ln\left(k_{ij}!\right) - \lambda_{ij}\right],
\end{eqnarray}
\begin{eqnarray}
 \label{eq29}
\lambda_{ij} = \frac{N\iint_{\nu_{\rm{\mathell}} <
\nu_{\rm{h}}}{f_N \left( \nu_{\rm{\mathell}}, \nu_{\rm{h}}; \theta
\right){\rm{d}} \nu_{\rm{\mathell}}{\rm{d}}
\nu_{\rm{h}}}}{\iint_{\left(\nu_{\rm{\mathell}} < \nu_{\rm{h}}
\right)\land\left(\nu_{\rm{\mathell}}<f_{\rm{h}}
\right)\land\left(\nu_{\rm{h}}
> f_{\rm{\mathell}}\right)}{f_N \left(\nu_{\rm{\mathell}},
\nu_{\rm{h}}; \theta\right){\rm{d}} \nu_{\rm{\mathell}}{\rm{d}}
\nu_{\rm{h}}}}\int_{Y_i}^{Y_{i+1}}{\int_{X_j}^{X_{j + 1}}{f_N
\left(\nu_{\rm{\mathell}}, \nu_{\rm{h}}; \theta\right){\rm{d}}
\nu_{\rm{\mathell}}}{\rm{d}}\nu_{\rm{h}}},
\end{eqnarray}
\begin{eqnarray}
 \label{eq30}
X_j = \left\{
    \begin{array}{lc}
         \mu_{\nu_{\rm{\mathell}}} - 5 \sigma_{ \nu_{\rm{\mathell}}}, \quad{j}=1, \\
         f_{\rm{\mathell}} + \frac{\left(j-1\right)\left(f_{\rm{h}}
         -f_{\rm{\mathell}}\right)}{n}, \quad2\le j\le n+1,
    \end{array}
\right.
\end{eqnarray}
\begin{eqnarray}
 \label{eq31}
Y_i = \left\{
    \begin{array}{lc}
         f_{\rm{\mathell}} + \frac{\left(i-1\right)\left(f_{\rm{h}}
         -f_{\rm{\mathell}}\right)}{n}, \quad1\le i\le n,  \\
         \mu_{\nu_{\rm{h}}}+5\sigma_{\nu_{\rm{h}}}, \quad{i} = n+1.
    \end{array}
\right.
\end{eqnarray}

\begin{figure}[!htbp]
\centering
\includegraphics[width=1\textwidth]{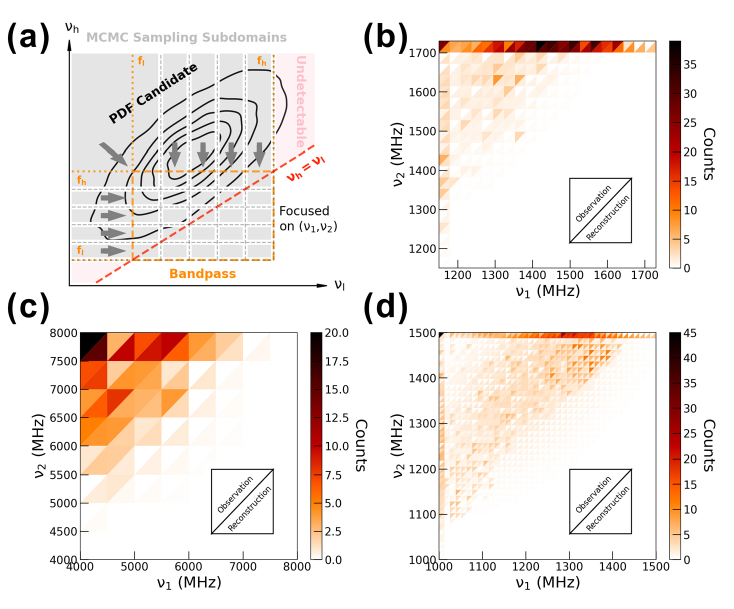}
\caption{Grid-based inverse modeling of 2D Gaussian
Distribution of $\left(\nu_{\rm{\mathell}},\nu_{\rm{h}}\right)$
for FRB 20121102A. $\nu_{\rm{\mathell}}$ and $\nu_{\rm{h}}$ denote
the proper lower and upper spectral boundaries of the burst,
respectively, while $\nu_1$ and $\nu_2$ are the
observed lower and upper spectral boundaries of the burst.
(a) A schematic illustration of how the
$\nu_{\rm{h}}$-$\nu_{\rm{\mathell}}$ plane is divided into subregions for
integrating the probability density function (PDF) of the intrinsic
$\left(\nu_{\rm{\mathell}},\nu_{\rm{h}}\right)$ distribution, computed
from a candidate trial parameter set. These subregions enable comparisons
between the trial theoretical counts and the observed counts of bursts on
the $\nu_{\rm{h}}$-$\nu_{\rm{\mathell}}$ plane, providing direct feedback
for MCMC sampling. (b) A direct comparison between the reconstructed
theoretical counts and the observed counts on the $\nu_2$-$\nu_1$ plane
for the Arecibo sample (observed between December 2015 and October
2016). The observed counts are derived  from the
Arecibo data \citep{3}, and the theoretical counts are computed from
the corresponding reconstructed $(\nu_{\rm{\mathell}},\nu_{\rm{h}})$
distribution shown in Figure \ref{Figure5}(a). In each subregion, the
top-left corner stands for the observed count, while the bottom-right
corner stands for the theoretical count. All counts are
color-coded, and the corresponding values are indicated by the color bar
on the right. (c) Same as panel (b), but for the GBT sample (observed on
August 26, 2017). The observed counts are from \citet{9}, and the
theoretical counts are from the reconstructed intrinsic
$\left(\nu_{\rm{\mathell}},\nu_{\rm{h}}\right)$ distribution shown in
Figure \ref{Figure7}(a). (d) Same as panel (b), but for the FAST sample
(observed between August 29 and October 29, 2019). The observed counts are
derived from the observed spectral boundaries that we extracted by
reprocessing the FAST data  \citep{1}. The theoretical counts are from the
reconstructed intrinsic $\left(\nu_{\rm{\mathell}},\nu_{\rm{h}}\right)$
distribution shown in Figure \ref{Figure8}(a).}
 \label{FigureD2}
\end{figure}

In Figure \ref{FigureD2}(a), which illustrates how the MCMC sampling
is performed to compare theoretical counts with observed counts
on the $\nu_{\rm{h}}$-$\nu_{\rm{\mathell}}$ plane,
we divide the $\nu_{\rm{h}}$-$\nu_{\rm{\mathell}}$ plane into
$n\times n$ grids as MCMC sampling subdomains. Here, the region
enclosed by orange dashed lines represents the operating band.
For the grids in the $j$-th column, both the left boundary
$X_j$ and the right boundary $X_{j+1}$ are obtained
using Equation (\ref{eq30}).
Similarly, for the grids in the $i$-th row, both the lower
boundary $Y_i$ and the upper boundary $Y_{i+1}$ are calculated
using Equation (\ref{eq31}). Note that due to the
operating-band cutoff in observations, the probability density
within the detectable region but outside the operating band region is
squeezed toward the boundaries of the operating band region (along the
direction indicated by the dark gray arrows), as displayed
intuitively in Figure \ref{FigureD2}(a). Therefore, to establish the
connection between the observed 2D distribution and the intrinsic
2D distribution of $\left(\nu_{\rm{\mathell}},\nu_{\rm{h}}\right)$,
when computing the theoretical counts, we need to extend those MCMC
sampling subdomains that lie near the orange dashed lines in the
opposite direction of the arrows; the resulting subdomains
are displayed as gray cells separated by gray dashed lines in
Figure \ref{FigureD2}(a).

For the extended MCMC sampling subdomains, as described in
Equation (\ref{eq30}), $X_j$ should be $-\infty$ when $j=1$,
but to avoid non-physical situations in practice, we
set $X_j = \mu_{\nu_{\rm{\mathell}}}-5\sigma_{\nu_{\rm{\mathell}}}$ for a
trial parameter set $\left(\mu_{\nu_{\rm{\mathell}}},
\sigma_{\nu_{\rm{\mathell}}}\right)$ of the $\left(\nu_{\rm{\mathell}},
\nu_{\rm{h}}\right)$ distribution.
Similarly, in Equation (\ref{eq31}), we set
$Y_{i=n+1}=\mu_{\nu_{\rm{h}}}+5\sigma_{\nu_{\rm{h}}}$ for a trial
parameter set $\left(\mu_{\nu_{\rm{h}}}, \sigma_{\nu_{\rm{h}}}\right)$.

$\lambda_{ij}$ in Equation (\ref{eq29}) represents the
theoretical count of data points within the grid of the $i$-th row
and $j$-th column for the PDF of the $\left(\nu_{\rm{\mathell}},
\nu_{\rm{h}}\right)$ distribution under a trial parameter set $\theta$.
Correspondingly, $k_{ij} = N_{ij}$ is the observed
count. $N = \sum_{i}\sum_{j} N_{ij}$ is the total count of
detected bursts in the sample whose intensity exceed the fluence
threshold that lie (at least partially) within the operating band,
corresponding to the region of $\left(
\nu_{\rm{\mathell}} < \nu_{\rm{h}}
\right)\land\left(\nu_{\rm{\mathell}} < f_{\rm{h}}
\right)\land\left(\nu_{\rm{h}} > f_{\rm{\mathell}} \right)$ on the
$\nu_{\rm{h}}$-$\nu_{\rm{\mathell}}$ plane in Figure \ref{FigureD2}(a). The
region of $\nu_{\rm{\mathell}} < \nu_{\rm{h}}$ on the
$\nu_{\rm{h}}$-$\nu_{\rm{\mathell}}$ plane corresponds to the
physically valid region, where both detectable and undetectable
events could exist. The region below the line of
$\nu_{\rm{h}}=\nu_{\rm{\mathell}}$ is the non-physical region,
which also needs to be considered to provide feedback to the PDF
candidate (i.e., the trial parameter set $\theta$ of the
$\left(\nu_{\rm{\mathell}}, \nu_{\rm{h}}\right)$ distribution) in order
to obtain a desired PDF that avoids this region
as much as possible. This implies that the theoretical count in
this region should be equal to the observed count, which is 0. It
is worth noting that, for the $n \times n$ MCMC sampling
subdomains, we can still empirically determine $n$ through $n =
\left\lceil\sqrt N\right\rceil$.

Using the above method, we can obtain the maximum likelihood
estimation of $\theta$ for the 2D Gaussian distribution of
$\left(\nu_{\rm{\mathell}}, \nu_{\rm{h}} \right)$. The derived
model parameters are shown in Figure \ref{Figure5}(b) for the Arecibo
sample, in Figure \ref{Figure7}(b) for the
GBT sample, and in Figure \ref{Figure8}(b)
for the FAST sample.

Note that, for the GBT bursts, the true peak frequency
$\nu_{\rm{p}}$ is available (rather than the observed value
$\nu_{\rm{p,obs}} = (\nu_1+\nu_2)/2$),
but $\nu_{\rm{h}} - \nu_{\rm{p}}$ and
$\nu_{\rm{p}} - \nu_{\rm{\mathell}}$ are usually not equal for the
in-band bursts. The reason is that the lower fluence of the
spectral wings is more affected by noise, causing different
degrees of signal masking in the noise on the two sides of the
spectral wings. This is especially true for weak bursts found
through machine learning in the GBT sample. To account for this
effect, adjustments were made to $\nu_1$ and $\nu_2$ based on
$\nu_{\rm{p}}$ to ensure integrality in the spectral frequency
range. Concretely, for a in-band burst in the GBT sample,
for the two parameters of $\nu_1$ and $\nu_2$, the one closer to
$\nu_{\rm{p}}$ was adjusted away from $\nu_{\rm{p}}$ (within the
range of the operating band) to maximize the symmetry of $\nu_1$
and $\nu_2$ with respect to $\nu_{\rm{p}}$. The model parameters
constrained from the adjusted sample are
broadly consistent with those from the original sample, with
relative differences of $\sim$5--15\% for most parameters; the
only notable difference is in $\rho_{\rm{GBT}}$ (changing from $-0.105$ for the
original sample to $0.189$ for the adjusted sample, though both values
indicate weak correlation; see Figure \ref{Figure7}(b) \& (c)).

For the FAST sample, the observed spectral boundaries $\nu_1$ and
$\nu_2$ were measured across the full 1--1.5 GHz operating band, with
higher precision within the central 1.05--1.45 GHz band. Outside this
central band, the measured boundaries are more subject to an uncertainty
of $\sim$20 MHz associated with the scintillation bandwidth. To account
for this, any $\nu_1$ or $\nu_2$ falling within 20 MHz of the operating
band edges ($f_{\rm{\mathell,FAST}}=1000$ MHz and $f_{\rm{h,FAST}}=1500$
MHz) was treated as band-chipped and replaced by the corresponding edge
value. The resulting sample, referred to as the original FAST sample, was
used to constrain the model parameters of the intrinsic
$\left(\nu_{\rm{\mathell}}, \nu_{\rm{h}} \right)$ distribution.
Additionally, considering that the central band of FAST offers more
reliable burst measurements, we constructed a truncated sample by
artificially clipping all $\nu_1$ and $\nu_2$ to the 1.05--1.45 GHz range,
mimicking observations taken with a narrower operating band. The
parameters inferred from this truncated sample are broadly consistent with
those from the original one, with relative differences of $\sim$1--15\%
for most parameters. The only notable difference lies in
$\rho_{\rm{FAST}}$, which changes from $0.561$ (original sample; Figure
\ref{Figure8}(b)) to $0.713$ (truncated sample; Figure \ref{Figure8}(c)),
though both values indicate a strong positive correlation.

To visually evaluate the
reconstruction model, the reconstructed theoretical count and the
observed count are directly compared in each MCMC sampling
subdomain on the $\nu_{\rm{h}}$-$\nu_{\rm{\mathell}}$ plane, as shown in
Figure \ref{FigureD2}(b) for the Arecibo sample, in Figure
\ref{FigureD2}(c) for the GBT sample, and in Figure
\ref{FigureD2}(d) for the FAST sample. These three figures
correspond to Figure \ref{Figure5}(a), Figure \ref{Figure7}(a), and Figure \ref{Figure8}(a),
respectively.

Additionally, such a grid-based counting approach, as employed
in the MCMC sampling process, can accommodate to some extent the variation
of the fluence threshold across bursts within the same sample, which may
in effect define the proper spectral boundaries $\nu_{\rm{\mathell}}$ and
$\nu_{\rm{h}}$ at slightly different monochromatic fluence ($F_\nu$)
levels and thus introduce some fluctuation in the observed counts within
the MCMC sampling subdomains on the $\nu_{\rm{h}}$-$\nu_{\rm{\mathell}}$
plane. In this regard, substituting Equation (\ref{eq5}) into  Equation
(\ref{eq3}), we have
\begin{eqnarray}
 \label{eq32}
F_{\nu,{\rm{thre}}} = \frac{F}{\sqrt{2\pi} \sigma_\nu}{\rm{e}}^{-
\frac{\left( \nu_{\rm{\mathell}} - \nu_{\rm{p}}
\right)^2}{2\sigma_\nu^2}},
\end{eqnarray}
where $\nu_{\rm{\mathell}}$ can also be replaced by $\nu_{\rm{h}}$. Across
the bursts, when $F_{\nu,{\rm{thre}}}$ changes by a mount of
$\Delta F_{\nu,{\rm{thre}}}$, we have
\begin{eqnarray}
 \label{eq33}
F_{\nu,{\rm{thre}}} + \Delta F_{\nu,{\rm{thre}}} = \frac{F}{\sqrt{2\pi}
\sigma_\nu}{\rm{e}}^{-\frac{\left(\nu_{\rm{\mathell}} + \Delta\nu_{\rm{\mathell}}
- \nu_{\rm{p}}\right)^2}{2\sigma_\nu^2}}.
\end{eqnarray}
Dividing Equation (\ref{eq33}) by Equation (\ref{eq32}), and
requiring $\Delta\nu_{\rm{\mathell}} \le \frac{\delta f}{n}$ (so that
the threshold-induced drift in $\nu_{\rm{\mathell}}$ never moves a
burst out of its original MCMC sampling grid), where
$\frac{\delta f}{n}$ is the grid spacing, we get
\begin{eqnarray}
\label{eq34}
\frac{\Delta F_{\nu,{\rm{thre}}}}{F_{\nu,{\rm{thre}}}} \le
\frac{\delta f}{2n}\frac{\delta\nu}{\sigma_\nu^2},
\end{eqnarray}
where we have taken the first-order Maclaurin approximation for
the exponential term and neglected other smaller terms. In this
way, we can estimate the maximum allowable relative change in
$F_{\nu,{\rm{thre}}}$ for grid-based counting, by considering the
sample conditions and the grid spacing.

\clearpage

\clearpage

\section{Transforming  $\left(\nu_{\rm{\mathell}},
\nu_{\MakeLowercase{\rm{h}}}\right)$ Distribution to
$\left(\nu_{\MakeLowercase{\rm{p}}},\delta\nu\right)$ Distribution}
\label{appendix E:Transforming}

\numberwithin{equation}{section}
\renewcommand{\theequation}{\Alph{section}\arabic{equation}}

Let $\left(x,y\right)$ be a 2D random variable with a joint
probability density of $f\left(x,y\right)$, and $\left(u,v\right)$
be a 2D random variable with a joint probability density of
$F\left(u,v\right)$. Assuming that $\left(x,y\right)$ and
$\left(u,v\right)$ are connected by the functions of
$u=u\left(x,y\right)$, $v=v\left(x,y\right)$ and the unique inverse
functions of $x=x\left(u,v\right)$, $y=y\left(u,v\right)$. Then
the Jacobian determinant of the transformation is
\begin{eqnarray}
 \label{eq35}
J = \frac{\partial \left(x,y\right)}{\partial \left(u,v\right)} =
\left|
\begin{matrix}\partial x/\partial u&\partial x/\partial
 v \\ \partial y/\partial u&\partial y/\partial v \\
\end{matrix} \right| = \left( \frac{\partial \left(u,v \right)}{ \partial
 \left(x,y\right)}\right)^{-1} = \left(\left|
\begin{matrix}\partial
u/\partial x&\partial v/\partial x\\\partial v/\partial x&\partial
v/\partial y\\
\end{matrix}\right|\right)^{-1}.
\end{eqnarray}
Here we have assumed that all the relevant partial derivatives are
continuous. If the Jacobian determinant is non-zero, then the
joint probability density $F\left(u,v\right)$ can be calculated as
\begin{eqnarray}
 \label{eq36}
F\left(u,v\right) = f\left(x\left(u,v\right), y\left(u,v\right)
\right) \left|J\right|.
\end{eqnarray}
In Equation (\ref{eq36}), it is not necessary for $x$ and $y$
to be mutually independent.

In this study, the distribution of
$\left(\nu_{\rm{\mathell}},\nu_{\rm{h}}\right)$ follows ${N}
\left( \nu_{\rm{\mathell}}, \nu_{\rm{h}};
\mu_{\nu_{\rm{\mathell}}}, \mu_{\nu_{\rm{h}}},
\sigma_{\nu_{\rm{\mathell}}}, \sigma_{\nu_{\rm{h}}}, \rho\right)$.
Also, the variable of $\left(\nu_{\rm{p}}, \delta\nu\right)$ is
connected with $\left(\nu_{\rm{\mathell}}, \nu_{\rm{h}}\right)$ as
$\nu_{\rm{p}}=\left(\nu_{\rm{\mathell}} + \nu_{\rm{h}}\right)/2$
and $\delta\nu = \nu_{\rm{h}} - \nu_{\rm{\mathell}}$, with the
corresponding unique inverse functions given by
$\nu_{\rm{\mathell}} = \nu_{\rm{p}} - \delta\nu/2$ and $\nu_{\rm{h}} =
\nu_{\rm{p}} + \delta\nu/2$. Then, according to Equations
(\ref{eq15}) and (\ref{eq36}), we can derive the joint
probability density of $\left(\nu_{\rm{p}}, \delta\nu\right)$ as
\begin{eqnarray}
 \label{eq37}
\begin{aligned}
f_N \left(\nu_{\rm{p}}, \delta\nu\right) & =
\frac{1}{2 \pi \sigma_{\nu_{\rm{\mathell}}} \sigma_{\nu_{\rm{h}}}
\sqrt{1 - \rho^2}} \exp \left\{-\frac{1}{2\left(1 - \rho^2 \right)}
\left[\frac{\left(\nu_{\rm{p}} - \delta \nu/2 -
\mu_{\nu_{\rm{\mathell}}} \right)^2}{\sigma_{\nu_{\rm{\mathell}}}^2} \right. \right.\\
& \quad\left.\left. - \frac{2 \rho \left(\nu_{\rm{p}} - \delta\nu/2
- \mu_{\nu_{\rm{\mathell}}}\right) \left(\nu_{\rm{p}} + \delta\nu/2
- \mu_{\nu_{\rm{h}}} \right)}{\sigma_{\nu_{\rm{\mathell}}}
\sigma_{\nu_{\rm{h}}}} + \frac{\left(\nu_{\rm{p}} + \delta\nu/2
- \mu_{\nu_{\rm{h}}}\right)^2}{\sigma_{\nu_{\rm{h}}}^2}\right]\right\}\ \\
& = \frac{1}{2\pi\sigma_{\nu_{\rm{\mathell}}}
 \sigma_{\nu_{\rm{h}}} \sqrt{1-\rho^2}}
\exp \left\{-\frac{\mathcal{A}\nu_{\rm{p}}^2 + \mathcal{B}
\left(\delta\nu\right)^2 + \mathcal{C}\nu_{\rm{p}} \delta \nu +
\mathcal{D} \nu_{\rm{p}} + \mathcal{E} \delta\nu + \mathcal{F}}{2
\left(1 - \rho^2 \right)} \right\}.
\end{aligned}
\end{eqnarray}
Here $\mathcal{A} = \frac{1}{\sigma_{\nu_{\rm{\mathell}}}^2} +
\frac{1}{\sigma_{\nu_{\rm{h}}}^2} -
\frac{2\rho}{\sigma_{\nu_{\rm{\mathell}}} \sigma_{\nu_{\rm{h}}}}$,
$\mathcal{B} = \frac{1}{4}
\left(\frac{1}{\sigma_{\nu_{\rm{\mathell}}}^2} +
\frac{1}{\sigma_{\nu_{\rm{h}}}^2} +
\frac{2\rho}{\sigma_{\nu_{\rm{\mathell}}}
\sigma_{\nu_{\rm{h}}}}\right)$, $\mathcal{C} =
\frac{1}{\sigma_{\nu_{\rm{h}}}^2} -
\frac{1}{\sigma_{\nu_{\rm{\mathell}}}^2}$, $\mathcal{D} = 2
\left[-\frac{\mu_{\nu_{\rm{\mathell}}}}{\sigma_{\nu_{\rm{\mathell}}}^2}
- \frac{\mu_{\nu_{\rm{h}}}}{\sigma_{\nu_{\rm{h}}}^2} +
\frac{\rho\left(\mu_{\nu_{\rm{\mathell}}} +
\mu_{\nu_{\rm{h}}}\right)}{\sigma_{\nu_{\rm{\mathell}}}
\sigma_{\nu_{\rm{h}}}} \right]$, $\mathcal{E} =
\frac{\mu_{\nu_{\rm{\mathell}}}}{\sigma_{\nu_{\rm{\mathell}}}^2} -
\frac{\mu_{\nu_{\rm{h}}}}{\sigma_{\nu_{\rm{h}}}^2} +
\frac{\rho\left(\mu_{\nu_{\rm{\mathell}}} -
\mu_{\nu_{\rm{h}}}\right)}{\sigma_{\nu_{\rm{\mathell}}}
\sigma_{\nu_{\rm{h}}}}$, and $\mathcal{F} =
\frac{\mu_{\nu_{\rm{\mathell}}}^2}{\sigma_{\nu_{\rm{\mathell}}}^2}
+ \frac{\mu_{\nu_{\rm{h}}}^2}{\sigma_{\nu_{\rm{h}}}^2} -
\frac{2\rho\mu_{\nu_{\rm{\mathell}}}
\mu_{\nu_{\rm{h}}}}{\sigma_{\nu_{\rm{\mathell}}}
\sigma_{\nu_{\rm{h}}}}$. Since $\left(\nu_{\rm{p}},
\delta\nu\right)$ is a linear transformation of
$\left(\nu_{\rm{\mathell}},\nu_{\rm{h}}\right)$, it should also
follow a 2D Gaussian distribution, denoted as
$f_N\left(\nu_{\rm{p}}, \delta\nu\right) \sim{N}
\left(\nu_{\rm{p}}, \delta\nu; \mu_{\nu_{\rm{p}}},
\mu_{\delta\nu}, \right.$ $\left.\sigma_{\nu_{\rm{p}}},
\sigma_{\delta\nu}, k\right)$. Integrating the PDF of a 2D
Gaussian distribution over one of the two random variables yields
the PDF of a 1D Gaussian distribution for the other variable, i.e.,
$f_N (\delta \nu) \sim{N} \left(\delta\nu; \mu_{\delta\nu},
\sigma_{\delta\nu} \right)$ and $f_N (\nu_{\rm{p}}) \sim{N}
\left(\nu_{\rm{p}}; \mu_{\nu_{\rm{p}}}, \sigma_{\nu_{\rm{p}}}
\right)$. As a result, we have
\begin{eqnarray}
 \label{eq38}
f_N \left(\delta\nu\right) = \int_{-\infty}^{\infty}{f_N
\left(\nu_{\rm{p}}, \delta \nu \right){\rm{d}}\nu_{\rm{p}}} =
\frac{1}{\sqrt{2\pi} \sigma_{\delta\nu}}{\rm{e}}^{ - \frac{\left(
\delta \nu - \mu_{\delta\nu} \right)^2}{2 \sigma_{\delta\nu}^2}},
\end{eqnarray}
\begin{eqnarray}
 \label{eq39}
f_N \left( \nu_{\rm{p}}\right) = \int_{-\infty}^{\infty}{f_N
\left( \nu_{\rm{p}}, \ \delta \nu \right){\rm{d}} \delta \nu} =
\frac{1}{\sqrt{2\pi} \sigma_{\nu_{\rm{p}}}}{\rm{e}}^{ - \frac{
\left(\nu_{\rm{p}} - \mu_{\nu_{\rm{p}}}
\right)^2}{2\sigma_{\nu_{\rm{p}}}^2}}.
\end{eqnarray}
This can help us determine the model parameters $\left(\mu_{\nu_{\rm{p}}},
\mu_{\delta\nu}, \sigma_{\nu_{\rm{p}}}, \sigma_{\delta\nu},
k\right)$ in the expression
of $N\left(\nu_{\rm{p}}, \delta\nu; \mu_{\nu_{\rm{p}}},
\mu_{\delta\nu}, \sigma_{\nu_{\rm{p}}}, \sigma_{\delta\nu},
k\right)$. We can separate the variable $\nu_{\rm{p}}$ in the expression
of $f_N(\nu_{\rm{p}},\delta\nu)$ by substituting the final expression
from Equation (\ref{eq37}) into Equation (\ref{eq38}):
\begin{eqnarray}
\label{eq40} \frac{1}{\sqrt{2 \pi}
\sqrt{\sigma_{\nu_{\rm{\mathell}}}^2 + \sigma_{\nu_{\rm{h}}}^2 - 2
\rho \sigma_{\nu_{\rm{\mathell}}} \sigma_{\nu_{\rm{h}}}}}
\exp{\left[-\frac{\left(\mathcal{B} -
\frac{\mathcal{C}^2}{4\mathcal{A}} \right) \left(\delta \nu
\right)^2 + \left(\mathcal{E} - \frac{\mathcal{CD}}{2 \mathcal{A}}
\right) \delta \nu + \left(\mathcal{F} - \frac{\mathcal{D}^2}{4
\mathcal{A}} \right)}{2 \left(1 - \rho^2 \right)} \right]} =
\frac{1}{\sqrt{2\pi} \sigma_{\delta\nu}}{\rm{e}}^{ -
\frac{\left(\delta \nu - \mu_{\delta\nu} \right)^2}{2
\sigma_{\delta\nu}^2}}.
\end{eqnarray}
Here, utilizing the commonly-used method of factorization in polynomial
division, we can easily derive $\mathcal{B} -
\frac{\mathcal{C}^2}{4\mathcal{A}} =
\frac{1-\rho^2}{\sigma_{\nu_{\rm{\mathell}}}^2 +
\sigma_{\nu_{\rm{h}}}^2 - 2 \rho \sigma_{\nu_{\rm{\mathell}}}
\sigma_{\nu_{\rm{h}}}}$, $\mathcal{E} - \frac{\mathcal{CD}}{2
\mathcal{A}} = - \frac{2 \left( \mu_{\nu_{\rm{h}}} -
\mu_{\nu_{\rm{\mathell}}} \right) \left( 1 - \rho^2
\right)}{\sigma_{\nu_{\rm{\mathell}}}^2 + \sigma_{\nu_{\rm{h}}}^2
- 2 \rho \sigma_{\nu_{\rm{\mathell}}} \sigma_{\nu_{\rm{h}}}}$ and
$\mathcal{F} - \frac{\mathcal{D}^2}{4 \mathcal{A}} =
\frac{\left(\mu_{\nu_{\rm{h}}} - \mu_{\nu_{\rm{\mathell}}}
\right)^2 \left(1 - \rho^2 \right)}{\sigma_{\nu_{\rm{\mathell}}}^2
+ \sigma_{\nu_{\rm{h}}}^2 - 2 \rho \sigma_{\nu_{\rm{\mathell}}}
\sigma_{\nu_{\rm{h}}}}$. Thus, Equation (\ref{eq40}) can be
simplified as
\begin{eqnarray}
 \label{eq41}
\frac{1}{\sqrt{2\pi} \sqrt{\sigma_{\nu_{\rm{\mathell}}}^2 +
\sigma_{\nu_{\rm{h}}}^2 - 2 \rho \sigma_{\nu_{\rm{\mathell}}}
\sigma_{\nu_{\rm{h}}}}} \exp{\left\{ - \frac{\left[\delta\nu -
\left(\mu_{\nu_{\rm{h}}} - \mu_{\nu_{\rm{\mathell}}} \right)
\right]^2}{2 \left( \sigma_{\nu_{\rm{\mathell}}}^2 +
\sigma_{\nu_{\rm{h}}}^2 - 2 \rho \sigma_{\nu_{\rm{\mathell}}}
\sigma_{\nu_{\rm{h}}} \right)} \right\}} = \frac{1}{\sqrt{2\pi}
\sigma_{\delta\nu}}{\rm{e}}^{-\frac{\left(\delta\nu -
\mu_{\delta\nu} \right)^2}{2 \sigma_{\delta\nu}^2}}.
\end{eqnarray}
By matching the terms on both sides of Equation (\ref{eq41}),
we get $\mu_{\delta\nu} = \mu_{\nu_{\rm{h}}} -
\mu_{\nu_{\rm{\mathell}}}$ and $\sigma_{\delta\nu} =
\sqrt{\sigma_{\nu_{\rm{\mathell}}}^2 + \sigma_{\nu_{\rm{h}}}^2 - 2
\rho \sigma_{\nu_{\rm{\mathell}}} \sigma_{\nu_{\rm{h}}}}$.

Similarly, by separating the variable $\delta \nu$
in the expression of $f_N \left(\nu_{\rm{p}}, \delta \nu
\right)$ through substituting the final expression from Equation
(\ref{eq37}) into Equation (\ref{eq39}), we ultimately get
\begin{eqnarray}
 \label{eq42}
\frac{1}{\sqrt{2 \pi} \sqrt{\left( \sigma_{\nu_{\rm{\mathell}}}^2
+ \sigma_{\nu_{\rm{h}}}^2 + 2\rho\sigma_{\nu_{\rm{\mathell}}}
\sigma_{\nu_{\rm{h}}} \right)/4}} \exp{\left\{ -
\frac{\left(\nu_{\rm{p}} - \frac{\mu_{\nu_{\rm{\mathell}}} +
\mu_{\nu_{\rm{h}}}}{2} \right)^2}{2 \left[\left(
\sigma_{\nu_{\rm{\mathell}}}^2 + \sigma_{\nu_{\rm{h}}}^2 + 2 \rho
\sigma_{\nu_{\rm{\mathell}}} \sigma_{\nu_{\rm{h}}}\right) / 4
\right]} \right\}} = \frac{1}{\sqrt{2 \pi}
\sigma_{\nu_{\rm{p}}}}{\rm{e}}^{ - \frac{\left(\nu_{\rm{p}} -
\mu_{\nu_{\rm{p}}}\right)^2}{2 \sigma_{\nu_{\rm{p}}}^2}}.
\end{eqnarray}
Hence, we have $\mu_{\nu_{\rm{p}}} =
\frac{\mu_{\nu_{\rm{\mathell}}} + \mu_{\nu_{\rm{h}}}}{2}$ and
$\sigma_{\nu_{\rm{p}}} =
\sqrt{\left(\sigma_{\nu_{\rm{\mathell}}}^2 +
\sigma_{\nu_{\rm{h}}}^2 + 2 \rho \sigma_{\nu_{\rm{\mathell}}}
\sigma_{\nu_{\rm{h}}}\right)/4}$. By matching the exponential
terms of Equations (\ref{eq37}) and (\ref{eq16}), we get
$\mathcal{A} = \frac{1-\rho^2}{1-k^2}
\frac{1}{\sigma_{\nu_{\rm{p}}}^2}$ and $\mathcal{C} = \frac{1 -
\rho^2}{1 - k^2} \left( - \frac{2k}{\sigma_{\nu_{\rm{p}}}
\sigma_{\delta\nu}} \right)$. Take the ratio of $\mathcal{C}$ to
$\mathcal{A}$ and substitute the definitions of $\mathcal{A} =
\frac{1}{\sigma_{\nu_{\rm{\mathell}}}^2} +
\frac{1}{\sigma_{\nu_{\rm{h}}}^2} -
\frac{2\rho}{\sigma_{\nu_{\rm{\mathell}}} \sigma_{\nu_{\rm{h}}}}$
and $\mathcal{C} = \frac{1}{\sigma_{\nu_{\rm{h}}}^2} -
\frac{1}{\sigma_{\nu_{\rm{\mathell}}}^2}$ into the resulting
equation, combining the obtained expressions of
$\sigma_{\nu_{\rm{p}}} =
\sqrt{\left(\sigma_{\nu_{\rm{\mathell}}}^2 +
\sigma_{\nu_{\rm{h}}}^2 + 2 \rho \sigma_{\nu_{\rm{\mathell}}}
\sigma_{\nu_{\rm{h}}} \right)/4}$ and $\sigma_{\delta\nu} =
\sqrt{\sigma_{\nu_{\rm{\mathell}}}^2 + \sigma_{\nu_{\rm{h}}}^2 - 2
\rho\sigma_{\nu_{\rm{\mathell}}} \sigma_{\nu_{\rm{h}}}}$, we can
derive
\begin{eqnarray}
 \label{eq43}
k = - \frac{\mathcal{C} \sigma_{\delta\nu}}{2 \mathcal{A}
\sigma_{\nu_{\rm{p}}}} = \frac{\sigma_{\nu_{\rm{h}}}^2 -
\sigma_{\nu_{\rm{\mathell}}}^2}{\sqrt{\left(\sigma_{\nu_{\rm{\mathell}}}^2
+ \sigma_{\nu_{\rm{h}}}^2\right)^2 - 4 \rho^2
\sigma_{\nu_{\rm{\mathell}}}^2 \sigma_{\nu_{\rm{h}}}^2}}.
\end{eqnarray}

In this way, based on the model parameters of the
$\left(\nu_{\rm{\mathell}},\nu_{\rm{h}}\right)$ distribution, we
can get the expressions for all the five model parameters of
$\left(\nu_{\rm{p}}, \delta\nu\right)$ distribution.

\section{Estimation of Integrated Fluence for Band-chipped Bursts}
\label{appendix F:Estimation of Integrated Fluence for Band-chipped Bursts}

Given the proper spectral boundaries $\nu_{\rm{\mathell}}$
and $\nu_{\rm{h}}$, the values of
$\nu_1$ and $\nu_2$ can be determined through Equations
(\ref{eq1}) and (\ref{eq2}) respectively, which in turn
yield $\delta\nu_{\rm{obs}} = \nu_2 - \nu_1$.
Also, $\nu_{\rm{p}}$ can be
obtained using Equation (\ref{eq4}). Furthermore, combining
$F_{\nu,{\rm{obs}}}^{\rm{equiv}}$, Equation (\ref{eq7}) can
be solved, and the solution can be substituted into Equation
(\ref{eq8}) to obtain the burst's integrated fluence
$F\left(\nu_{\rm{\mathell}},\nu_{\rm{h}}, F_{\nu,
{\rm{obs}}}^{\rm{equiv}}, F_{\nu,{\rm{thre}}}\right)$.

In the case of a band-chipped burst, condidering
$\nu_{\rm{\mathell}}$ and/or $\nu_{\rm{h}}$ are unknown, we can,
for all possible combinations of
$\left(\nu_{\rm{\mathell}},\nu_{\rm{h}}\right)$ that can lead to
the observed spectral boundaries $\left(\nu_1,\nu_2\right)$,
compute the corresponding integrated fluence ($F$) values and
convolve them with the reconstructed PDF $f_N
\left(\nu_{\rm{\mathell}},\nu_{\rm{h}}\right)$ (i.e., Equation
(\ref{eq15})) to obtain the expected value of the integrated
fluence, $E\left(F\right)$ (i.e., Equation \ref{eq17}). Using the
expectation value as an estimator is a well-established practice,
which is routinely adopted in cases where a probabilistic
description is necessary. For instance, in equilibrium statistical
mechanics, macroscopic observables are ensemble averages (i.e.,
expectation values) of microscopic quantities, and in quantum
mechanics the predicted result of measuring an observable is its
expectation value in the given state. We therefore take
$E\left(F\right)$ as the estimate of the intrinsic burst energy
for band-chipped bursts.

Note that band-chipped bursts are classified into three categories
based on their positions on the $\nu_{\rm{h}}$-$\nu_{\rm{\mathell}}$
plane: lower-chipped ($\nu_{\rm{\mathell}} < f_{\rm{\mathell}} < \nu_{\rm{h}} < f_{\rm{h}}$), upper-chipped
($f_{\rm{\mathell}} < \nu_{\rm{\mathell}} < f_{\rm{h}} < \nu_{\rm{h}}$), and over-band ($\nu_{\rm{\mathell}} < f_{\rm{\mathell}} <
f_{\rm{h}} < \nu_{\rm{h}}$), corresponding to the three conditions in Equation
(\ref{eq17})). For lower-chipped or upper-chipped bursts, since one of
$\nu_{\rm{\mathell}}$ and $\nu_{\rm{h}}$ is known, we only need to perform
the convolution over the other variable.

Additionally, the three
denominator terms in Equation (\ref{eq17}) represent the conditional
probabilities guided by the potential distribution domains of possible
combinations of $\left(\nu_{\rm{\mathell}},\nu_{\rm{h}}\right)$ under three different
conditions on the $\nu_{\rm{h}}$-$\nu_{\rm{\mathell}}$ plane.

\section{Table of Observed and Reconstructed Burst Properties for Arecibo Bursts}
\label{appendix G:Table of Observed and Reconstructed Burst Properties for Arecibo Bursts}

For the 478 bursts detected with the Arecibo telescope from FRB
20121102A between December 2015 and October 2016, as reported in
\citet{3}, we present the observed and derived quantities analyzed in this
work in Table \ref{Table1:Arecibo Sample}.

\begin{deluxetable*}{ccccccccccccc}[h!]
\renewcommand{\arraystretch}{1.8}
\tabletypesize{\scriptsize} \tabcolsep=3.5pt
\tablecaption{
Observed and Reconstructed Burst Properties of FRB 20121102A Detected by Arecibo
\label{Table1:Arecibo Sample}
}
\tablehead{
\colhead{ID$^{\rm a}$} & \colhead{$\nu_1$} & \colhead{$\nu_2$} & \colhead{$\delta\nu_{\rm obs}$} & \colhead{$W_{\rm obs}^{\rm equiv}$} & \colhead{$F_{\nu,{\rm obs}}^{\rm equiv}$} & \colhead{$E_{\rm obs}$$^{\rm b}$} & \colhead{$\nu_{\rm p,obs}$$^{\rm c}$} & \colhead{$F_{\nu{\rm ,thre}}$$^{\rm d}$} & \colhead{$\sigma_\nu$} & \colhead{$F$} & \colhead{$E$$^{\rm e}$} & \colhead{Type$^{\rm f}$} \\
\colhead{} & \colhead{(MHz)} & \colhead{(MHz)} & \colhead{(MHz)} & \colhead{(ms)} & \colhead{(Jy ms)} & \colhead{($10^{37}$ erg)} & \colhead{(MHz)} & \colhead{(Jy ms)} & \colhead{(MHz)} & \colhead{(Jy ms MHz)} & \colhead{($10^{37}$ erg)} & \colhead{}
}
\startdata
1  & 1184 & 1730 & 546 & 0.68 & 0.08 & 3.95 & 1457 & 0.024 & -- & 51.88 & 4.69 & upper-chipped \\
2  & 1349 & 1730 & 381 & 4.7  & 0.38 & 13.08 & 1539.5 & 0.062 & -- & 190.84 & 17.24 & upper-chipped \\
3  & 1184 & 1730 & 546 & 2.27 & 0.29 & 14.30 & 1457 & 0.043 & -- & 174.85 & 15.79 & upper-chipped \\
4  & 1425 & 1730 & 305 & 2.2  & 0.11 & 3.03 & 1577.5 & 0.042 & -- & 60.05 & 5.42 & upper-chipped \\
5  & 1311 & 1730 & 419 & 3.4  & 0.27 & 10.22 & 1520.5 & 0.053 & -- & 142.43 & 12.86 & upper-chipped \\
6  & 1463 & 1730 & 267 & 2.0  & 0.19 & 4.58 & 1596.5 & 0.040 & -- & 99.50 & 8.99 & upper-chipped \\
7  & 1222 & 1730 & 508 & 6.9  & 0.38 & 17.44 & 1476 & 0.075 & -- & 223.01 & 20.14 & upper-chipped \\
8  & 1399 & 1730 & 331 & 2.7  & 0.30 & 8.97 & 1564.5 & 0.047 & -- & 147.67 & 13.34 & upper-chipped \\
9  & 1513 & 1730 & 217 & 1.1  & 0.09 & 1.76 & 1621.5 & 0.030 & -- & 52.17 & 4.71 & upper-chipped \\
10 & 1513 & 1730 & 217 & 0.8  & 0.10 & 1.96 & 1621.5 & 0.025 & -- & 58.71 & 5.30 & upper-chipped \\
\enddata
\tablecomments{
The data were obtained with the Arecibo telescope between December 2015
and October 2016, as reported in \citet{3}. Columns (1)--(13): burst ID;
observed lower spectral boundary; observed upper spectral boundary;
observed bandwidth; boxcar-equivalent pulse width; observed
frequency-equivalent fluence; observed energy; observed peak frequency;
burst-specific fluence threshold; Gaussian spectral width of the intrinsic
spectrum; intrinsic integrated fluence; intrinsic energy; burst
classification. Columns (2)--(7) can be directly adopted from \citet{3};
columns (8)--(13) list the quantities derived in this work. A dash (--)
indicates that the quantity is not available (i.e., $\sigma_\nu$, $F$, and
$E$ for discarded bursts; $\sigma_\nu$ for band-chipped bursts).  \\
$^{\rm a}$ The burst IDs are the same as those in \citet{1}.  \\
$^{\rm b}$ Recalculated from $\delta\nu_{\rm obs}$ and $F_{\nu,{\rm
obs}}^{\rm equiv}$; the values are consistent with those in \citet{3}.  \\
$^{\rm c}$ Calculated as $\nu_{\rm p,obs}=\left(\nu_1+\nu_2\right)/2$.
For in-band bursts, $\nu_{\rm p,obs}$ equals the true peak frequency
$\nu_{\rm p}$.  \\
$^{\rm d}$ Calculated as $F_{\nu,{\rm thre}}=0.057\sqrt{W_{\rm obs}^{\rm
equiv}/4\,{\rm ms}}$ Jy ms, scaled from the reference fluence sensitivity
of $\sim$0.057 Jy ms given in \citet{3} for a typical pulse
width of 4 ms.  \\
$^{\rm e}$ For band-chipped bursts (lower-chipped, upper-chipped, and
over-band), the intrinsic energy is the expected value $E\left(F\right)$
obtained from the population-based reconstruction (Section
\ref{sec4:Operating-band Cutoff}).  \\
$^{\rm f}$ Classification of the detected bursts: in-band, lower-chipped,
upper-chipped, over-band, and discarded. The discarded category indicates
bursts with $F_{\nu,{\rm obs}}^{\rm equiv}$ below $F_{\nu,{\rm thre}}$
that were excluded from the analysis.  \\
\\
(This table is available in its entirety in machine-readable form online.)
}
\end{deluxetable*}

\clearpage

\section{Table of Original and Adjusted Frequency Quantities for GBT Bursts}
\label{appendix H:Table of Original and Frequency Quantities for GBT Bursts}

For the 93 bursts detected with GBT from FRB 20121102A on August 26,
2017, as reported in \citet{9}, we present the original and adjusted
frequency quantities analyzed in this work in Table
\ref{Table2:GBT Sample}. The original observed spectral boundaries $\nu_1$
and $\nu_2$ were often asymmetric with respect to the true peak frequency
$\nu_{\rm p}$, because many of these bursts were identified via
machine-learning methods \citep{9}. To
preserve the expected symmetry of the intrinsic Gaussian spectrum about
$\nu_{\rm p}$, we adjusted these boundaries following the procedure
detailed Appendix
\ref{appendix D:Reconstructing the 2D Gaussian Distribution}. These
quantities, together with the burst classification derived in this work,
are listed below.

\begin{deluxetable*}{ccccccc}[h!]
\renewcommand{\arraystretch}{1.8}
\tabletypesize{\scriptsize} \tabcolsep=3.5pt
\tablecaption{
Original and Adjusted Frequency Quantities of FRB 20121102A Detected by GBT
\label{Table2:GBT Sample}
}
\tablehead{
\colhead{ID$^{\rm a}$} & \colhead{$\nu_1$} & \colhead{$\nu_2$} & \colhead{$\nu_{\rm p}$$^{\rm b}$} & \colhead{$\nu_1^{\prime}$$^{\rm c}$} & \colhead{$\nu_2^{\prime}$$^{\rm c}$} & \colhead{Type$^{\rm d}$} \\
\colhead{} & \colhead{(MHz)} & \colhead{(MHz)} & \colhead{(MHz)} & \colhead{(MHz)} & \colhead{(MHz)} & \colhead{}
}
\startdata
1  & 4100 & 8000 & 7020 & 4100 & 8000 & upper-chipped \\
2  & 5600 & 7900 & 7050 & 5600 & 8000 & upper-chipped \\
3  & 5100 & 7500 & 7020 & 5100 & 8000 & upper-chipped \\
4  & 5000 & 7500 & 6170 & 4840 & 7500 & in-band \\
5  & 5000 & 7500 & 5560 & 4000 & 7500 & lower-chipped \\
6  & 5000 & 6400 & 5470 & 4540 & 6400 & in-band \\
7  & 5100 & 7500 & 7020 & 5100 & 8000 & upper-chipped \\
8  & 4000 & 7900 & 5550 & 4000 & 7900 & lower-chipped \\
9  & 4000 & 6900 & 5450 & 4000 & 6900 & lower-chipped \\
10 & 4500 & 7500 & 5440 & 4000 & 7500 & lower-chipped \\
\enddata
\tablecomments{
The data were obtained with GBT on August 26, 2017, as reported in
\citet{9}. Columns (1)--(7): burst ID; original observed lower spectral
boundary; original observed upper spectral boundary; true peak frequency;
adjusted lower spectral boundary; adjusted upper spectral boundary; burst
classification. Columns (2)--(4) are adopted from \citet{9}; columns
(5)--(7) are the quantities derived in this work.  \\
$^{\rm a}$ The burst IDs are the same as those in \citet{9}.  \\
$^{\rm b}$ True peak frequency directly provided in \citet{9}.  \\
$^{\rm c}$ Adjusted spectral boundaries that actually serve as the
observed spectral boundaries in our analysis, obtained by symmetrizing
$\nu_1$ and $\nu_2$ with respect to $\nu_{\rm p}$: $\nu_1^{\prime} =
\max[f_{\rm{\mathell,GBT}}, \nu_{\rm{p}} - \max(\nu_{\rm{p}} - \nu_1,
\nu_2 - \nu_{\rm{p}})]$, $\nu_2^{\prime} = \min[f_{\rm{h,GBT}},
\nu_{\rm{p}} + \max(\nu_{\rm{p}} - \nu_1,\nu_2 - \nu_{\rm{p}})]$.  \\
$^{\rm d}$ Classification of the detected bursts: in-band, lower-chipped,
upper-chipped, and over-band. The latter three types are collectively
referred to as band-chipped bursts.  \\
\\
(This table is available in its entirety in machine-readable form online.)
}
\end{deluxetable*}

\clearpage

\section{Table of Observed and Reconstructed Burst Properties for FAST Bursts}
\label{appendix I:Table of Observed and Reconstructed Burst Properties for FAST Bursts}

For the 1652 bursts detected with FAST from FRB 20121102A between
29 August and 29 October 2019, as reported in \citet{1}, we present the
observed and derived quantities analyzed in this work in Table
\ref{Table3:FAST Sample}. The observed lower and upper spectral boundaries
$\nu_1$ and $\nu_2$ were extracted from our reprocessed data and corrected
for band-edge measurement uncertainties associated with the $\sim$20 MHz
scintillation bandwidth, as described in Appendix
\ref{appendix D:Reconstructing the 2D Gaussian Distribution}.

\begin{deluxetable*}{ccccccccccccc}[h!]
\renewcommand{\arraystretch}{1.8}
\tabletypesize{\scriptsize} \tabcolsep=3.5pt
\tablecaption{
Observed and Reconstructed Burst Properties of FRB 20121102A Detected by FAST
\label{Table3:FAST Sample}
}
\tablehead{
\colhead{ID$^{\rm a}$} & \colhead{$W_{\rm obs}^{\rm equiv}$$^{\rm b}$} & \colhead{$F_{\nu,{\rm obs}}^{\rm equiv}$$^{\rm b}$} & \colhead{$\nu_1$$^{\rm c}$} & \colhead{$\nu_2$$^{\rm c}$} & \colhead{$\delta\nu_{\rm obs}$$^{\rm d}$} & \colhead{$\nu_{\rm p,obs}$$^{\rm e}$} & \colhead{$E_{\rm obs}$$^{\rm f}$} & \colhead{$F_{\nu{\rm ,thre}}$$^{\rm g}$} & \colhead{$\sigma_\nu$} & \colhead{$F$} & \colhead{$E$$^{\rm h}$} & \colhead{Type$^{\rm i}$} \\
\colhead{} & \colhead{(ms)} & \colhead{(Jy ms)} & \colhead{(MHz)} & \colhead{(MHz)} & \colhead{(MHz)} & \colhead{(MHz)} & \colhead{($10^{37}$ erg)} & \colhead{(Jy ms)} & \colhead{(MHz)} & \colhead{(Jy ms MHz)} & \colhead{($10^{37}$ erg)} & \colhead{}
}
\startdata
1  & 6.13 & 0.079 & 1069 & 1143 & 74 & 1106 & 3.55 & 0.037 & 14.2 & 39.72 & 3.59 & in-band \\
2  & 4.00 & 0.113 & 1253 & 1469 & 216 & 1361 & 5.10 & 0.030 & 45.4 & 57.51 & 5.19 & in-band \\
3  & 2.83 & 0.761 & 1027 & 1468 & 442 & 1248 & 34.39 & 0.025 & 74.5 & 381.86 & 34.49 & in-band \\
4  & 6.35 & 0.045 & 1350 & 1428 & 78 & 1389 & 2.01 & 0.038 & 17.0 & 22.73 & 2.05 & in-band \\
5  & 8.51 & 0.053 & 1350 & 1413 & 64 & 1381 & 2.41 & 0.044 & 13.2 & 27.09 & 2.45 & in-band \\
6  & 4.25 & 0.019 & 1376 & 1468 & 92 & 1422 & 0.86 & 0.031 & 25.4 & 10.22 & 0.92 & in-band \\
7  & 5.49 & 0.069 & 1245 & 1391 & 146 & 1318 & 3.09 & 0.035 & 32.6 & 35.15 & 3.17 & in-band \\
8  & 7.99 & 0.063 & 1374 & 1433 & 59 & 1403 & 2.87 & 0.042 & 11.6 & 32.09 & 2.90 & in-band \\
9  & 6.52 & 0.058 & 1346 & 1500 & 154 & 1423 & 2.64 & 0.038 & --   & 42.97 & 3.88 & upper-chipped \\
10 & 3.50 & 0.074 & 1073 & 1175 & 102 & 1124 & 3.33 & 0.028 & 19.9 & 37.22 & 3.36 & in-band \\
\enddata
\tablecomments{
The data were obtained with FAST between 29 August and 29 October 2019,
as reported in \citet{1}. Columns (1)--(13): burst ID; boxcar-equivalent
pulse width; observed frequency-equivalent fluence; observed lower
spectral boundary; observed upper spectral boundary; observed bandwidth;
observed peak frequency; observed energy; burst-specific fluence
threshold; Gaussian spectral width of the intrinsic spectrum; intrinsic
integrated fluence; intrinsic energy; burst classification. Columns
(2)--(3) are adopted from \citet{1}; columns (4)--(13) list the quantities
newly obtained in this work. A dash (--) indicates that the quantity is
not available (i.e., $\sigma_\nu$, $F$, and $E$ for discarded bursts;
$\sigma_\nu$ for band-chipped bursts).  \\
$^{\rm a}$ The burst IDs are the same as those in \citet{1}.  \\
$^{\rm b}$ These quantities remain unchanged in our reprocessed data.  \\
$^{\rm c}$ We reprocessed the FAST data to precisely extract
$\nu_1$ and $\nu_2$, and corrected them for band-edge measurement
uncertainties (see Appendix \ref{appendix D:Reconstructing the 2D Gaussian
Distribution}).  \\
$^{\rm d}$ Recalculated from the $\nu_1$ and $\nu_2$ that we
extracted; the values are broadly comparable to those in
\citet{1}, with differences arising from the more precise
temporal windowing applied in our spectral extraction. \\
$^{\rm e}$ Calculated as $\nu_{\rm p,obs}=\left(\nu_1+\nu_2\right)/2$.
For in-band bursts, $\nu_{\rm p,obs}$ serves as the true peak frequency
$\nu_{\rm p}$.  \\
$^{\rm f}$ Calculated from $\delta\nu_{\rm obs}$ and $F_{\nu,{\rm
obs}}^{\rm equiv}$.  \\
$^{\rm g}$ Calculated as $F_{\nu,{\rm thre}}=0.015\sqrt{W_{\rm obs}^{\rm
equiv}/1\,{\rm ms}}$ Jy ms, scaled from the reference fluence sensitivity
of 0.015 Jy ms adopted for the FAST sample \citep{1}.  \\
$^{\rm h}$ For band-chipped bursts (lower-chipped, upper-chipped, and
over-band), the intrinsic energy is the expected value $E\left(F\right)$
obtained from the population-based reconstruction (Section
\ref{sec4:Operating-band Cutoff}). This estimate of the intrinsic
energy differs from the alternative estimate obtained by replacing
the observed bandwidth with the telescope's center frequency, as
adopted in \citet{1}.  \\
$^{\rm i}$ Classification of the detected bursts: in-band, lower-chipped,
upper-chipped, over-band, and discarded. The discarded category
indicates bursts with $F_{\nu,{\rm obs}}^{\rm equiv} \times
500\,\rm{MHz}/\delta\nu_{\rm obs}$ below
$F_{\nu,{\rm thre}}$ that were excluded from the analysis. In
\citet{1}, $F_{\nu,{\rm obs}}^{\rm equiv}$ was computed by
averaging the observed integrated fluence over the full 500 MHz FAST
operating band; multiplying by $500\,\rm{MHz}/\delta\nu_{\rm obs}$
converts it to the spectral frequency-equivalent fluence, which is the
quantity directly compared with the burst-specific fluence threshold.  \\
\\
(This table is available in its entirety in machine-readable form online.)
}
\end{deluxetable*}


\bibliography{Ref}{}
\bibliographystyle{aasjournal}

\end{CJK*}

\end{document}